# Fundamental Limits of Communications in Interference Networks

## Part I: Basic Structures

Reza K. Farsani[1]

Email: reza_khosravi@alum.sharif.ir

***Abstract:*** **In these series of multi-part papers, a systematic study of fundamental limits of communications in interference networks is established. Here, *interference network* is referred to as a general single-hop communication scenario with arbitrary number of transmitters and receivers, and also arbitrary distribution of messages among transmitters and receivers. It is shown that the information flow in such networks follows similar derivations from many aspects.**

**This systematic study is launched by considering the basic building blocks in Part I. The Multiple Access Channel (MAC), the Broadcast Channel (BC), the Classical Interference Channel (CIC) and the Cognitive Radio Channel (CRC) are proposed as the main building blocks for all interference networks. First, a brief review of existing results regarding these basic structures is presented. New observations are also presented in this regard. Specifically, it is shown that the well-known strong interference conditions for the two-user CIC do not change if the inputs are dependent. Next, new capacity outer bounds are established for the basic structures with two receivers. These outer bounds are all derived based on a unified framework. By using the derived outer bounds, some new capacity results are proved for the CIC and the CRC; a *mixed interference regime* is identified for the two-user CIC where the sum-rate capacity is established. This result contains the previously obtained result regarding the Gaussian mixed interference channel as a special case. Also, a *noisy interference regime* is derived for the one-sided CIC in which treating interference as noise is sum-rate optimal. For the CRC, a full characterization of the capacity region for a class of *more-capable* channels is obtained. Moreover, it is shown that the derived outer bounds are useful to study the channels with one-sided receiver side information wherein one of the receivers has access to the non-intended message; capacity bounds are also discussed in details for such scenarios. Our results lead to new insights regarding the nature of information flow in these basic interference networks.**

***Index Terms: Network Information Theory; Fundamental Limits; Single-Hop Communication Networks; Interference Networks; Broadcast Channels; Cognitive Radio Channels;***

---

[1] Reza K. Farsani was with the department of electrical engineering, Sharif University of Technology. He is by now with the school of cognitive sciences, Institute for Research in Fundamental Sciences (IPM), Tehran, Iran.

**To:**

**THE MAHDI**

# Contents

# I. INTRODUCTION

In his landmark paper [1], Shannon derived the fundamental limits of communications for a point-to-point channel. This fundamental limit called *channel capacity* represents the maximum rate of information which can be reliably transmitted over a communication system. Channel capacity theorem, as the fundamental theorem of information theory, states that in a transmitter/receiver communication setup if the information rate is equal to or less than the channel capacity, there exist encoding and decoding schemes for the transmitter and the receiver which enable transmission with arbitrarily small error probability in the presence of noise. On the other hand, if the information rate is greater than the channel capacity, regardless of the coding scheme used in the system, the error probability is close to one.

Despite the single user channel for which the capacity is known, the problem still remains unsolved for many multiuser communication networks. Up to now, capacity bounds have been widely studied for different multiuser settings. Nevertheless, we are still far from a unified theory for the subject. Most of existing research concentrates on the study of certain networks individually, and there are few works which deal with the information theoretic issues for the networks systematically from a unified point of view. The multiuser networks can be categorized based on their essential features such as *interference*, *relay or interactive user cooperation*. Each of these key features may appear in different topologies, but its effect on the behavior of information flow does not change in essence from one model to another. In other words, different network topologies which are common in the essential features exhibit similarities from many aspects. Now if we recognize these similarities, then it is possible to adapt them to create a unified mathematical language for re-describing the behavior of the networks in a systematic approach. Thus, we can have powerful tools which are applicable not only for studying some given specific scenarios but also for arbitrary large topologies. In these multi-part papers, we will try to build such a systematic study for the *interference networks*. In communication systems, the interference arises, in principle, when the signal of a transmitter during transmission interferes with the other signals in the network before being received by the desired user. The interference signals are not noise and indeed they convey information for the other users. Therefore, from the viewpoint of a particular user, a transmitting signal may be interference or contain information. Such description of the interference is suitable more for the Gaussian networks in which a linear combination of the network input signals and Gaussian noise is received at each receiver. More precisely, one can state that in an arbitrary communication network, a received signal at a certain user has been interfered if it contains information regarding both desired and undesired messages for that user. Hence, the interference is essentially unavoidable in almost all multiuser communication scenarios. However, in these multi-part papers, the term "interference network" is referred to the general single-hop communications scenario with arbitrary number of transmitters and receivers with arbitrary distribution of messages among transmitters and receivers but without any interactive (relay) node [2]. In fact, the essential feature, which uniformly appears in these networks, is the interference element. Moreover, as we will discuss in the following, the classical interference channel is one of the main building blocks of these networks. Hence, we generally refer to them as the interference networks. We intend to show that the information flow in such networks follows similar directions in many aspects.

## ➢ Common Elements of Our Systematic Study

The key elements of our systematic study are given as follows.

- ***Fundamental Mathematical Tools:*** In network information theory, a coding theorem, which yields the characterization of the capacity region, is composed of two parts:

1) *Achievability scheme:* In this step the encoding and decoding strategies to achieve a rate region is described. In other words, it is shown that if the information rates belong to a given rate region, then a reliable communication with arbitrary small probability of error is achievable. Achievability proofs are extensively derived using random coding arguments in which the typical sequences and their properties [3] have a central role. Shannon also proved the channel capacity theorem using the random coding technique [1]. For multi-terminal networks, there exist two main random coding strategies: The *superposition coding* and the *random binning*.

   The superposition coding technique was originally devised for the broadcast channels [4]. It was shown in [5, 6] that this coding strategy is useful to achieve the capacity region of degraded BCs. Let $P_{XY}(x,y)$ be a given probability distribution on the product set $\mathcal{X}\times\mathcal{Y}$, where $\mathcal{X}$ and $\mathcal{Y}$ are arbitrary sets, and let $P_X(x)$ and $P_{Y|X}(y|x)$ be its marginal and conditional distributions. Let also $n$ be an arbitrary natural number and $(R_X, R_Y)$ be a non-negative pair of real numbers. Assume that a collection of $2^{n(R_X+R_Y)}$ codewords of length-$n$ are randomly generated in the following style:

- ✓ First, $2^{nR_X}$ independent codewords $X^n$ are generated according to $Pr(x^n) = \prod_{t=1}^{n} P_X(x_t)$. Label these codewords as $X^n(m_X)$, where $m_X$ belongs to the set $\{1, \dots, 2^{nR_X}\}$.
- ✓ Next, with respect to each codeword $X^n(m_X)$, $2^{nR_Y}$ independent codewords $Y^n$ are generated according to $Pr(y^n) = \prod_{t=1}^{n} P_{Y|X}(y_t|x_t)$. Label such codewords $Y^n(m_X, m_Y)$, where $m_Y$ belongs to the set $\{1, \dots, 2^{nR_Y}\}$.

In this case, it is said that the codewords $Y^n$ are superimposed on the codewords $X^n$. In other words, the pair $\big(X^n(m_X), Y^n(m_X, m_Y)\big)$ constitutes a superposition structure where $X^n(m_X)$ is called the *cloud center* codeword and $Y^n(m_X, m_Y)$ the *satellite* codeword. Note that the index $m_X \in \{1, \dots, 2^{nR_X}\}$ is contained in both the cloud center and the satellite while the index $m_Y \in \{1, \dots, 2^{nR_Y}\}$ is contained only in the satellite codeword. In this paper, we develop a graphical tool to illustrate a superposition structure among the generated codewords for a random coding scheme as shown in Fig. 1.

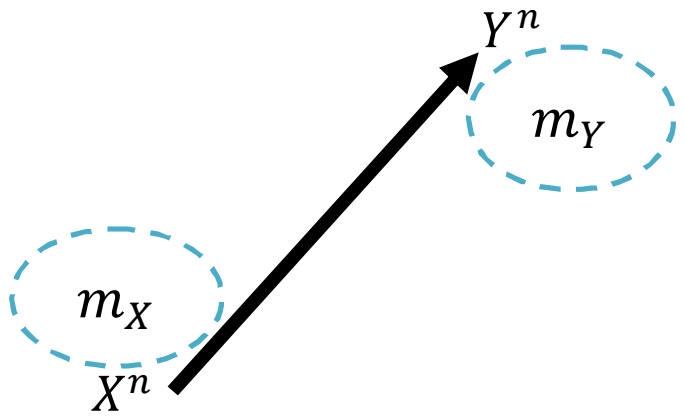

Figure 1. A superposition structure. The codeword $X^n$ is the cloud center and the codeword $Y^n$ is the satellite. These two codewords are connected by a directed edge. The ellipse beside the codeword at the beginning of the directed edge indicates the index which is contained in the cloud center. The ellipse beside the codeword at the end of the directed edge indicates the index which is contained in the satellite but not in the cloud center.

In this illustration, the cloud center is connected by a directed edge to the satellite. The ellipse beside the codeword at the beginning of the directed edge indicates the index which is contained in the cloud center. Also, the ellipse beside the codeword at the end of the directed edge indicates the index which is contained in the satellite but not in the cloud center. As demonstrated in our previous paper [7] presented in [8], such graphical illustration is very useful not only to describe a given achievability scheme but also to design coding schemes with satisfactory performance for arbitrary large networks. We will extensively utilize these graphical illustrations to *describe* and *design* the achievability schemes throughout our multi-part papers.

The second random coding strategy is the binning. In this technique, some bins of codewords are independently generated. These bins are then explored against each other to find a jointly typical tuple of codewords. The sizes of the bins are selected large enough to guarantee the existence of such a jointly typical tuple. The chosen typical codewords are designated for transmission. For communication channels, the binning strategy was first used by Gel'fand and Pinsker [9] to derive the capacity of the single-user channel with non-causal side information at the transmitter. In the Gel'fand-Pinsker scheme, a bin of codewords is generated and then a codeword from this bin is picked for transmission which is jointly typical with the sequence of the side information. Later, it was shown in [10] that for the Gaussian channel with additive interference, known non-causally at the transmitter, the binning strategy achieves the interference free capacity. In other words, the interference is perfectly canceled. Due to the name of the paper [10], the binning strategy for the Gaussian scenario is also referred to as the "*dirty paper coding*". In recent years, dirty paper coding is extensively used to pre-cancel the known interference. The most significant result in this regard is that the dirty paper coding achieves the capacity region of the Multiple-Input Multiple-Output (MIMO) broadcast channels [11].

2) *The converse theorem:* In this step, it is proved that if there exists a coding scheme for reliable communications with arbitrary small probability of error over a certain network, then it is required that the respective information rates belong to a given rate region. If this rate region is also achievable, the capacity region has been established; otherwise, the derived rate region constitutes an outer bound on the capacity region. The main tool for proving a converse theorem (or deriving a capacity outer bound for a given network) is the Fano's inequality [3]. Given a random variable $Y$, assume that we intend to estimate a correlated random variable $X$ with a distribution $P_X(.)$ such that $Y$ is related to $X$ by the conditional distribution $P_{Y|X}(.|.)$. For any estimator $\hat{X}$, where $X \to Y \to \hat{X}$ forms a Markov chain, the Fano's inequality bound the probability that $\hat{X} \neq X$. If we denote $P_e \triangleq Pr\{\hat{X} \neq X\}$, the Fano's inequality states [3]:

$$P_e \geq \frac{H(X|Y) - 1}{\log\|\mathcal{X}\|} \qquad \text{(I~1)}$$

where $H(.|.)$ is the conditional entropy [3] and $\|\mathcal{X}\|$ is the cardinality of the alphabet set of $X$. Therefore, the Fano's inequality is crucial to bound the probability of incorrect decoding at the receivers of a network for a given code.

- ***Basic Structures:***

Before studying networks with large topologies, it is essential to understand the basic building blocks. The interference networks, in principle, have two infrastructures: the *transmitters* and the *receivers*. Accordingly, one may consider the Multiple Access Channel (MAC) and the Broadcast Channel (BC) as the main building blocks of all interference networks. From an abstract point of view, this is indeed the case. Nonetheless, in order to obtain a better insight of the behavior of information flow in arbitrary large networks, we prefer to add two other simple topologies in this category of building blocks, i.e., the Classical Interference Channel (CIC) [12] and the Cognitive Radio Channel (CRC) [13]. In the CIC, some fundamental concepts such as *strong interference* [14, 15] are remarkable. Moreover, the CIC is the simplest topology in which both the MAC and the BC characteristics occur simultaneously. The CRC includes the element of "*cognition*", i.e., the knowledge of a user from the information available at the other one. The detailed descriptions of these basic building blocks are given below.

1) *MAC:* This is a communication network wherein some transmitters transmit independent messages to a single receiver. The MAC is the only multi-user communication scenario for which a full characterization of the capacity region (in the general case) is available [16]. Nevertheless, by the approach of [16], characterizing the capacity region for the MAC with a large number of transmitters is difficult. We will revisit this problem in Part II of our work [23] and demonstrate that our graphical tool in Fig. 1 is very useful to describe the capacity region of the MAC with arbitrary messages. It is worth nothing that the element of interference does not appear actually in the MAC because there exists only one receiver; accordingly, all signals in the network from the viewpoint of this receiver contain information regarding desired messages. Nonetheless, according to our general definition, the MAC still is categorized as an interference network because it is a single-hop communication network with no interactive (relay) node.

2) *BC:* The BC is a communication scenario in which a transmitter broadcasts independent messages to several receivers. Each receiver decodes its respective messages. The capacity region of the BC, even for the two-user case, is still an open problem except for some special cases. The BC has a central role in network information theory since many ideas used in developing this field have been originally devised for this channel. In fact, many of the results in our multi-part papers have been developed during our attempts to find capacity results for the multi-user BC.

3) *CIC:* In this scenario, a number of transmitters separately send independent messages to their respective receivers via a common media, while causing interference to the other receivers. The capacity region of the CIC for the general case is also an open problem. In recent years, numerous researches in network information theory have studied these channels because they are very useful models for many practical wireless networks.

4) *CRC:* This channel was presented in [13] (see also [17]) as an information theoretic model for a class of software defined radios [18] called as "cognitive radio" [19]. These are wireless communication systems proposed to promote the efficient use of the scarce radio spectrum by allowing some users called "*cognitive users*" to access the spectrum licensed out to the "*primary users*". In fact, the cognitive users may utilize the under-utilized licensed spectrum to the primary users. The idealized information theoretic model presented in [13] for this setup differs from the two-user CIC in the encoding scheme; for the CRC, one transmitter, i.e., the *cognitive* transmitter, has access non-causally to the message of the other transmitter, i.e., the *primary* transmitter. One may look at the CRC as a combination of the CIC and the BC. Nonetheless, the new element of *cognition* appeared in this configuration deserves to be considered as a building block because many multi-user interference networks may include this feature.

In the present part of our multi-part papers, we shall investigate in details capacity bounds for the above basic building blocks.

- ***Characteristic Interference Regimes:***

Some parts of the results in this paper are concerned to identify *characteristic interference regimes* for all interference networks, regardless of their topologies. Clearly speaking, a characteristic interference regime is typically associated with the optimality of a certain encoding\decoding strategy. The main interference regimes which are discussed throughout our multi-part papers are as follows:

1) *Degraded Networks:* The notation of "*degraded*" was first used in [4, 5, 6] to identify a type of BC for which the superposition coding achieves the capacity region. This concept has been generalized to many multiuser settings. Two types of degradedness are commonly recognized: The *physical* and the *stochastic* degradedness. The physically degraded networks are those for which the receivers can be arranged in a successive order from stronger to weaker such that the signal with respect to each receiver is statistically independent of the input signals conditioned on the signal of a stronger receiver. Stochastically degraded networks are those for which there exists an equivalent physically degraded network, i.e., the marginal distributions of the transition probability function for both networks are identical. For interference networks, since the two equivalent networks have the same capacity region, it is not required to distinguish between the two types of degradedness; hence, we generally refer these two as degraded interference networks. We will investigate the degraded networks in Part II [23] of our paper where a full characterization of the sum-rate capacity is derived for networks with any possible configurations, i.e., any number of transmitters, any number of receivers and any distribution of messages among transmitters and receivers. We will prove that for all degraded networks, a successive decoding scheme is sum-rate optimal. In this decoding strategy, each receiver decodes its messages as well as all the messages with respect to receivers weaker than itself in a successive order from weaker receivers to the stronger ones and lastly its own messages.

2) *Noisy Interference Networks:* In some network scenarios, the interference level (the power of the interference signals) at the receivers is low enough to allow treating it as noise. In other words, no part of the interferences (other user's messages) is decoded at any receiver. If this "*treating interference as noise*" strategy is optimal in the sense of achieving the sum-rate capacity, it is said that the network has *noisy interference*. For the first time, a noisy interference regime was identified for the two-user CIC by [20], [21] and [22], independently. In these papers, the authors derived outer bounds for the capacity region of the two-user CIC using genie-aided techniques and showed that when the interference level is below a certain threshold, treating interference as noise is sum-rate optimal. Moreover, in [21] this result was extended to some CICs with more than two users. In Parts II [23] and IV [28] of our multi-part papers, we will identify the noisy interference regime for several new network topologies.

3) *Strong Interference Networks:* For the two-user Gaussian CIC, Sato in [14] determined a regime wherein decoding both messages at both receivers is optimal and achieves the capacity. Later, the Sato's result was also extended to the discrete channel in [15]. This regime is called "*strong interference regime*" because the interference level is so high that each receiver can decode and remove it from its received signal without breaking the optimality. In [25], such a strong interference regime was also identified for some variations of the two-user CIC. In these references, the definition of the strong interference regime is based on the conditions under which the capacity region is derived by allowing each receiver to decode all messages. In Part III of our paper [27], we will demonstrate the incompleteness\inconsistency of such definitions and show that the concept is required to be re-defined more precisely. Clearly, we look at the strong interference as a *characteristic* of the networks and re-define it based on the optimal decoding scheme associated to this regime. In our definition, "*a network is said to be in the strong interference regime when the coding scheme achieving its capacity region is that each receiver decodes all messages transmitted by its connected transmitters*". We remark that, for a given network, a transmitter is connected to a receiver if the received signal statistically depends on the input signal which is sent by that transmitter. We then prove that the strong interference regime is always identifiable for any network topology. We will explicitly determine sufficient conditions for any arbitrary interference network (specifically for the multi-user CIC) under which the strong interference regime holds.

4) *Networks with a Sequence of Less-Noisy Receivers:* In [20], a regime for the two-user Gaussian CIC named the "*mixed interference regime*" was identified in which the optimal strategy to achieve the sum-rate capacity is for the one of the receivers to decode both messages while the other receiver decodes only its own message. The receiver which decodes both messages experiences strong interference, while the one which treats interference as noise and decodes only its own message experiences weak interference; hence, the name of mixed interference. In the present part of our work, we derive a similar result for the discrete channel. Our derivation reveals that the mixed interference regime actually represents a situation where a receiver is *less noisy* than the other one such that the stronger (less noisy) receiver, without violating the optimality in achieving the sum-rate, can decode the message of the weaker one as well. This behavior is exactly similar to the less-noisy BCs [26]. These are a class of BCs strictly larger than the degraded ones but with the same optimal coding strategy. Therefore, when we explore the problem from the viewpoint of interference level, we have a mixed interference regime; whereas if we consider the channel from the viewpoint of the decoding side, we have a less-noisy channel. In Part IV [28], we will identify classes of interference networks with less noisy receivers and derive a full characterization of their sum-rate capacity. These networks strictly contain the degraded ones as a subset but the optimal strategy to achieve their sum-rate capacity is still a successive interference cancellation scheme.

In addition to the above regimes, in Parts II [23], we will also identify networks for which incorporation of successive decoding and treating interference as noise strategies is sum-rate optimal.

- ***Graphical Approach for Coding Schemes:***

To present a subject systematically, it is essential to have a suitable descriptive language. Here, we are involved in achievability schemes as well as capacity outer bounds for large multi-message network topologies. To describe capacity bounds for such networks, rate regions with numerous constraints are unavoidable. Without an expressive language describing the philosophy behind derivations, it is almost impossible to understand the behavior of information flow in these networks.

Regarding the achievability schemes, some authors [29]-[31] have used diagrammatical illustrations to explain the process of random codebook generation for certain channels. Although such illustrations are useful to describe a given coding strategy to some extent, yet they are not efficient to design achievability schemes with satisfactory performance for arbitrary large networks. In our multi-part papers, we will develop novel graphical illustrations based on directed graphs to depict achievability schemes, where the directed edges are used to represent the superposition structures among the generated codewords for random coding. The key feature of these graphical tools is shown in Fig. 1. As demonstrated in our previous paper [7], which was also presented in [8], such graphical illustrations have the following advantages:

- ✓ They clearly describe given achievability schemes for arbitrary networks. The directed edges depict the superposition structures among the generated codewords; meanwhile, using the ellipses beside by the codewords, without introducing any complexity, the binning steps applied in the process of random codebook generation are also easily understood. Moreover, these ellipses indicate messages and bin indices conveyed by the codewords.
- ✓ They are powerful tools to systematically design achievability schemes with satisfactory performance for arbitrary network topologies based on the best known coding strategies for the basic building blocks. This procedure has been explained in details in our previous papers [7, 8].
- ✓ They can be used to easily derive the joint probability distributions with respect to the random coding schemes. See our discussion on Page 6 of [7, 8].
- ✓ They are helpful to extract adapted rate regions for networks with specific characteristics from a given general achievability scheme. Clearly, for special cases such as deterministic networks, to obtain either capacity region or a suitable achievable rate region from a general achievability scheme, it is required to substitute some auxiliary random variables of that general scheme with specific parameters. Our graphical illustrations are helpful to simplify and evaluate general achievable rate regions for special cases. We will demonstrate this capability of the proposed graphical tools for the CRC in Subsection III.A.4 below. Detailed discussions can also be found in Part III [27, Sec. VI] and Part V [32] of our multi-part papers.
- ✓ They are useful tools to explain the behavior of information flow in large network topologies, specifically, in characteristic interference regimes. For this purpose, we extensively use our graphical tools in all parts of our paper.
- ✓ They enable us to derive general expressions describing the constraints with respect to achievability schemes for arbitrary networks, as given in our previous paper [7, P. 12]. Clearly, given the graphical illustration with respect to any achievability scheme together with the encoding\decoding strategies, the constraints required for vanishing the error probability of the corresponding random coding are intuitively derived.

In addition, using this graphical approach, we shall present a random coding scheme for arbitrary interference networks in Part V of our multi-part papers [32].

- ***Characterizing Outer Bounds Based on Messages***

Next, let us deal with approaches for deriving capacity outer bounds. In network information theory, a capacity outer bound is usually derived from inequalities of the form below:

$$R \leq I(M; Y^n) + n\epsilon_n \tag{I~2}$$

where $M$ is a random variable representing a given message which is required to be decoded at the receiver $Y$ and the real number $R$ is the respective communication rate, the natural number $n$ is the length of the code and the random vector $Y^n$ is the received sequence at the receiver $Y$. Also, the operator $I(.;.)$ is the mutual information function [3], and $\epsilon_n$ is a parameter with $\epsilon_n \to 0$ as $n \to \infty$, (due to the Fano's inequality). For multi-user scenarios, we are involved in several inequalities of this form as well as their linear combinations. In the literature, it is common to manipulate such constraints using the chain rule [3] as well as some technical lemmas such as Csiszar-Korner identity [33, Lemma 7]. In this procedure, tuples of random variables such as $(M,Y^{t-1}),(M,Y^{t-1},Y_{t+1}^{n})$, and $(M,Y_{t+1}^{n})$ are represented by new random variables such as $W_t,U_t,V_t$ for $t=1,\dots,n$. Then, by applying a time-sharing argument desired constraints are derived where mutual information functions with only single letter random variables appear; these single letter variables are auxiliary random variables such as $W,U,V$, random variables representing the input signals such as $X$ and random variables representing the output signals such as $Y$. This process is called *single-letterization*. By this approach, we obtain computable capacity outer bounds; however, the concealment of messages with a cover of auxiliary random variables removes any sense from the structure of the outer bound. Clearly, the derived constraints are not sufficiently expressive to perceive their relation with the corresponding network topology. This approach is never efficient to describe the behavior of information flow, specifically for large topologies with multiple messages.

In this paper, we present another approach. In our procedure, to derive capacity outer bounds for the interference networks, we only look at the decoding order at the receivers and ignore the encoding order at the transmitters. Note that for writing inequalities such as (I~2), it is necessary to know that the message $M$ is decoded at the receiver $Y$ without requiring further information regarding the network topology. Also, during the process of single letterization, the messages are exactly preserved in the constraints without being represented by new auxiliary random variables. In other words, auxiliary random variables are used only to represent tuples such as $Y^{t-1}$ and $Y_{t+1}^{n}$, $t=1,\dots,n$, (or pairs of such tuples). Thus, the encoding order at the transmitters is introduced in the problem only when we intend to determine the joint probability distribution with respect to the outer bound. As a result, we establish capacity outer bounds with a unified framework which are uniformly applicable for any network topology. This approach is very useful, because it clearly depicts how the structure of the constraints included in the outer bound varies by network topology. Moreover, for many networks, the outer bound can be easily transformed into computable bounds with a few auxiliary random variables. We will demonstrate this fact in the present part and also Parts II-IV [23, 27, 28] of our multi-part papers. As we will see, our methodology for representing outer bounds is a powerful strategy to describe the behavior of information flow in large multi-message networks.

- ***Systematic Design:***

Given a network topology of arbitrary dimensions, an essential question is how to design a suitable achievability scheme. For large multi-message networks, there exist numerous methods to build achievability schemes. But what is really the best strategy? To illustrate the importance of the problem, let us discuss an example. The CRC was introduced in 2006 in [13] where the authors established an achievable rate region for the channel by a combination of the Gel'fand-Pinsker binning technique and the Han-Kobayashi rate splitting scheme. Following up this paper, many other researchers [17], [25], [29], [31], [34]-[41] studied information theoretic bounds for this channel. Specifically, several achievability schemes were proposed for the channel in these papers. But it was not clear which of the proposed achievable rate regions is better than the others because their comparisons seem to be difficult. Recently in [29], the authors presented a new achievability scheme and showed that it includes the previous results. Even for the case of [29], due to the lack of systematic derivation, some extra complexities have been introduced in the achievability scheme. In their proposed achievability scheme, the primary transmitter splits its message into *three parts* and then one of these sub-messages is ignored and relegated to the cognitive transmitter. In Subsection III.A.4, we will demonstrate that this strategy, i.e., splitting the primary message into three parts, not only is superfluous, but also it may cause rate loss. In fact, the authors in [29] have derived their achievability scheme by a combination of previous results. As the CRC is rather a simple topology, it is clear that for larger networks, the problem is more intricate. Therefore, it is essential to have suitable criteria to build achievability schemes. In our multi-part papers, we systematically design achievability schemes based on the following criteria:

- ✓ Given a certain network, our achievability design is such that, when it is specialized for the basic building blocks contained in the network, it does work essentially similar to the best known coding strategies.
- ✓ To reduce the complexity, to the extent possible without violating the previous criterion, the achievability scheme is designed with the lowest number of message splitting. In the general case, in Parts III [27, Sec. VI] and V [32] of our paper, we will prove that for an intelligent achievability deign for a network with $K$ receivers, each message is required to be split at most into $2^{K-1}$ parts (each sub-message is designated to be transmitted to the desired receiver as well as a subset of non-desired receivers). Specially, for networks with two receivers, splitting each message into more than two parts is always redundant.
- ✓ The superposition structures among the generated codewords are such that each receiver is required to decode only its designated codewords. Clearly, when two codewords build a superposition structure, for correct decoding of the satellite codeword it is required to decode the cloud center first. Now if a satellite codeword is designated for some specific receivers but its cloud center is not, then those receivers should necessarily decode some non-desired codewords, which irrevocably causes rate loss.

Based on the above criteria, in Part V of our multi-part papers [32], we will design a unique achievability scheme for all interference networks with arbitrary topologies. This design is derived by a systematic combination of the best achievability schemes for the MAC and the BC with common messages. Please refer to [7, 8] for details.

We follow our systematic view also in establishing capacity outer bounds. Almost all capacity outer bounds throughout different parts of our papers are derived based on unified frameworks which are uniformly applicable to a broad range of network topologies. Even different constraints of these outer bounds are derived by a unique style such that they can be represented using a unique expression. By this approach, we obtain very useful capacity outer bounds which are systematically adaptable to different network scenarios. Moreover, they can easily be adapted to include features such as secrecy constraints, see [115]. The main benefit of such general expressions for constraints of the outer bounds is that we can flexibly make use of them to prove capacity results for networks in the characteristic interference regimes. Many new network topologies with characteristic interference regimes, such as noisy interference regime, are identified by this approach, e.g., Many-to-One Interference Networks.

- **Pervious Works:**

Let us now present a review of existing literature regarding capacity bounds for different interference networks. For the MAC, wherein each transmitter sends an independent message to the receiver, the capacity region was derived by Ahlswede [42] and Liao [43]. The capacity of the MAC with Common Messages (MACCM) among transmitters was determined by Slepian and Wolf in [16]. They showed that for this scenario, the superposition coding is optimal. The two-user MAC with partial cooperation between transmitters was studied in details in [44].

The BC was introduced in [4]. This channel has been widely studied: a review of the subject can be found in [45]. For the degraded channel, wherein one output is a noisy version of the other one, it was shown in [4, 5] that superposition coding achieves the capacity region. This result was extended to the scalar Gaussian BC [46]; in fact, the scalar Gaussian BC is stochastically degraded. To date, the best capacity inner bound for the two-user BC is due to Marton [47, 48] which derived using a binning scheme. This inner bound is optimal in all special cases for which the capacity region is known, i.e., the degraded BC [4, 5], the less noisy [26] and the more capable BC [49], the semi-deterministic BC [50], and also the vector Gaussian BC [11]. Recently, a computable characterization of the Marton's inner bound has been provided in [51]. Several capacity outer bounds have been also established for the BC in [47], [52-54]. The outer bound developed in [53] by Nair-El Gamal, which is strictly tighter than those of [47] and [52], is optimal for the two-user BC without common message in all special cases for which the capacity region is known. Also, in [55], it was shown that this outer bound (for the channel without common message) is equivalent to those developed in [54]. Therefore, the Nair-El Gamal outer bound is the tightest known capacity outer bound for the two-user BC. However, in a recent work [56], it was shown that Marton's inner bound and the Nair-El Gamal outer bound for the BC do not agree in the general case, and the outer bound is strictly suboptimal. The most recent results regarding capacity bounds for the two-user BC can be found in [57, 58]. Some of the results for the two-user channel are extendable to the multi-receiver case. Specifically, the capacity region of the multi-user MIMO BC [11], the multi-user degraded BC [5], and the multi-user deterministic BC [59] are known. Nevertheless, the capacity region for the general case is still an open problem.

The two-user CIC was initially studied in [60], [61] and [12]. This channel has been also investigated in many papers where capacity inner and outer bounds are derived. The capacity region remains unknown in the general case. However, a single-letter characterization of the capacity region is available for some special cases such as the CIC with strong interference [14, 15], a subclass of deterministic CIC [62], a class of discrete degraded CIC [63] (which includes the discrete additive degraded CIC studied by Benzel [116]) and a subclass of one-sided CIC [64]. A limiting expression for the capacity region in the general case was derived in [65]; however, computing the capacity through this expression is excessively difficult [20]. For the Gaussian channel, the capacity region was approximately determined to within one bit in [66]. To date, the best capacity inner bound for the two-user CIC is the well-known Han-Kobayashi (HK) achievable rate region established in 1981 [67]. A compact description of the HK rate region was presented in [68]. Also, a full characterization of the HK rate region for the Gaussian channel with Gaussian input distributions was provided in [20]. In the concurrent works [20], [21] and [22], the authors established new capacity outer bounds for the Gaussian CIC. Using the derived outer bounds, they also identified a noisy interference regime for the channel. In this regime, treating interference as noise at each receiver is sum-rate optimal. The outer bounds in the latter papers, which all were derived using genie-aided techniques, are tighter than the previous results [66] and [69-71]. For the discrete CIC, capacity outer bounds can be found in [71-74]. All these outer bounds are sub-optimal in the general case.

As discussed before, for the CRC capacity inner bounds were derived in [13], [29], and [34-38]. Capacity outer bounds were also proposed for this channel in [34] and [36]. Moreover, explicit capacity results have been established in some special cases. Specifically, a strong interference regime was identified for the channel in [25]. In [34], the authors showed that a simple binning scheme achieves the capacity region for a class of Gaussian CRCs. They also derived the capacity region for a class of discrete channels [34, Th. 3.4]. In a recent work [75], the capacity region of the semi-deterministic CRC as well as a class of discrete channels coined as "better cognitive decoding channels" was derived. Capacity regions were also derived for special cases of the one-sided Gaussian CRC in the concurrent works [40], [41], and [76]. Also, for a class of one-sided CRCs with one noiseless component the capacity region was derived in [39]. The capacity region for the CRC in the general case is still an open problem. Capacity bounds for multi-user/multi-message cognitive radio networks were developed in [119-122].

The multi-user\multi-message interference networks have been also studied widely in recent years. A large body of studies in this regard relies on the new idea of *interference alignment* which was introduced for the MIMO X-Channel by Maddah-Ali, *et al* in [77] and for the multi-user CIC by Cadambe and Jafar in [78]. In the latter paper, the authors utilized the interference alignment technique to approximately characterize the capacity of the multi-user Gaussian time-varying CIC in high signal to noise regime. For a detailed discussion regarding the interference alignment approaches refer to [79]. The so-called many-to-one and one-to-many CICs -special cases where interference is experienced, or is caused, by only one user- were studied in [24] and [80]. Capacity bounds were presented in [81] for a class of three user deterministic CICs. In [82] the authors derived a family of genie-aided MAC type outer bounds for the multi-user CIC and by using of them established the sum-rate capacity of the Gaussian degraded channel. An outer bound was presented in [83] for the three-user CIC. Also, the sum-rate capacity was established for some certain Gaussian channels. A class of multi-user cyclic Gaussian CIC was studied in [84] where the capacity region was approximately determined to within a constant gap; the authors also identified a strong interference regime for the channel. Capacity region of a class of multi-user symmetric Gaussian CICs in the very strong interference regime was determined in [117] using the lattice coding. In this regime, the interference can be perfectly cancelled by all the receivers without any rate penalty. The sum-rate capacity has been recently studied in [118] for a class of multi-user Gaussian CICs wherein each user, except the first one, experiences interference only from the previous one. Capacity bounds for network scenarios composed of interfering MACs were explored in [100-104].

The two-user CICs with cooperative transmitters were studied in [25] for which capacity regions were established in the strong interference regime. The two-user CIC with common information, in which the transmitters send a common message to both receivers beside sending their private messages to their respective receivers, was investigated in [85]; an achievable rate region was obtained for the channel using a superposition coding scheme that consists of successive encoding and jointly decoding. The capacity region for a class of semi-deterministic CICs with common information was characterized in [86]. Also, capacity outer bounds were derived for the Gaussian CIC with common message in [87]. As a variation of the CRC, the cognitive interference channel with degraded message sets was considered in [88]; the difference of this scenario with the CRC is that here one receiver decodes the primary message while the other one decodes both the primary and the cognitive messages. The authors in [88] derived a full characterization of the capacity region for this channel. The references [38] and [89] considered the BC with cognitive relays, i.e., a scenario wherein a transmitter (broadcasting node) sends two private messages to two receivers while being assisted by two cognitive relays each of which has access to one of the messages non-casually. Achievability schemes were presented in these papers for the channel; moreover, in [89] an outer bound on the capacity region and the degrees of freedom were presented. The so-called X interference channel as a two-transmitter/two-receiver communication scenario, where each transmitter separately broadcasts two private messages to the receivers, was studied in [77] and [90]. Also, an achievable rate region was established for this channel in [91]. The authors in [92] considered a variation of the two-user CIC coined as the interference-multiple-access channel. This model differs from the two-user CIC in the decoding scheme at the receivers so that one receiver is required to decode both transmitted messages. The authors presented an achievable rate region for the channel and showed that for the Gaussian case it is to within one bit of the capacity region. In a recent work [93], a network scenario wherein a BC generates interference to a single-user channel has been studied and some capacity results were derived in the strong interference regime. The one-sided CIC, in which the signal of one receiver is statistically independent of its non-respective transmitter, is sometimes referred to as the Z Interference Channel (ZIC) because its topology is similar to the letter Z. As a generalization of the ZIC, the Z-channel was introduced in [94]. This is a channel which has the same topology as the ZIC (from the view point of channel transition probability function) but with an additional message. In this channel, one transmitter sends only a private message to its respective receiver and it has no effect on the signal of its non-receptive receiver. The other transmitter broadcasts two private messages to both receivers. The Gaussian Z-channel was studied in [95]. Explicit capacity results were established for a class of degraded Z-channels in [96] and for a class of deterministic Z-channels in [97]. New results have been recently derived in [98] and [99] for this channel.

This was a detailed review of the existing literature in which Shannon theoretic capacity limits are derived for single-hop communication networks.

## ➢ Our Contribution in This Part of the Paper:

In this first part of our multi-part papers, we first provide a brief review of existing results in literature regarding capacity bounds for the basic building blocks. This review is given in Subsection III.A. New observations are also presented in this regard. Specifically, we revisit the strong interference conditions derived in [15] for the two-user CIC and prove that dependence between the inputs does not change these conditions. In Part III [27], we will demonstrate that this remarkable fact leads to a deep insight regarding the behavior of information flow in large multi-message networks in the strong interference regime. Moreover, this result constitutes a key element to prove some new capacity results and re-derive some of the existing results for the two-user CIC in Subsection III.C. Also, we revisit the achievability scheme of the recent paper [29] for the CRC and show that, unlike this work, splitting the primary message into three parts may cause some rate loss besides being superfluous. In fact, we demonstrate that by a simpler achievability scheme than that of [29], a larger (potentially strictly) rate region is derived.

Then in Subsection III.B, we establish new capacity outer bounds for the two-user BC, the two-user CIC, and the two-user CRC. All these outer bounds are derived based on a unified framework. In fact, our approach is very useful to recognize the existing similarities among different network topologies while developing capacity bounds for them. We then demonstrate that how the derived outer bounds can be easily translated to simpler forms. Also, we show that for channels with specific properties, such as one-sided channels, tighter outer bounds can be derived by establishing additional constraints.

Next in Subsection III.C, we demonstrate that the derived outer bounds are very efficient to establish important capacity results for the channels. Specifically, three new capacity results are proved. In the first one, we identify a ***mixed interference regime*** for the two-user CIC. Clearly, we determine conditions for the channel under which decoding the interference at one receiver and treating it as noise at the other one is sum-rate optimal. The determined conditions contain the mixed interference regime identified in [20] for the Gaussian channel as a special case. The second result is that we indentify a ***noisy interference regime*** for the one-sided CIC. This result indeed generalizes the noisy interference regime obtained in [109] for the Gaussian one-sided CIC to any arbitrary (not necessarily Gaussian) channel. As a third result, we consider the ***more-capable*** CRCs and derive a full characterization of the capacity region for a class of these channels. We also prove that the more-capable CRCs strictly include the class of "cognitive better decoding channels" for which the capacity region has been recently determined in [75]. Moreover, we show that some of known results for the two-user CIC and the CRC can be easily re-derived using the presented outer bounds.

In Subsection III.D, we study the two-user CIC, the CRC, and the two-user BC with one-sided receiver side information. In these scenarios, it is assumed that one of the receivers has access to the message designated to the other receiver such that it can use of this side information to decode its own message. We present capacity inner and outer bounds for each network. Specifically, we demonstrate that the derived outer bounds in Subsection III.B are also useful to study such scenarios. In some special cases, we derive explicit capacity results, as well. Our results in this subsection indeed give us some useful insights regarding the nature of information flow in the basic structures (accordingly, in all interference networks).

In the following subsection, we provide notations, definitions for the general interference networks and also some useful technical lemmas. These preliminaries are needed throughout various parts of our multi-part papers.

# II. PRELIMINARIES AND NETWORK MODEL

## II.A) Preliminaries

We use notations, definitions, and preliminaries that are common to Parts I-V of our multi-part papers. Also, a broad range of various interference network scenarios are studied in these parts. To avoid repetition, we provide the preliminaries and network model definitions for all parts in this section.

***Notations:*** Random Variables (RV) are denoted by upper case letters (e.g. $X$) and lower case letters are used to show their realization (e.g. $x$). The range set of $X$ is represented by $\mathcal{X}$. The Probability Distribution Function (PDF) of $X$ is denoted by $P_X(x)$, and the conditional PDF of $X$ given $Y$ is denoted by $P_{X|Y}(x|y)$; also, in representing PDFs, the arguments are sometimes omitted for brevity. A sequence of random variables $(X_t,\dots,X_n)$ with the same range $\mathcal{X}$ is denoted by $X_t^n$, where $x_t^n$ indicates its realization that is the sequence $(x_t,\dots,x_n)$. For the special case of $X_1^n$ and $x_1^n$, the subscript is dropped for brevity. The probability of the event $A$ is expressed by $Pr(A)$. The notations $\mathbb{E}[.]$, $|.|$, $||.||$ stand for the expectation operator, absolute value and cardinality, respectively. As a common notation, $H(.)$ is used to represent both the entropy function and the differential entropy. The set of real numbers, nonnegative real numbers and natural numbers are denoted by $\mathbb{R}$, $\mathbb{R}_+$, and $\mathbb{N}$, respectively. The notation $[1:K]$, where $K$ is a positive integer, represents the set $\{1,\dots,K\}$. The function $\psi(x)$ is defined as: $\psi(x) \equiv \frac{1}{2}\log(1+x)$, for $x \in \mathbb{R}_+$.

***Definition II.1:*** *Let $A = \{A_1,\dots,A_K\}$ be an arbitrary indexed set with K elements, where $K \in \mathbb{N}$. Let $\Omega$ be an arbitrary subset of A, i.e., $\Omega \subseteq A$. The identification of the set $\Omega$, denoted by $\underline{id}_\Omega$, is defined as follows:*

$$\underline{id}_\Omega \triangleq \{l \in [1:K] : A_l \in \Omega\} \tag{II~1}$$

***Definition II.2:*** *Given arbitrary random variables A and B, define:*

$$\mathbb{E}^2[A|B=b] \triangleq \mathbb{E}[A^2|B=b] - (\mathbb{E}[A|B=b])^2 \tag{II~2}$$

In fact, $\mathbb{E}^2[A|B=b]$ is the variance of $A$ with respect to the distribution $P_{A|B}(a|b)$, and hence a positive quantity.

***Definition II.3:*** *Let $\mathcal{X}_1,\dots,\mathcal{X}_K$ be arbitrary finite sets. Given a PDF $P_{X_1\dots X_K}(x_1,\dots,x_K)$ on the set $\mathcal{X}_1\times\dots\times\mathcal{X}_K$, the notation $p_{min}(P_{X_1\dots X_K})$ is defined as follows:*

$$p_{min}(P_{X_1\dots X_K}) \triangleq \min_{\substack{(x_1,\dots,x_K):\\ 0<P_{X_1\dots X_K}(x_1,\dots,x_K)}} P_{X_1\dots X_K}(x_1,\dots,x_K) \tag{II~3}$$

***Lemma II.1)*** *Consider the random variables $A, B$ and $C$, where A and C are real-valued and $A \to B \to C$ form a Markov chain. The following relations hold:*

$$\mathbb{E}^2[A+C|B=b] = \mathbb{E}^2[A|B=b] + \mathbb{E}^2[C|B=b] \tag{II~4}$$

$$|\mathbb{E}[AC]| \le \sqrt{\mathbb{E}[(\mathbb{E}[A|B])^2]}\sqrt{\mathbb{E}[(\mathbb{E}[C|B])^2]} \tag{II~5}$$

*Proof of Lemma II.1)* The equality (II~4) is derived by direct computation. To prove the inequality (II~5), we can write:

$$
\begin{aligned}
|\mathbb{E}[AC]| &= \left|\mathbb{E}\left[\mathbb{E}[AC|B]\right]\right| \\
&= \left|\mathbb{E}\left[\mathbb{E}[A|B] \times \mathbb{E}[C|B]\right]\right| \\
&\overset{(a)}{\leq} \sqrt{\mathbb{E}[(\mathbb{E}[A|B])^2]}\sqrt{\mathbb{E}[(\mathbb{E}[C|B])^2]}
\end{aligned}
$$

where (a) is due to the Cauchy-Schwarz inequality. ■

***Lemma II.2)*[3], *Entropy Power Inequality (EPI)***

*Let $X$ and $Y$ be independent random variables with continuous densities, then we have:*

$$2^{2H(X+Y)} \geq 2^{2H(X)} + 2^{2H(X)} \tag{II~6}$$

In the following, we also briefly recall the concept of typical sequences and their basic properties. These sequences and their properties have a key role in deriving random coding arguments for different networks.

***Definition II.4: Typical sequences* [105]*;***

*Let $\mathcal{X},\mathcal{Y}$ be arbitrary finite sets. Let $N(a,b|x^n,y^n)$ be the number of times that the pair $(a,b)$ occurs in $(x^n,y^n)$. Consider a joint PDF $P_{XY}(x,y)$ defined on the set $\mathcal{X}\times\mathcal{Y}$. For $\epsilon \geq 0$, the set of all jointly $\epsilon$-letter typical $n$-sequences $(x^n,y^n)$ with respect to the PDF $P_{XY}(x,y)$ is defined as follows:*

$$\mathcal{T}_\epsilon^n(P_{XY}) \triangleq \left\{(x^n,y^n) : \left|\frac{1}{n}N(a,b|x^n,y^n) - P_{XY}(a,b)\right| \leq \epsilon . P_{XY}(x,y) \text{ for all } (a,b) \in \mathcal{X}\times\mathcal{Y}\right\} \tag{II~5}$$

Note that based on (II~5), one can derive the sets $\mathcal{T}_\epsilon^n(P_X)$ and $\mathcal{T}_\epsilon^n(P_Y)$ immediately. Typical sequences have a fundamental role in network information theory as they are widely used to prove coding theorems for multi-terminals and also rate-distortion problems. In the following, we will present the main properties of these sequences. But first, we need to define the concept of conditionally typical sequences.

Let $\mathcal{X},\mathcal{Y}$ be arbitrary finite sets. Consider a joint PDF $P_{XY}(x,y)$ defined on the set $\mathcal{X}\times\mathcal{Y}$. Given an arbitrary sequence $x^n \in \mathcal{X}^n$, for $\epsilon \geq 0$, the set of all conditionally $\epsilon$-letter typical $n$-sequence $y^n$ with respect to the PDF $P_{XY}(x,y)$ is defined as follows:

$$\mathcal{T}_\epsilon^n(P_{XY}|x^n) \triangleq \{y^n : (x^n,y^n) \in \mathcal{T}_\epsilon^n(P_{XY})\} \tag{II~6}$$

It can be easily verified that if $(x^n,y^n) \in T_\epsilon^n(P_{XY})$, then $x^n \in \mathcal{T}_\epsilon^n(P_X)$ and $y^n \in \mathcal{T}_\epsilon^n(P_Y)$. Therefore, by (II~6) if $x^n \notin \mathcal{T}_\epsilon^n(P_X)$, then $\|\mathcal{T}_\epsilon^n(P_{XY}|x^n)\| = 0$.

The following lemmas and corollaries proved in [105] contain the basic properties of typical sequences.

***Lemma II.3)*** *Consider a joint PDF $P_{XY}(x,y)$. Let $0 < \epsilon_1 < \epsilon_2 \leq p_{min}(P_{XY})$, and also $(x^n,y^n) \in \mathcal{T}_{\epsilon_1}^n(P_{XY})$. Suppose that the random sequences $X^n,Y^n$ are jointly generated according to $(X^n,Y^n)\sim \prod_{t=1}^n P_{XY}(x_t,y_t)$. Then, we have:*

- $$2^{-n(1+\epsilon_1)H(X,Y)} \leq P(X^n = x^n, Y^n = y^n) \leq 2^{-n(1-\epsilon_1)H(X,Y)} \tag{II~7}$$
- $$(1-\delta_{\epsilon_1}^n)2^{n(1-\epsilon_1)H(X,Y)} \leq \left\|\mathcal{T}_{\epsilon_1}^n(P_{XY})\right\| \leq 2^{n(1+\epsilon_1)H(X,Y)} \tag{II~8}$$
-

$$1-\delta_{\epsilon_1}^n \leq Pr\left((X^n,Y^n)\in\mathcal{T}_{\epsilon_1}^n(P_{XY})\right)\leq 1$$

(II~9)

- $$2^{-n(1+\epsilon_1)H(X|Y)}\leq P(X^n=x^n|Y^n=y^n)\leq 2^{-n(1-\epsilon_1)H(X|Y)}$$

(II~10)

- $$\left(1-\delta_{\epsilon_1,\epsilon_2}^n\right)2^{n(1-\epsilon_2)H(X|Y)}\leq \left\|\mathcal{T}_{\epsilon_2}^n(P_{XY}|y^n)\right\|\leq 2^{n(1+\epsilon_2)H(X|Y)}$$

(II~11)

- $$1-\delta_{\epsilon_1,\epsilon_2}^n\leq Pr\left(X^n\in\mathcal{T}_{\epsilon_2}^n(P_{XY}|y^n)\middle|Y^n=y^n\right)\leq 1$$

(II~12)

*where* $\delta_{\epsilon_1}^n\to 0$ *and* $\delta_{\epsilon_1,\epsilon_2}^n\to 0$ *as* $n\to\infty$.

***Lemma II.4)*** *Consider a joint PDF* $P_{XY}(x,y)$ *with its marginal PDFs* $P_X(x)$ *and* $P_Y(y)$. *Let* $0<\epsilon_1<\epsilon_2\leq p_{min}(P_{XY})$. *Suppose that the random sequence* $X^n$ *is generated independent and identically distributed (i.i.d.) according to* $X^n\sim\prod_{t=1}^n P_X(x_t)$. *Given a sequence* $y^n\in\mathcal{T}_{\epsilon_1}^n(P_Y)$, *we have:*

$$\left(1-\delta_{\epsilon_1,\epsilon_2}^n\right)2^{-n\left(I(X;Y)+2\epsilon_2H(X)\right)}\leq Pr\left(X^n\in\mathcal{T}_{\epsilon_2}^n(P_{XY}|y^n)\right)\leq 2^{-n\left(I(X;Y)-2\epsilon_2H(X)\right)}$$

(II~13)

*where* $\delta_{\epsilon_1,\epsilon_2}^n\to 0$ *as* $n\to 0$.

***Corollary II.1)*** *Consider a joint PDF* $P_{XUY}(x,u,y)$ *with its marginal PDFs* $P_{X|U}(x|u)$ *and* $P_{UY}(u,y)$. *Suppose that* $0<\epsilon_1<\epsilon_2\leq p_{min}(P_{XYU})$. *Let* $U^n$ *be a random sequence generated i.i.d. according to* $U^n\sim\prod_{t=1}^n P_U(u_t)$. *Given a pair of sequences* $(u^n,y^n)\in\mathcal{T}_{\epsilon_1}^n(P_{UY})$, *let* $X^n$ *be also a random sequence generated i.i.d. according to* $X^n\sim\prod_{t=1}^n P_{X|U}(x_t|u_t)$. *We have:*

$$\left(1-\delta_{\epsilon_1,\epsilon_2}^n\right)2^{-n\left(I(X;Y|U)+2\epsilon_2H(X|U)\right)}\leq Pr\left(X^n\in\mathcal{T}_{\epsilon_2}^n(P_{XUY}|u^n,y^n)\middle|U^n=u^n\right)\leq 2^{-n\left(I(X;Y|U)-2\epsilon_2H(X|U)\right)}$$

(II~14)

*where* $\delta_{\epsilon_1,\epsilon_2}^n\to 0$ *as* $n\to 0$.

## II.B) General Memoryless Interference Networks

In these multi-part papers, we intend to study capacity bounds for different interference networks. In the following, we define the General Interference Network (GIN) based on a unified view in encoding and decoding schemes such that it contains all models previously studied in the literature as a special case.

***Definition II.5: General Discrete Interference Network;***

*A discrete memoryless interference network with* $K_1$ *transmitters and* $K_2$ *receivers denoted by* $\left\{\mathcal{X}_1,\dots,\mathcal{X}_{K_1},\mathcal{Y}_1,\dots,\mathcal{Y}_{K_2},\mathbb{P}_{Y_1\dots Y_{K_2}|X_1\dots X_{K_1}}\right\}$, *is a network which is organized by* $K_1$ *input alphabet sets* $\mathcal{X}_1,\dots,\mathcal{X}_{K_1}$, $K_2$ *output alphabet sets* $\mathcal{Y}_1,\dots,\mathcal{Y}_{K_2}$, *and a transition probability function* $\mathbb{P}_{Y_1\dots Y_{K_2}|X_1\dots X_{K_1}}$ *that describes the relation between the inputs and outputs of the network.*

The network is assumed to be memoryless, i.e.,

$$Pr(y_1^n,\dots,y_{K_2}^n|x_1^n,\dots,x_{K_1}^n)=\prod_{t=1}^{n}\mathbb{P}_{Y_1\dots Y_{K_2}|X_1\dots X_{K_1}}(y_{1,t},\dots,y_{K_1,t}|x_{1,t},\dots,x_{K_2,t}),\qquad n\geq 1$$

(II~15)

Fig. 2 depicts the network model.

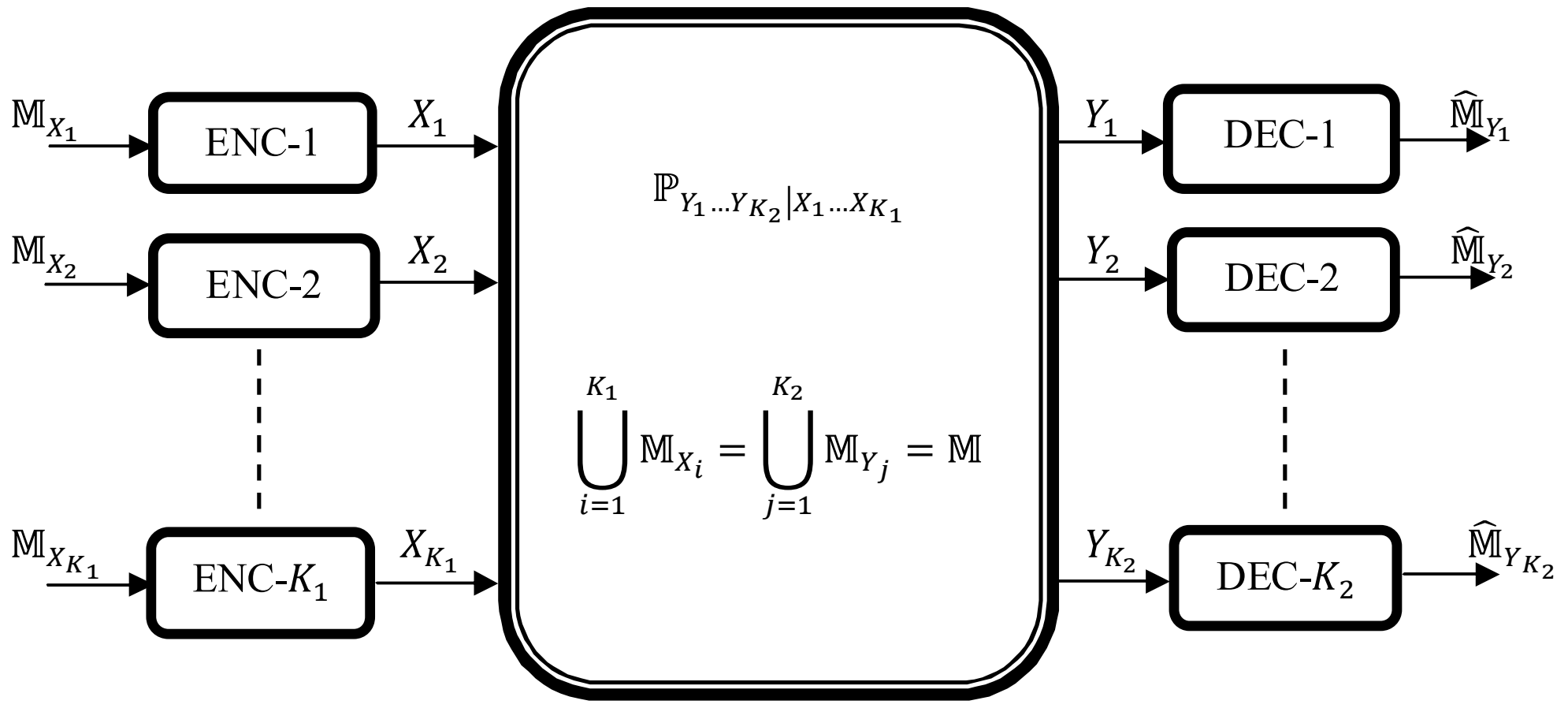

Figure 2. The General Interference Netwrok (GIN).

***Encoding and decoding scheme:*** For the general interference network (II~15), a length-$n$ code $\mathfrak{C}^n(R_1,\dots,R_K)$ is defined as follows: Let $\mathcal{M}_l=[1:2^{nR_l}]$ for $l=1,\dots,K$. Consider the independent random variables $M_1,\dots,M_K$ with the range sets $\mathcal{M}_1,\dots,\mathcal{M}_K$, respectively. The messages $M_1,\dots,M_K$ are intended to be transmitted over the network. Let $\mathfrak{M}\triangleq\{\mathcal{M}_1,\dots,\mathcal{M}_K\}$ and $\mathbb{M}\triangleq\{M_1,\dots,M_K\}$. Also, let $\mathcal{M}_{X_1},\dots,\mathcal{M}_{X_{K_1}}$ and $\mathcal{M}_{Y_1},\dots,\mathcal{M}_{Y_{K_2}}$ be nonempty arbitrary subsets of $\mathfrak{M}$ such that:

$$\bigcup_{i\in[1:K_1]}\mathcal{M}_{X_i}=\bigcup_{j\in[1:K_2]}\mathcal{M}_{Y_j}=\mathfrak{M}$$

(II~16)

Define:

$$\begin{cases}\mathbb{M}_{X_i}\triangleq\left\{M_l\,:\,l\in \underline{id}_{\mathcal{M}_{X_i}}\right\}, & i\in[1:K_1]\\ \mathbb{M}_{Y_j}\triangleq\left\{M_l\,:\,l\in \underline{id}_{\mathcal{M}_{Y_j}}\right\}, & j\in[1:K_2]\end{cases}$$

(II~17)

The $i^{th}$ transmitter sends the messages $\mathbb{M}_{X_i}$ over the network, $i\in[1:K_1]$; therefore, the encoder functions are given by:

$$\mathfrak{E}_i:\mathcal{M}_{X_i}\to\mathcal{X}_i^n,\qquad i\in[1:K_1]$$

(II~18)

where the generated codewords are as $X_i^n=\mathfrak{E}_i(\mathbb{M}_{X_i})$. On the other hand, the $j^{th}$ receiver decodes the messages $\mathbb{M}_{Y_j}$for $j\in[1:K_2]$; hence, the decoder functions are given by:

$$\mathfrak{D}_j:\mathcal{Y}_j^n\to\mathcal{M}_{Y_j},\qquad j\in[1:K_2]$$

(II~19)

where the messages in $\mathbb{M}_{Y_j}$ are estimated as $\widehat{\mathbb{M}}_{Y_j}=\mathfrak{D}_j(Y_j^n)$.

The rate of the code is the $K$-tuple $(R_1,\dots,R_K)$. Also, the average probability of decoding, denoted by $P_e^{\mathfrak{C}^n}$, is given below:

$$P_e^{\mathfrak{C}^n} \triangleq Pr\left(\bigcup_{j\in[1:K_2]}\left\{\mathfrak{D}_j(Y_j^n) \neq \mathbb{M}_{Y_j}\right\}\right)$$

(II~20)

***Definition II.6:*** *A K-tuple rate* $(R_1, \dots, R_K) \in \mathbb{R}_+^K$ *is said to be achievable for the general interference network in Fig. 2 if for every* $\epsilon > 0$ *and for all sufficiently large n, there exists a length-n code* $\mathfrak{C}^n(R_1, \dots, R_K)$ *such that* $P_e^{\mathfrak{C}^n} < \epsilon$.

***Definition II.7:*** *The capacity region of the general interference network in Fig. 2, denoted by* $\mathcal{C}^{GIN}$*, is the closure of the set of all achievable K-tuple* $(R_1, \dots, R_K)$.

Sometimes, in a certain network, it is desirable to determine the sum-rate capacity or the partial sum-rate capacity with respect to a specific subset of messages.

***Definition II.8:*** *Consider the general interference network in Fig. 2. Assume that* $\Omega$ *is an arbitrary subset of* $\mathbb{M}$. *The partial sum-rate capacity with respect to* $\Omega$*, denoted by* $\mathcal{C}_{\Sigma_\Omega}^{GIN}$*, is defined as follows:*

$$\mathcal{C}_{\Sigma_\Omega}^{GIN} \triangleq \max\left\{\boldsymbol{R}_{\Sigma_\Omega} : \exists\ (R_1, \dots, R_K) \in \mathcal{C}^{GIN}\ \ with \sum_{l\in\underline{id}_\Omega} R_l = \boldsymbol{R}_{\Sigma_\Omega}\right\}$$

(II~21)

*Also, the sum-rate capacity of the whole network, which we denote by* $\mathcal{C}_{sum}^{GIN}$*, is defined as:*

$$\mathcal{C}_{sum}^{GIN} \triangleq \mathcal{C}_{\Sigma_{\mathbb{M}}}^{GIN} = \max\left\{\boldsymbol{R}_{sum} : \exists\ (R_1, \dots, R_K) \in \mathcal{C}^{GIN}\ \ with \sum_{l\in[1:K]} R_l = \boldsymbol{R}_{sum}\right\}$$

(II~22)

Based on Definition II.8, the following lemma is easily derived.

***Lemma II.5)*** *Consider the general interference network in Fig.2. Assume that* $\Omega$ *is an arbitrary subset of* $\mathbb{M}$. *We have:*

$$\mathcal{C}_{\Sigma_\Omega}^{GIN} \triangleq max\left\{\boldsymbol{R}_{\Sigma_\Omega} : \exists\ (R_1, \dots, R_K) \in \mathcal{C}^{GIN}\ \ with \sum_{l\in\underline{id}_\Omega} R_l = \boldsymbol{R}_{\Sigma_\Omega},\quad R_l = 0\ if\ l \notin \underline{id}_\Omega\right\}$$

(II~23)

*In other words,* $\mathcal{C}_{\Sigma_\Omega}^{GIN}$ *is equal to the sum-rate capacity of the network where only the messages of* $\Omega$ *are aimed to be transmitted.*

*Proof of Lemma II.5)* To demonstrate this statement, it is sufficient to note that since the capacity region of memoryless networks is a convex set, if $(R_1, \dots, R_K) \in \mathcal{C}^{GIN}$, then $(\acute{R}_1, \dots, \acute{R}_K) \in \mathcal{C}^{GIN}$ where $\acute{R}_l$ is equal to $R_l$ or zero. ■

It should be noted that the capacity region of the interference network depends only on the marginal PDFs $\mathbb{P}_{Y_1|X_1\dots X_{K_1}}, \dots, \mathbb{P}_{Y_{K_2}|X_1\dots X_{K_1}}$, which are computed using the network transition probability function $\mathbb{P}_{Y_1\dots Y_{K_2}|X_1\dots X_{K_1}}$ as:

$$\mathbb{P}_{Y_j|X_1\dots X_{K_1}} = \sum_{\substack{y_l \\ l\in[1:K_2], l\neq j}} \mathbb{P}_{Y_1\dots Y_{K_2}|X_1\dots X_{K_1}},\qquad j \in [1:K_2]$$

(II~24)

This fact can be easily proved using the definition of coding for the network.

The general interference network introduced here contains the MAC, the BC, the CIC, the CRC, and other single-hop communication scenarios studied in the literature as special cases. In these multi-part papers, we will try to present a systematic study of capacity bounds for different interference networks. In some interference network scenarios, for example the ZIC, the receiving signal at certain receivers is not affected by all the transmitted signals but depends only on specific ones. In these models, it is useful to define connected and unconnected transmitters with respect to each receiver.

***Definition II.9: Connected and Unconnected Transmitters with Respect to Receivers;***

*Consider the general interference network in Fig. 2 with the marginal PDFs* $\mathbb{P}_{Y_1|X_1\ldots X_{K_1}},\ldots,\mathbb{P}_{Y_{K_2}|X_1\ldots X_{K_1}}$ *as computed by (II~24). For the* $j^{th}$ *receiver,* $j=1,\ldots,K_2$*, assume that* $\mathbb{P}_{Y_j|X_1\ldots X_{K_1}}$ *is as follows:*

$$\mathbb{P}_{Y_j|X_1\ldots X_{K_1}} \equiv \mathbb{P}_{Y_j|\mathbb{X}_{c\to Y_j}} \tag{II~25}$$

*where* $\mathbb{X}_{c\to Y_j}$ *is a specific subset of* $\{X_1,\ldots,X_{K_1}\}$*. In other words, the following Markov chain holds:*

$$\mathbb{X}_{c\nrightarrow Y_j} \to \mathbb{X}_{c\to Y_j} \to Y_j \tag{II~26}$$

*where* $\mathbb{X}_{c\nrightarrow Y_j} \triangleq \{X_1,\ldots,X_{K_1}\} - \mathbb{X}_{c\to Y_j}$*. In this case, the connected transmitters of the receiver* $Y_j$ *are* $\mathbb{X}_{c\to Y_j}$ *and its unconnected transmitters are* $\mathbb{X}_{c\nrightarrow Y_j}, j=1,\ldots,K_2$*.*

It should be mentioned that a connected transmitter to a certain receiver does not necessarily send messages to that receiver, either separately or cooperatively with other transmitters. On the one hand, the knowledge of messages intended for a certain receiver at its unconnected transmitters may affect the capacity region. Nonetheless, one can derive the following statement.

***Lemma II.6)*** *Consider the general interference network in Fig. 2. Let* $j$ *be an arbitrary index belonging to* $[1:K_2]$*. Denote the set of connected transmitters to receiver* $Y_j$ *by* $\mathbb{X}_{c\to Y_j}$*. Assume that there exists* $M_l \in \{M_1,\ldots,M_K\}$ *such that:*

$$\begin{cases} M_l \in \mathbb{M}_{Y_j} \\ , \\ M_l \notin \bigcup_{X_i \in \mathbb{X}_{c\to Y_j}} \mathbb{M}_{X_i} \end{cases} \tag{II~27}$$

*In other words,* $M_l$ *is required to be decoded at receiver* $Y_j$ *(and possibly at some other receivers), but it is not transmitted by any transmitter connected to* $Y_j$*. In this case, we have:*

$$\mathcal{C}_{\Sigma_{\{M_l\}}}^{GIN} = 0 \tag{II~28}$$

*where* $\mathcal{C}_{\Sigma_{\{M_l\}}}^{GIN}$ *is given by (II~21). Equivalently, the maximum achievable rate* $R_l$ *is zero and the message* $M_l$ *cannot be sent over the network by any positive rate. Thereby, the network is reduced to the scenario in which only the messages* $\{M_1,\ldots,M_K\}-\{M_l\}$ *are transmitted.*

*Proof of Lemma II.6)* Note that based on Lemma II.5, to evaluate $\mathcal{C}_{\Sigma_{\{M_l\}}}^{GIN}$ we can consider the network in which only the message $M_l$ is sent to the receivers. In this case, it is clear that since the message $M_l$ is not transmitted by any transmitter connected to $Y_j$, no positive rate can be achieved. ■

Based on the result of Lemma II.6, in any interference network, without loss of generality, one can assume that for each receiver there is no message known only at its unconnected transmitters, as we will treat in the rest of the paper.

***Definition II.10) General Gaussian Interference Network;***

*The general Gaussian interference network with real-valued input and output signals is given as follows:*

$$\boldsymbol{Y} = \boldsymbol{A}\boldsymbol{X} + \boldsymbol{Z}$$

(II~29)

*where*

$$\boldsymbol{Y} = \begin{bmatrix} Y_1 \\ \vdots \\ Y_{K_2} \end{bmatrix}, \quad \boldsymbol{A} = \left[a_{ji}\right]_{K_2 \times K_1} = \begin{bmatrix} a_{11} & \cdots & a_{1K_1} \\ \vdots & \ddots & \vdots \\ a_{K_2 1} & \cdots & a_{K_2 K_1} \end{bmatrix}, \quad \boldsymbol{X} = \begin{bmatrix} X_1 \\ \vdots \\ X_{K_1} \end{bmatrix}, \quad \boldsymbol{Z} = \begin{bmatrix} Z_1 \\ \vdots \\ Z_{K_2} \end{bmatrix}$$

(II~30)

- *The parameters $\{a_{ji}\}_{\substack{j=1,\dots,K_2 \\ i=1,\dots,K_1}}$ are (fixed) real-valued numbers.*
- *The RVs $\{X_i\}_{i=1}^{K_1}$ are the input symbols: the $i^{th}$ transmitter transmits $X_i$.*
- *The RVs $\{Y_j\}_{j=1}^{K_2}$ are the output symbols: the $j^{th}$ receiver receives $Y_j$.*
- *The noises $\{Z_j\}_{j=1}^{K_2}$ are zero-mean unit-variance Gaussian RVs. Since the capacity region of the network depends only on the marginal PDFs, without loss of generality we assume that the RVs $\{Z_j\}_{j=1}^{K_2}$ are pairwise independent.*

The encoding and decoding scheme for the Gaussian interference network is the same as the discrete network, discussed before. However, in the Gaussian case an average power constraint is also imposed on the codewords of each encoder; thus, the $i^{th}$ encoder is subject to an average power constraint: $\mathbb{E}[X_i^2] \leq P_i$, where $P_i \in \mathbb{R}_+$, $i \in [1:K_1]$. The capacity region of the general Gaussian interference network in (II~29) is denoted by $\mathcal{C}^{GIN\sim G}$.

By these preliminaries, we are ready to launch our study for the basic building blocks.

# III. BASIC STRUCTURES

In this section, we investigate the MAC, the BC, the CIC and the CRC as the basic building blocks of interference networks. First, we briefly review existing results concerning capacity bounds for these channels. We provide some new insights as well. Specifically, we shall prove that dependence between the inputs does not change the well-known strong interference conditions for the two-user CIC. Also, we show that, unlike the recent work [29], to derive achievable rate region for the CRC, splitting the messages into three sub-messages may cause some rate loss besides being superfluous. Next we derive unified capacity outer bounds for the channels with two-receivers based on a unified framework. Subsequently, we establish new capacity results for the two-user CIC and the CRC using the derived outer bounds. Finally, the effect of one-sided receiver side information on the capacity region of the two-user BC, the two-user CIC, and the CRC will be discussed.

## III.A) Channel Models: a Brief Review of Existing Results and New Insights

### III.A.1) The Multiple Access Channel (MAC)

The two-user MAC with common information is a channel in which two transmitters send two private messages as well as a common message to a receiver. Fig. 3 shows the channel model.

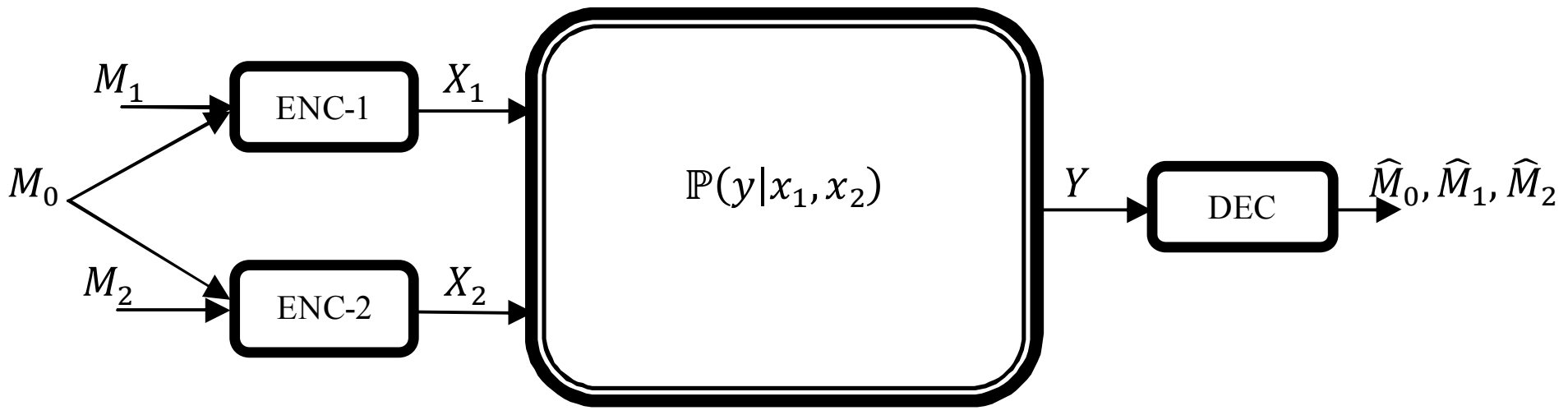

Figure 3. The two-user MAC with Common Message (MACCM).

This channel is obtained from the general interference network defined in Section II.B by setting $K_1 = 2$, $K_2 = 1$, $\mathbb{M} = \mathbb{M}_Y = \{M_0, M_1, M_2\}$, $\mathbb{M}_{X_1} = \{M_0, M_1\}$, and $\mathbb{M}_{X_2} = \{M_0, M_2\}$. The capacity region of the channel was explicitly characterized in [16]. The capacity derivation for the MACs with common information will be dealt with in details in Subsection II of Part II [23]. Here, we omit the discussion.

### III.A.2) The Broadcast Channel (BC)

Next, consider the two-user BC with common message. In this scenario a transmitter sends two private messages as well as a common message to two receivers. Each receiver decodes its respective private message together with the common message. Fig. 4 illustrates the channel.

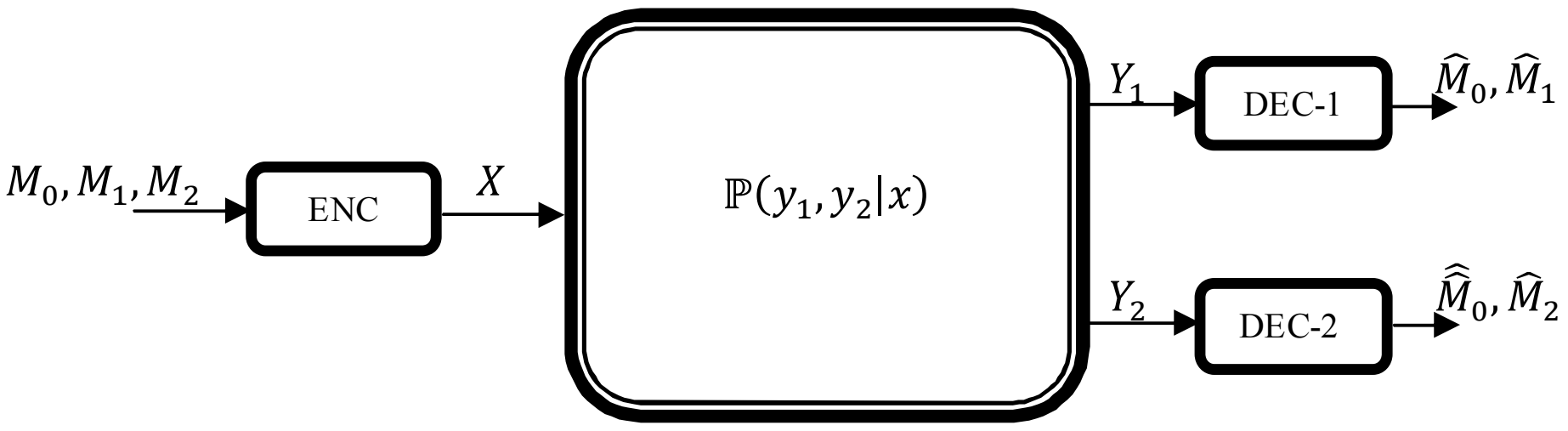

Figure 4. The two-user BC with common message.

This communication scenario is obtained from the general interference network defined in Section II.B by setting $K_1 = 1$, $K_2 = 2$, $\mathbb{M} = \{M_0, M_1, M_2\}$, $\mathbb{M}_X = \{M_0, M_1, M_2\}$, $\mathbb{M}_{Y_1} = \{M_0, M_1\}$, and $\mathbb{M}_{Y_2} = \{M_0, M_2\}$, (note that since there is only one transmitter for the BC, we omit the subscripts from $\mathcal{X}_1$, $X_1$, $x_1$). The Gaussian channel is formulated as follows:

$$\begin{cases} Y_1 = aX + Z_1 \\ Y_2 = bX + Z_2 \end{cases} \tag{III~1}$$

where $Z_1, Z_2$ are zero-mean unit-variance Gaussian RVs and $\mathbb{E}[X^2] \leq P$. Moreover, without loss of generality, we can assume that $|a| \geq |b|$.

The capacity region of the BC is still an open problem. To date, the best achievability scheme for this channel is due to Marton [47-48] as stated in the next proposition.

***Proposition III.1)*** **[47], [106]**, *Define the rate region $\mathfrak{R}_i^{Marton}$ as follows:*

$$\mathfrak{R}_i^{Marton} \triangleq \bigcup_{P_{WUVX}(w,u,v,x)} \left\{ \begin{array}{l} (R_0, R_1, R_2) \in \mathbb{R}_+^3 : \\ R_0 + R_1 \leq I(W, U; Y_1) \\ R_0 + R_2 \leq I(W, V; Y_2) \\ R_0 + R_1 + R_2 \leq I(W, U; Y_1) + I(V; Y_2|W) \\ \qquad -I(U; V|W) \\ R_0 + R_1 + R_2 \leq I(U; Y_1|W) + I(W, V; Y_2) \\ \qquad -I(U; V|W) \\ 2R_0 + R_1 + R_2 \leq I(W, U; Y_1) + I(W, V; Y_2) \\ \qquad -I(U; V|W) \end{array} \right\}$$

(III~2)

*The rate region $\mathfrak{R}_i^{Marton}$ is achievable for the two-user BC with common message.*

We discuss in details the Marton's achievability scheme in Subsection IV.A2. The Marton's rate region (III~2) is optimal in all special cases for which the capacity region is known.

***Definition III.1:*** *The two-user BC is said to be physically degraded if*

$$\mathbb{P}(y_1, y_2|x) = \mathbb{P}(y_1|x)\mathbb{P}(y_2|y_1)$$

(III~3)

*Also, it is said to be stochastically degraded if there exists some transition probability function $\tilde{\mathbb{P}}(y_2|y_1)$ with:*

$$\mathbb{P}(y_2|x) = \sum_{y_1} \mathbb{P}(y_1|x)\tilde{\mathbb{P}}(y_2|y_1)$$

(III~4)

For the degraded BC defined by either (III~3) or (III~4), the receiver $Y_2$ is said to be the degraded component. Note that due to symmetry of the channel, the degraded BC could also be defined by exchanging the indices 1 and 2 in (III~3)-(III~4). Moreover, since the capacity region of the interference channels, specially the BC, depends only on the marginal PDFs of the probability transition function of the channel, no distinction is required to be considered between stochastically degraded and physically degraded BCs (also for other interference networks). We address both cases as the degraded BCs. Note that the scalar Gaussian BC in (III~1) is stochastically degraded; it is indeed equivalent to the following channel:

$$\begin{cases} Y_1 = aX + Z_1 \\ \tilde{Y}_2 = \frac{b}{a}Y_1 + \sqrt{\left(1 - \frac{b^2}{a^2}\right)}\tilde{Z}_2 \end{cases}$$

(III~5)

where $\tilde{Z}_2$ is a zero-mean unit-variance Gaussian RV independent of $Z_2$.

***Definition III.2:*** *The two-user BC is said to be less-noisy if*

$$I(U; Y_2) \leq I(U; Y_1) \qquad \textit{for all PDFs } P_{UX}(u, x)$$

(III~6)

*Also, it is said to be more-capable if*

$$I(X; Y_2) \leq I(X; Y_1) \qquad \textit{for all PDFs } P_X(x)$$

(III~7)

Again, due to symmetry of the channel one could also define the less-noisy and the more-capable BCs by exchanging the indices 1 and 2 in (III~6) and (III~7). We remark that the more-capable BCs strictly contain the less-noisy BCs and the less-noisy BCs strictly contain the degraded BCs as special cases; see [49].

***Proposition III.2)*** **[5, 6], [26]**, *The capacity region of the degraded (III~3)-(III~4) and the less-noisy (III~6) BCs is achieved by using the superposition coding technique and is given by:*

$$\bigcup_{P_{WX}(w,x)} \left\{ \begin{array}{l} (R_0, R_1, R_2) \in \mathbb{R}_+^3: \\ R_1 \leq I(X; Y_1|W) \\ R_0 + R_2 \leq I(W; Y_2) \end{array} \right\} \tag{III~8}$$

*Also, the capacity region of the scalar Gaussian BC (III~1) is given by:*

$$\bigcup_{0\leq\alpha\leq 1} \left\{ \begin{array}{l} (R_0, R_1, R_2) \in \mathbb{R}_+^3 : \\ R_1 \leq \psi(a^2\bar{\alpha}P + 1) \\ R_0 + R_2 \leq \psi\left(\dfrac{b^2\alpha P}{b^2\bar{\alpha}P + 1}\right) \end{array} \right\} \tag{III~9}$$

***Proposition III.3)*** **[49]**, *The capacity region of the more-capable BC (III~7) is given by:*

$$\bigcup_{P_{WX}(w,x)} \left\{ \begin{array}{l} (R_0, R_1, R_2) \in \mathbb{R}_+^3: \\ R_0 + R_2 \leq I(W; Y_2) \\ R_0 + R_1 + R_2 \leq I(X; Y_1|W) + I(W; Y_2) \\ R_0 + R_1 + R_2 \leq I(X; Y_1) \end{array} \right\} \tag{III~10}$$

We remark that by setting $U \equiv X$ and $V \equiv \emptyset$ in (III~2), the Marton's coding is reduced to the superposition coding scheme which achieves the capacity region of the degraded as well as the more-capable channel.

***Definition III.3)*** *The two-user discrete BC is said to be semi-deterministic if the received signal at one of the receivers is a deterministic function of the input signal, i.e., there exists either a deterministic function $f_1(.)$ such that $Y_1 = f_1(X)$, or a deterministic function $f_2(.)$ such that $Y_2 = f_2(X)$.*

***Proposition III.4)*** **[50]**, *The capacity region of the semi-deterministic BC where there exists a deterministic function $f_1(.)$ such that $Y_1 = f_1(X)$ is given by:*

$$\bigcup_{P_{WVX}(w,v,x)} \left\{ \begin{array}{l} (R_0, R_1, R_2) \in \mathbb{R}_+^3 : \\ R_0 \leq min\{I(W; Y_1), I(W; Y_2)\} \\ R_0 + R_1 \leq H(Y_1) \\ R_0 + R_2 \leq I(W, V; Y_2) \\ R_0 + R_1 + R_2 \leq H(Y_1|W, V) + I(W, V; Y_2) \\ R_0 + R_1 + R_2 \leq I(W; Y_1) + H(Y_1|W, V) + I(V; Y_2|W) \end{array} \right\} \tag{III~11}$$

Capacity outer bounds have also been established for the two-user BC in several papers [47], [52-54]. The best known outer bound for this channel with common message was derived in [54, Th. 1], which has been recently simplified to a computable characterization in [55]. Inspired by its auxiliary RVs, this simplified outer bound is called "*UVW-outer bound*" [55]. It should be noted that the author in [55] derived the UVW-outer bound by simplifying that of [54]; however, it could also be deduced from the outer bound provided in [107, Th. 2] for the channel with confidential messages. In the next proposition, we describe the UVW-outer bound.

***Proposition III.5)*** **[55], [107]**, *Define the rate region $\mathfrak{R}_o^{UVW}$ as follows:*

$$\mathfrak{R}_o^{UVW\to BC} \triangleq \bigcup_{P_{UVWX}(u,v,w,x)} \left\{ \begin{array}{l} (R_0,R_1,R_2) \in \mathbb{R}_+^3 : \\ R_0 \leq \min\{I(W;Y_1), I(W;Y_2)\} \\ R_0 + R_1 \leq I(U;Y_1|W) + \min\{I(W;Y_1), I(W;Y_2)\} \\ R_0 + R_2 \leq I(V;Y_2|W) + \min\{I(W;Y_1), I(W;Y_2)\} \\ R_0 + R_1 + R_2 \leq I(X;Y_1|V,W) + I(V;Y_2|W) + \min\{I(W;Y_1), I(W;Y_2)\} \\ R_0 + R_1 + R_2 \leq I(X;Y_2|U,W) + I(U;Y_1|W) + \min\{I(W;Y_1), I(W;Y_2)\} \end{array} \right\}$$

(III~12)

*The rate region* $\mathfrak{R}_o^{UVW}$ *constitutes an outer bound for the two-user BC with common message.*

The UVW-outer bound is optimal (and coincides with the Marton's achievable rate region (III~2)) in all special cases of the BC for which the capacity region is known. Also, as demonstrated in [55], the UVW-outer bound for the case of no common information can be further simplified to the following characterization called the "*UV-outer bound*".

$$\mathfrak{R}_o^{UV\to BC} \triangleq \bigcup_{P_{UVX}(u,v,x)} \left\{ \begin{array}{l} (R_1,R_2) \in \mathbb{R}_+^2 : \\ R_1 \leq I(U;Y_1) \\ R_2 \leq I(V;Y_2) \\ R_1 + R_2 \leq I(X;Y_1|V) + I(V;Y_2) \\ R_1 + R_2 \leq I(X;Y_2|U) + I(U;Y_1) \end{array} \right\}$$

(III~13)

Therefore, based on the result of [55], the UV-outer bound is the tightest known capacity outer bound for the two-user BC without common message. However, recently it is shown that this outer bound is strictly sub-optimal [56].

We now intend to present a class of BCs with the interesting property that "*the capacity is achieved by decoding both messages at both receivers*". Let us first derive the achievable rate region from the coding scheme where both messages are required at both receivers. It can be easily seen that this coding scheme yields the following achievable rate region for the two-user BC with common message:

$$\bigcup_{P_X(x)} \left\{ \begin{array}{l} (R_0,R_1,R_2) \in \mathbb{R}_+^3 : \\ R_0 + R_1 + R_2 \leq I(X;Y_1) \\ R_0 + R_1 + R_2 \leq I(X;Y_2) \end{array} \right\}$$

(III~14)

Hence, we need to derive the conditions under which the rate region (III~14) is optimal. We present such conditions in the following lemma.

***Proposition III.6)*** *Consider the two-user BC satisfying the following condition:*

$$I(X;Y_2) = I(X;Y_1) \quad \textit{for all } P_X(x)$$

(III~15)

*The capacity region is given by (III~14).*

*Proof of Proposition III.6)* It is sufficient to prove the optimality of (III~14). To see this, let $(R_0,R_1,R_2) \in \mathbb{R}_+^3$ be an achievable rate triple for the channel. Note that the condition (III~15) can be restated as follows:

$$\begin{cases} I(X;Y_1) \leq I(X;Y_2) \\ I(X;Y_2) \leq I(X;Y_1) \end{cases}$$

(III~16)

for all PDFs $P_X(x)$. Now from the last two constraints of the WUV-outer bound in (III~12) we have:

$$R_0 + R_1 + R_2 \leq I(X;Y_1|V,W) + I(V,W;Y_2) \overset{(a)}{\leq} I(X;Y_2|V,W) + I(V,W;Y_2) = I(X;Y_2)$$
$$R_0 + R_1 + R_2 \leq I(X;Y_2|U,W) + I(U,W;Y_1) \overset{(b)}{\leq} I(X;Y_1|U,W) + I(U,W;Y_1) = I(X;Y_1)$$

where (a) is due the first inequality of (III~16) and (b) is due to the second one. ■

It is remarkable to note that the conditions (III~16) to some extent are reminiscent of those derived in [14, 15] (see conditions (III~21) below) for the two-user CIC under which decoding both messages at both receivers is optimal. In fact, by viewing the BC as an *interference network*, we can interpret the conditions (III~16) as the *strong interference regime*. This insight indeed constitutes a foundation to derive a strong interference regime for all interference networks with any arbitrary configuration. Our criterion to this development is that in the strong interference regime the decoding of all messages at all receivers is optimal. This will be treated in details in Part III of our multi-part papers [27].

## III.A.3) The Classical Interference Channel (CIC)

The third scenario that we consider as a main building block of the interference networks is the two-user CIC. In this channel, two transmitters send independent messages to their respective receivers. Fig. 5 illustrates the channel model.

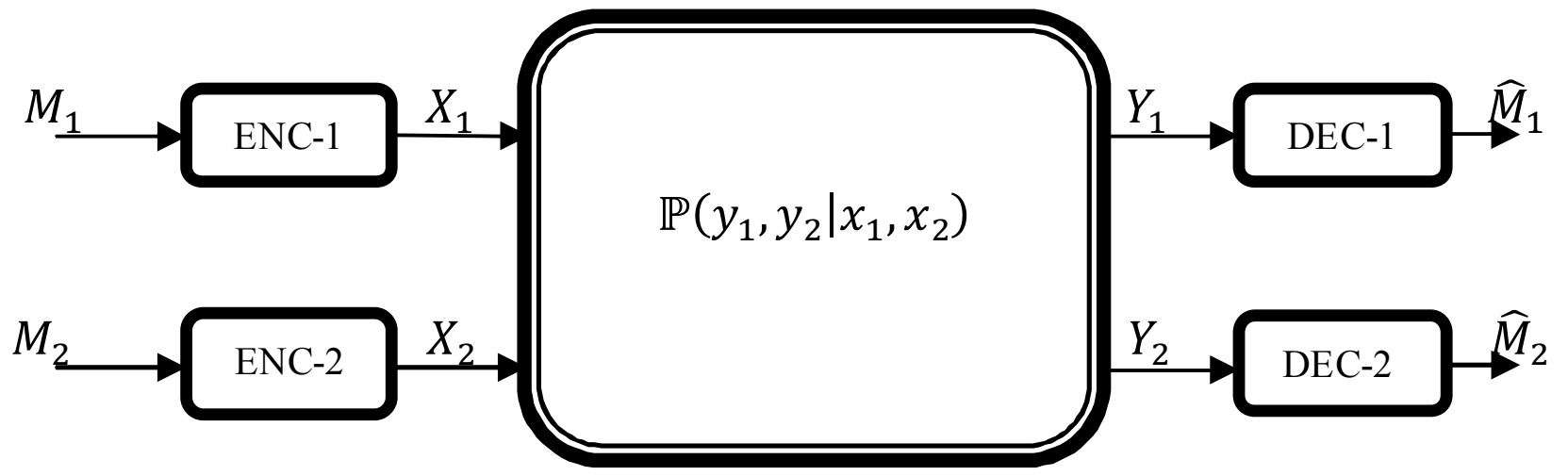

Figure 5. The two-user CIC.

The two-user CIC is obtained from the general interference network defined in Section II.B by setting $K_1 = K_2 = 2$, $\mathbb{M} = \{M_1, M_2\}$, $\mathbb{M}_{X_1} = \mathbb{M}_{Y_1} = \{M_1\}$, and $\mathbb{M}_{X_2} = \mathbb{M}_{Y_2} = \{M_2\}$. The Gaussian channel is usually formulated in the following standard form:

$$\begin{cases} Y_1 = X_1 + aX_2 + Z_1 \\ Y_2 = bX_1 + X_2 + Z_2 \end{cases}$$

(III~17)

where $Z_1, Z_2$ are zero-mean unit-variance Gaussian RVs, and $\mathbb{E}[X_i^2] \leq P_i, i = 1,2$.

As mentioned in the introduction, the best capacity inner bound for the CIC is due to Han and Kobayashi which was derived in 1981 [67]. A compact description of the HK rate region was provided in [68]. In the following, we describe this compact version.

***Proposition III.7)* [67, 68]**, *Define the rate region* $\mathfrak{R}_i^{HKc}$ *as follows*:

$$\mathfrak{R}_i^{HKc} \triangleq \bigcup_{\mathcal{P}_i^{HKc}} \left\{ \begin{array}{l} (R_1, R_2) \in \mathbb{R}_+^2 : \\ R_1 \leq I(X_1; Y_1|W_2, Q) \\ R_2 \leq I(X_2; Y_2|W_1, Q) \\ R_1 + R_2 \leq I(X_1, W_2; Y_1|Q) + I(X_2; Y_2|W_1, W_2, Q) \\ R_1 + R_2 \leq I(X_1; Y_1|W_1, W_2, Q) + I(X_2, W_1; Y_2|Q) \\ R_1 + R_2 \leq I(X_1, W_2; Y_1|W_1, Q) + I(X_2, W_1; Y_2|W_2, Q) \\ 2R_1 + R_2 \leq I(X_1, W_2; Y_1|Q) + I(X_1; Y_1|W_1, W_2, Q) \\ \qquad\qquad +I(X_2, W_1; Y_2|W_2, Q) \\ R_1 + 2R_2 \leq I(X_2; Y_2|W_1, W_2, Q) + I(X_2, W_1; Y_2|Q) \\ \qquad\qquad +I(X_1, W_2; Y_1|W_1, Q) \end{array} \right\}$$

(III~18)

*where* $\mathcal{P}_i^{HKc}$ *denotes the set of all joint PDFs* $P_{QW_1W_2X_1X_2}(q, w_1, w_2, x_1, x_2)$ *satisfying:*

$$P_{QW_1W_2X_1X_2}(q, w_1, w_2, x_1, x_2) = P_Q(q)P_{W_1|Q}(w_1|q)P_{W_2|Q}(w_2|q)P_{X_1|W_1Q}(x_1|w_1, q)P_{X_2|W_2Q}(x_2|w_2, q)$$

(III~19)

*The set* $\mathfrak{R}_i^{HKc}$ *is an achievable rate region for the two-user CIC.*

The achievable rate region $\mathfrak{R}_i^{HK_c}$ can be directly obtained using the rate splitting technique as well as the superposition coding strategy [68, Lemma 3]. This rate region has a compact description compared to the initial HK characterization [67]. In the initial HK scheme for the two-user CIC, each encoder splits its message into two parts named the *common part* and the *private part*, and then these two parts are encoded using independent codewords. The encoders send a deterministic function of these two independent codewords over the channel. By incorporating a jointly decoding technique, each receiver decodes the common part of the messages of both transmitters as well as the private part of its respective transmitter. The rate region due to this simple scheme, if the time-sharing is employed, is equivalent to $\mathfrak{R}_i^{HK_c}$ given in (III~18); see [68] for details. The HK coding strategy achieves the capacity region of the Gaussian channel (III~17) to within half a bit [66]. It is also optimal for a class of deterministic CICs [62].

Regarding capacity outer bounds for the two-user CIC, several results have been recently established for the Gaussian channel [20-22]. These outer bounds which are derived using the genie-aided techniques, are tighter than previous results derived in [66] and [69-71]; however, their characterization is rather complex. For the discrete channel, one of the earliest results is due to Sato derived in 1977. We state this outer bound in the next proposition.

***Proposition III.8)*** **[72]**, *Define the rate region* $\mathfrak{R}_o^{Sato}$ *as follows:*

$$\mathfrak{R}_o^{Sato} \triangleq \bigcap_{\mathcal{P}_{\widetilde{CIC}}} \left( \bigcup_{P_Q P_{X_1|Q} P_{X_2|Q}} \left\{ \begin{array}{l} (R_1, R_2) \in \mathbb{R}_+^2 : \\ R_1 \le I(X_1; Y_1 | X_2, Q) \\ R_2 \le I(X_2; Y_2 | X_1, Q) \\ R_1 + R_2 \le I(X_1, X_2; Y_1, Y_2 | Q) \end{array} \right\} \right) \tag{III~20}$$

*where* $\mathcal{P}_{\widetilde{CIC}}$ *denotes the set of all conditional PDFs* $\widetilde{\mathbb{P}}(y_1, y_2 | x_1, x_2)$ *which have the same marginal distributions as the transition probability function of the channel, i.e.,* $\mathbb{P}(y_1, y_2 | x_1, x_2)$.

The Sato's outer bound is obtained from the general cut-set outer bound for communication networks [4] and the fact that the capacity of the CIC depends only on the marginal distributions of $\mathbb{P}(y_1, y_2 | x_1, x_2)$. Other outer bounds could be found in [71-74]. Capacity results also have been obtained for the two-user CIC for some special cases [14, 15], [62, 63]. Specifically, the capacity region is known for the channels with strong interference.

***Definition III.4:*** **[14, 15],** *The two-user CIC is said to have strong interference if*

$$\begin{cases} I(X_1; Y_1 | X_2) \le I(X_1; Y_2 | X_2) \\ I(X_2; Y_2 | X_1) \le I(X_2; Y_1 | X_1) \end{cases} \tag{III~21}$$

*for all joint PDFs* $P_{X_1}(x_1) P_{X_2}(x_2)$. *Also, the two-user Gaussian CIC (III~17) is said to have strong interference if*

$$|a| \ge 1, \qquad |b| \ge 1 \tag{III~22}$$

Note that the conditions (III~22) can be obtained from (III~21) by setting the input distributions to be Gaussian. The inequalities (III~21) and (III~22) represent the case where the amount of information flowing in the undesired directions (referred to as interference) is greater than those flowing in the desired direction, hence the *strong interference* regime.

***Proposition III.9)*** **[14, 15]**, *The capacity region of the discrete CIC with strong interference (III~21) is given by:*

$$\bigcup_{P_Q P_{X_1|Q} P_{X_2|Q}} \left\{ \begin{array}{l} (R_1, R_2) \in \mathbb{R}_+^2 : \\ R_1 \le I(X_1; Y_1 | X_2, Q) \\ R_2 \le I(X_2; Y_2 | X_1, Q) \\ R_1 + R_2 \le \min\big(I(X_1, X_2; Y_1 | Q), I(X_1, X_2; Y_2 | Q)\big) \end{array} \right\} \tag{III~23}$$

*Also, the capacity region of the Gaussian channel (III~17) with strong interference (III~21) is given by:*

$$\bigcup\left\{\begin{array}{l}(R_1,R_2)\in\mathbb{R}_+^2:\\ R_1\leq\psi(P_1)\\ R_2\leq\psi(P_2)\\ R_1+R_2\leq\min\big(\psi(P_1+a^2P_2),\psi(b^2P_1+P_2)\big)\end{array}\right\}$$

(III~24)

For both the discrete and the Gaussian CICs with strong interference, it was shown that the simple coding scheme based on which each receiver jointly decodes both messages (equivalently by viewing the channel as a *compound MAC* [108]) is optimal and achieves the capacity region. The HK rate region (III~18) is also optimal in the strong interference regime.

Here, we prove two important lemmas regarding the conditions in (III~21) and (III~22). First, note that the strong interference regime was determined by holding the inequalities (III~21) for all product input distributions. It may be thought that this regime will be changed if we impose that (III~21) holds for all correlated input distributions. However, this is not the case. In the following, we prove if the two-user CIC satisfies (III~21) for all product distributions then it also satisfies these conditions for all correlated distributions.

***Lemma III.1)*** *For the two-user CIC, the dependence between the input variables, $X_1$ and $X_2$, does not change the strong interference condition, which means if the inequalities (III~21) hold for all product input distributions $P_{X_1}(x_1)P_{X_2}(x_2)$, they also hold for all correlated input distributions $P_{X_1X_2}(x_1,x_2)$.*

*Proof of Lemma III.1)* Let us assume that the inequalities (III~21) hold for all product distributions $P_{X_1}(x_1)P_{X_2}(x_2)$. Consider the first inequality that is $I(X_1;Y_1|X_2)\leq I(X_1;Y_2|X_2)$. As remarked in [15, App], this yields the following inequality:

$$I(X_1;Y_1|X_2,U)\leq I(X_1;Y_2|X_2,U)$$

(III~25)

for all $P_U(u)P_{X_1|U}(x_1|u)P_{X_2|U}(x_2|u)$. Note that such PDFs imply the Markov chain $X_1\to U\to X_2$. Consider a joint PDF $P_{VX_1X_2}(v,x_1,x_2)$, where $V$ is an arbitrary RV. Now, substitute $U\equiv(V,X_2)$ in (III~25). Note that, by this choice for $U$, the Markov relation $X_1\to U=(V,X_2)\to X_2$ is clearly guaranteed. We obtain:

$$I(X_1;Y_1|X_2,V)\leq I(X_1;Y_2|X_2,V)$$

(III~26)

for all joint PDFs $P_{VX_1X_2}(v,x_1,x_2)$. This is a stronger result than the desired one. ■

***Remark III.1)*** The inequality (III~26) holds for those random variables $V$ such that $V\to X_1,X_2\to Y_1,Y_2$ forms a Markov chain. For example, it is wrong to set $V\equiv Y_2$ in (III~26) and deduce that the first inequality of (III~21) implies $I(X_1;Y_1|X_2,Y_2)=0$. In general, (III~21) does not imply the latter equality. See Part III [27, Remark 1] for a detailed discussion.

In Part III [27, Sec II], we will provide a generalization of Lemma III.1 to the multivariate case. The remarkable fact included in this lemma leads to a deep insight regarding the behavior of information flow in large interference networks with strong interference. The details can be found in Part III [27].

Next, consider the strong interference conditions given by (III~22) for the Gaussian channel. As mentioned before, these conditions are derived by setting the input distributions in (III~21) to be Gaussian. Now the natural question is that given (III~22), whether the conditions (III~21) hold for all joint PDFs $P_{X_1}(x_1)P_{X_2}(x_2)$ that are not necessarily Gaussian? In the next lemma, we prove this is the case.

***Lemma III.2)*** *For the two-user Gaussian CIC in (III~17), the strong interference conditions (III~22) are equivalent to (III~21).*

*Proof of Lemma III.2)* It is sufficient to show that (III~22) implies the inequalities (III~21) for all joint PDFs $P_{X_1}(x_1)P_{X_2}(x_2)$. Consider the condition $|b|\geq 1$. Define:

$$\tilde{Y}_1\triangleq\frac{Y_2}{b}+\left(a-\frac{1}{b}\right)X_2+\sqrt{\left(1-\frac{1}{b^2}\right)}\tilde{Z}_1$$

(III~27)

where $\tilde{Z}_1$ is a Gaussian RV with zero mean and unit variance and independent of $Z_2$. By considering (III~27), it is readily derived that $\tilde{Y}_1 \approx Y_1$ in the sense that $\mathbb{P}(\tilde{y}_1|x_1,x_2) \equiv \mathbb{P}(y_1|x_1,x_2)$. Therefore, for all joint PDFs $P_{X_1}(x_1)P_{X_2}(x_2)$ we have:

$$\begin{aligned} I(X_1;Y_1|X_2) &= I(X_1;\tilde{Y}_1|X_2) \leq I(X_1;\tilde{Y}_1,Y_2|X_2) \\ &= I(X_1;Y_2|X_2) + I(X_1;\tilde{Y}_1|Y_2,X_2) \overset{(a)}{=} I(X_1;Y_2|X_2) \end{aligned}$$

where equality (a) holds because $X_1 \to Y_2, X_2 \to \tilde{Y}_1$ forms a Markov chain. In fact, for the channel with $|b| \geq 1$, given $X_2$, the receiver $Y_1$ is a (stochastically) degraded version of $Y_2$. Similarly, one can prove the second inequality of (III~21). ■

For the Gaussian CIC (III~17), in addition to the strong interference regime, the sum-rate capacity is known for two other cases: the mixed interference and the noisy interference regimes. We present these results in the next propositions.

***Proposition III.10)*** **[20]**, *The sum-rate capacity of the Gaussian CIC (III~17) in the mixed interference regime, determined by* $|b| < 1 \leq |a|$, *is given below:*

$$\mathcal{C}_{sum}^{CIC_{mi}\sim G} = \min\left\{\begin{array}{c} \psi(P_1 + a^2P_2), \\ \psi(P_1) + \psi\left(\dfrac{P_2}{b^2P_1+1}\right) \end{array}\right\} \tag{III\sim28}$$

In the mixed interference regime with $|b| < 1 \leq |a|$, the receiver $Y_1$ perceives strong interference while the receiver $Y_2$ perceives weak interference. In order to achieve the sum-rate capacity, the receiver $Y_1$ decodes both the interference and its corresponding signal, while the receiver $Y_2$ treats the interference as noise and only decodes its own signal. Note that a similar result is derived symmetrically for channels with $|a| < 1 \leq |b|$.

***Definition III.5)*** *The two-user Gaussian CIC (III~17) is said to have noisy interference provided that:*

$$|a(b^2P_1+1)| + |b(a^2P_2+1)| \leq 1 \tag{III\sim29}$$

***Proposition III.11)*** **[20-22]**, *The sum-rate capacity of the Gaussian CIC (III~17) in the noisy interference regime (III~29) is given by:*

$$\mathcal{C}_{sum}^{CIC_{ni}\sim G} = \psi\left(\frac{P_1}{a^2P_2+1}\right) + \psi\left(\frac{P_2}{b^2P_1+1}\right) \tag{III\sim30}$$

In the noisy interference regime (III~29), treating interference as noise at both receivers achieves the sum-rate capacity. This result is derived using the outer bounds established for the Gaussian channel using the genie-aided techniques.

A special case of the two-user CICs is the class of ZICs. In these channels, the signal of one receiver does not depend on the signal of the non-respective transmitter. Due to this reason, these channels are sometimes called the *one-sided* CICs.

***Definition III.6)*** *The two-user ZIC (the one-sided CIC) is a special case of the CIC for which either one of the conditions below:*

$$\begin{cases} \mathbb{P}(y_1,y_2|x_1,x_2) = \mathbb{P}(y_1|x_1,x_2)\mathbb{P}(y_2|x_2) & (a) \\ \mathbb{P}(y_1,y_2|x_1,x_2) = \mathbb{P}(y_1|x_1)\mathbb{P}(y_2|x_1,x_2) & (b) \end{cases} \tag{III\sim31}$$

*is satisfied. The Gaussian ZIC is also derived by setting either* $a = 0$ *or* $b = 0$ *in (III~17).*

In our multi-part papers, without loss of generality (due to symmetry of the channel), we only consider the one-sided CICs that satisfy the condition (III~31-a). Similarly, for the Gaussian channel, we only consider the case of $b = 0$ as follows:

$$\begin{cases} Y_1 = X_1 + aX_2 + Z_1 \\ Y_2 = X_2 + Z_2 \end{cases}$$

(III~32)

where other channel parameters (power constraints on the input signals and noise parameters) are the same as (III~17). The capacity region of Gaussian ZIC in (III~32) is known for the case of strong interference where $|a| \geq 1$. Also, for the weak interference regime where $|a| < 1$, the sum-rate capacity has been established. We present these results in the next proposition.

***Proposition III.12)*** **[109]**, *The capacity region of the Gaussian ZIC (III~32) with strong interference* $|a| \geq 1$ *is given by:*

$$\mathcal{C}^{ZIC_{si}\sim G} = \bigcup \left\{ \begin{array}{l} (R_1, R_2) \in \mathbb{R}_+^2 : \\ R_1 \leq \psi(P_1) \\ R_2 \leq \psi(P_2) \\ R_1 + R_2 \leq \psi(P_1 + a^2 P_2) \end{array} \right\}$$

(III~33)

*Also, for the case of weak interference, i.e.,* $|a| < 1$, *the sum-rate capacity is given by:*

$$\mathcal{C}_{sum}^{ZIC_{wi}\sim G} = \psi\left(\frac{P_1}{a^2 P_2 + 1}\right) + \psi(P_2)$$

(III~34)

To achieve the rate region (III~33) in the strong interference channel, the receiver $Y_1$ jointly decodes both messages. In the weak interference channel, this receiver treats interference as noise to achieve the sum-rate (III~34). Note that in both cases the receiver $Y_2$, which perceives no interference, decodes its own message.

## III.A.4) The Cognitive Radio Channel (CRC)

Finally, we consider the CRC. In this channel, two transmitters send independent messages to two receivers such that the first transmitter named the "*cognitive transmitter*" has access non-causally to the message of the second transmitter named the "*primary transmitter*". Each receiver decodes one of the messages. Fig. 6 illustrates the channel model.

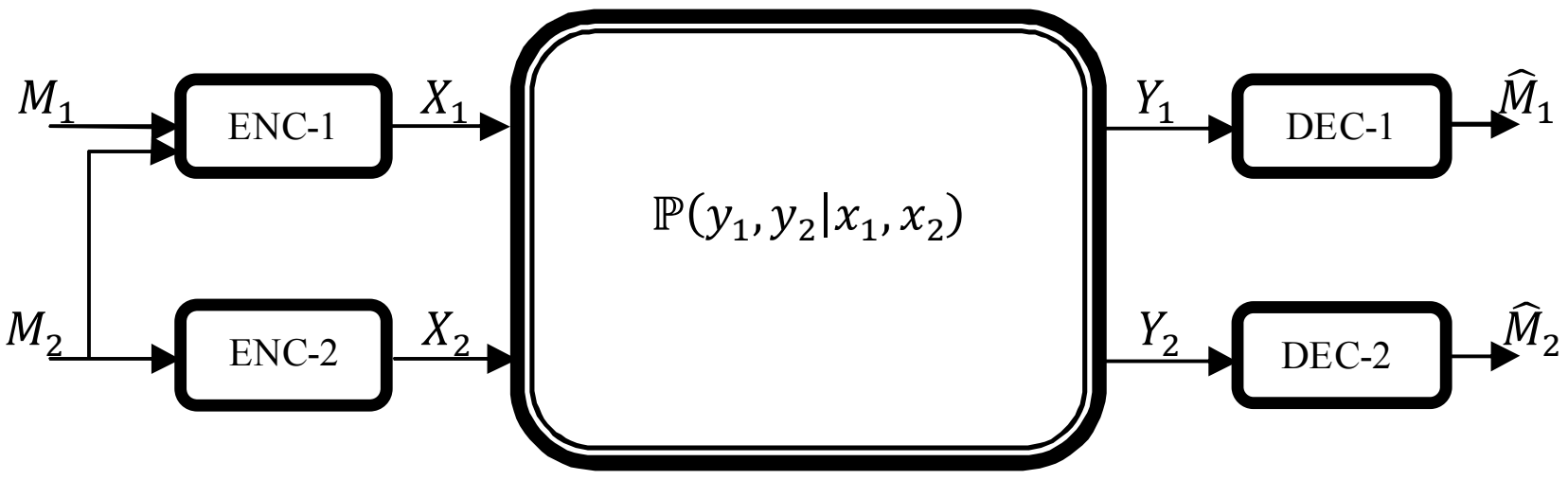

Figure 6. The CRC: $X_1$ is the cognitive transmitter, $X_2$ is the primary transmitter, $Y_1$ is the cognitive receiver and $Y_2$ is the primary receiver

The CRC is obtained from the general interference network defined in Section II.B by setting $K_1 = K_2 = 2$, $\mathbb{M} = \{M_1, M_2\}$, $\mathbb{M}_{X_1} = \{M_1, M_2\}$, $\mathbb{M}_{Y_1} = \{M_1\}$, and $\mathbb{M}_{X_2} = \mathbb{M}_{Y_2} = \{M_2\}$. Similar to the CIC, the Gaussian CRC is formulated by (III~17).

Since its introduction in 2006, several achievability schemes were proposed for the CRC [34-38]. Recently the authors in [29] presented a new scheme for the channel and showed that their achievable rate region encompasses the previously derived ones as special cases. Nevertheless, due to the lack of systematic derivation, some redundant complexities have been introduced in the achievability scheme. Let us first present this achievable rate region in the next proposition.

***Proposition III.13)*** **[29]**, *Define the rate region* $\mathfrak{R}_i^{RTD}$ *as follows:*

$$\mathfrak{R}_i^{RTD} \triangleq \bigcup_{\mathcal{P}_i^{RTD}} \left\{ \begin{array}{r} (R_1,R_2) \in \mathbb{R}_+^2 : \ \exists \left(\acute{R}_{1c},\acute{R}_{1pb},\acute{R}_{2pb},R_{1c},R_{1pb},R_{2c},R_{2pa},R_{2pb}\right) \in \mathbb{R}_+^8 : \\ R_1 = R_{1c} + R_{1pb} \\ R_2 = R_{2c} + R_{2pa} + R_{2pb} \\ \acute{R}_{1c} \geq I(X_2;U_{1c}|U_{2c}) \\ \acute{R}_{1c} + \acute{R}_{1pb} \geq I(X_2;U_{1c}|U_{2c}) + I\left(X_2;U_{1pb}\middle|U_{2c},U_{1c}\right) \\ \acute{R}_{1c} + \acute{R}_{1pb} + \acute{R}_{2pb} \geq I(X_2;U_{1c}|U_{2c}) + I\left(X_2;U_{1pb}\middle|U_{2c},U_{1c}\right) \\ +I\left(U_{1pb};U_{2pb}\middle|U_{2c},X_2,U_{1c}\right) \\ R_{1pb} + \acute{R}_{1pb} \leq I\left(U_{1pb};Y_1\middle|U_{2c},U_{1c}\right) \\ R_{1c} + \acute{R}_{1c} + R_{1pb} + \acute{R}_{1pb} \leq I\left(U_{1c},U_{1pb};Y_1\middle|U_{2c}\right) \\ R_{2c} + R_{1c} + \acute{R}_{1c} + R_{1pb} + \acute{R}_{1pb} \leq I\left(U_{2c},U_{1c},U_{1pb};Y_1\right) \\ R_{2pb} + \acute{R}_{2pb} \leq I\left(U_{2pb};Y_2\middle|U_{1c},X_2,U_{2c}\right) \\ R_{2pa} + R_{2pb} + \acute{R}_{2pb} \leq I\left(X_2,U_{2pb};Y_2\middle|U_{2c},U_{1c}\right) + I(X_2;U_{1c}|U_{2c}) \\ R_{1c} + \acute{R}_{1c} + R_{2pb} + \acute{R}_{2pb} \leq I\left(U_{2pb},U_{1c};Y_2\middle|X_2,U_{2c}\right) + I(X_2;U_{1c}|U_{2c}) \\ R_{2pa} + R_{1c} + \acute{R}_{1c} + R_{2pb} + \acute{R}_{2pb} \leq I\left(X_2,U_{1c},U_{2pb};Y_2\middle|U_{2c}\right) + I(X_2;U_{1c}|U_{2c}) \\ R_{2c} + R_{2pa} + R_{1c} + \acute{R}_{1c} + R_{2pb} + \acute{R}_{2pb} \leq I\left(U_{2c},X_2,U_{1c},U_{2pb};Y_2\right) + I(X_2;U_{1c}|U_{2c}) \end{array} \right\}$$

(III~35)

*where $\mathcal{P}_i^{RTD}$ denotes the set of all joint PDFs $P_{U_{1c}U_{2c}U_{1pb}U_{2pb}X_1X_2}(u_{1c},u_{2c},u_{1pb},u_{2pb},x_1,x_2)$. The set $\mathfrak{R}_i^{RTD}$ is an achievable rate region for the CRC.*

To derive this achievable rate region, the authors of [29] use a message splitting technique wherein the message $M_1$ and the message $M_2$ are split into two and three parts, respectively, as shown below:

$$\begin{cases} M_1 \triangleq \left(M_{1c},M_{1pb}\right) \\ M_2 \triangleq \left(M_{2c},M_{2pa},M_{2pb}\right) \end{cases}$$

(III~36)

One of the main features of the coding scheme provided in [29] is that the primary transmitter ignores some parts of its message, i.e., $M_{2pb}$ and relegates its transmission to the cognitive transmitter. The message $M_{2pb}$ then collaborates with the messages $M_{1c}, M_{1pb}$ to build a Marton's type coding scheme [47-48] at the cognitive transmitter. But, is this idea beneficial? At the end of this subsection, we will demonstrate that this is not the case at all and ignoring some parts of the message $M_2$ at the primary transmitter only makes the resulting achievable rate region weaker. In fact, we will present a simpler achievability scheme that results a potentially larger rate region.

Regarding capacity outer bounds for the CRC, the best bound is due to Maric *et al* [36] that is stated in the following proposition.

***Proposition III.14)*** **[36]**, *Define the rate region $\mathfrak{R}_o^{MGKS}$ as follows:*

$$\mathfrak{R}_o^{MGKS} \triangleq \bigcup_{\mathcal{P}_o^{MGKS}} \left\{ \begin{array}{l} (R_1,R_2) \in \mathbb{R}_+^2 : \\ R_1 \leq I(V,U_1;Y_1) \\ R_2 \leq I(V,U_2,X_2;Y_2) \\ R_1 + R_2 \leq I(X_1;Y_1|X_2,U_2,V) + I(V,U_2,X_2;Y_2) \\ R_1 + R_2 \leq I(X_1,X_2;Y_2|U_1,V) + I(V,U_1;Y_1) \end{array} \right\}$$

(III~37)

*where $\mathcal{P}_o^{MGKS}$ denotes the set of all joint PDFs $P_{VU_1U_2X_1X_2}(v,u_1,u_2,x_1,x_1)$ satisfying:*

$$P_{U_1U_2VX_1X_2}(v,u_1,u_2,x_1,x_1) = P_{U_1}(u_1)P_{U_2}(u_2)P_{V|U_1U_2}(v|u_1,u_2)P_{X_1|U_1U_2V}(x_1|u_1,u_2,v)P_{X_2|U_2V}(x_2|u_2,v)$$

(III~38)

*as well as, $P_{X_1|U_1U_2V}(x_1|u_1,u_2,v) \in \{0,1\}$, and $P_{X_2|U_2V}(x_2|u_2,v) \in \{0,1\}$; in other words, $X_1$ is a deterministic function of $(U_1,U_2,V)$, and $X_2$ is a deterministic function of $(U_2,V)$. Then the rate region $\mathfrak{R}_o^{MGKS}$ constitutes an outer bound on the capacity region of the CRC.*

For the CRC, the capacity region has been established for some special cases. Specifically, inspired by the CIC, a strong interference regime was defined for this channel in [25].

***Definition III.7:*** **[25]**, *The CRC is said to have strong interference if*

$$\begin{cases} I(X_1,X_2;Y_2) \leq I(X_1,X_2;Y_1) & (a) \\ I(X_1;Y_1|X_2) \leq I(X_1;Y_2|X_2) & (b) \end{cases}$$

(III~39)

*for all joint PDFs* $P_{X_1X_2}(x_1,x_2)$. *Also, the Gaussian CRC is said to have strong interference if*

$$\begin{cases} |1+a\alpha| \geq |b+\alpha| \\ |1-a\alpha| \geq |b-\alpha|, & \alpha \triangleq \sqrt{\dfrac{P_1}{P_2}} \\ |b| \geq 1 \end{cases}$$

(III~40)

Note that the conditions (III~39) are derived from (III~40) by setting the input distributions to be Gaussian. Similar to the CIC, it was shown in [25] that the capacity region of the CRC under the strong interference conditions in (III~39) and (III~40) is achieved by requiring that each receiver decodes both messages. In other words, the capacity region in this regime coincides with that of a cognitive channel where both messages should be decoded at both receivers.

***Proposition III.15)*** **[25]**, *The capacity region of the discrete CRC with strong interference (III~39) is given by:*

$$\bigcup_{P_{X_1X_2}(x_1,x_2)} \begin{Bmatrix} (R_1,R_2) \in \mathbb{R}_+^2 : \\ R_1 \leq I(X_1;Y_1|X_2) \\ R_1+R_2 \leq I(X_1,X_2;Y_2) \end{Bmatrix}$$

(III~41)

*Also, the capacity region of the Gaussian CRC, formulated as (III~17), in the strong interference regime (III~40) is given by:*

$$\bigcup_{0\leq\alpha\leq 1} \begin{Bmatrix} (R_1,R_2) \in \mathbb{R}_+^2 : \\ R_2 \leq \psi(\alpha P_1) \\ R_1+R_2 \leq \psi\left(b^2P_1+P_2+2|b|\sqrt{(1-\alpha)P_1P_2}\right) \end{Bmatrix}$$

(III~42)

It is remarkable that, unlike the Gaussian CIC for which the strong interference conditions (III~21) are equivalent to those in (III~22), for the Gaussian CRC, the conditions (III~39) *are not equivalent* to those in (III~40). To explain this fact, we need the following lemma.

***Lemma III.3)*** *Consider the CRC in Fig. 6. Assume that the channel satisfies* $I(X_1,X_2;Y_2) \leq I(X_1,X_2;Y_1)$ *for all joint PDFs* $P_{X_1X_2}(x_1,x_2)$.*Then, we have:*

$$I(X_1,X_2;Y_2|U) \leq I(X_1,X_2;Y_1|U)$$

(III~43)

*for all joint PDFs* $P_{UX_1X_2}(u,x_1,x_2)$. *Specifically, by setting either* $U \equiv X_1$ *or* $U \equiv X_2$ *in (III~43), we obtain:*

$$\begin{cases} I(X_1;Y_2|X_2) \leq I(X_1;Y_1|X_2) \\ I(X_2;Y_2|X_1) \leq I(X_2;Y_1|X_1) \end{cases} \quad \text{for all joint PDFs } P_{X_1X_2}(x_1,x_2)$$

(III~44)

*Proof of Lemma III.3)* We can write:

$$
\begin{aligned}
I(X_1,X_2;Y_2|U) &= \sum_u P_U(u) I(X_1,X_2;Y_2|U=u) \\
&= \sum_u P_U(u) I(X_1,X_2;Y_2)_{\langle P_{X_1X_2|u}(x_1,x_2|u)\rangle} \\
&\overset{(a)}{\leq} \sum_u P_U(u) I(X_1,X_2;Y_1)_{\langle P_{X_1X_2|u}(x_1,x_2|u)\rangle} \\
&= \sum_u P_U(u) I(X_1,X_2;Y_1|U=u) \\
&= I(X_1,X_2;Y_1|U)
\end{aligned}
$$

where the notation $I(A;B|C)_{\langle P(.)\rangle}$ indicates that the mutual information function $I(A;B|C)$ is evaluated by the distribution $P(.)$. Note that for each given $u$, the function $P_{X_1X_2|u}(x_1,x_2|u)$ is a probability distribution defined over the set $\mathcal{X}_1 \times \mathcal{X}_2$. The inequality (a) holds because $I(X_1,X_2;Y_2) \leq I(X_1,X_2;Y_1)$ for all joint PDFs on the set $\mathcal{X}_1 \times \mathcal{X}_2$, including $P_{X_1X_2|u}(x_1,x_2|u)$. ■

Now consider the conditions (III~40). As mentioned before, these conditions are derived by evaluating (III~39) for Gaussian input distributions; nevertheless, (III~39) do not imply that the inequalities (III~40) hold for all joint PDFs $P_{X_1X_2}(x_1,x_2)$. To see this, consider (III~39). As we showed in Lemma III.3, if the inequality (III~39-a) hold for all joint PDFs $P_{X_1X_2}(x_1,x_2)$, then we also have:

$$I(X_2;Y_2|X_1) \leq I(X_2;Y_1|X_1) \tag{III~45}$$

for all $P_{X_1X_2}(x_1,x_2)$. Therefore, the conditions (III~39) are indeed equivalent to the following:

$$\begin{cases} I(X_1,X_2;Y_2) \leq I(X_1,X_2;Y_1) & (a) \\ I(X_1;Y_1|X_2) = I(X_1;Y_2|X_2) & (b) \end{cases} \tag{III~46}$$

for all $P_{X_1X_2}(x_1,x_2)$. Now, for the equality in (III~46-b) to hold for the Gaussian channel formulated as (III~17), it is required that $|b| = 1$. But to satisfy the conditions (III~40), one can choose $|b| > 1$. The fact is that the conditions (III~40) imply the inequalities (III~39) hold for *Gaussian input distributions*; however, they never imply these inequalities hold for *all joint PDFs* $P_{X_1X_2}(x_1,x_2)$. We will present a detailed discussion in this regard in Part III of our multi-part papers [27, Sec. III].

The second capacity result was derived in [34] for the Gaussian CRC as (III~17) with $|b| \leq 1$. It was shown that in this case, the capacity region is achieved by using dirty paper coding to pre-cancel the interference effect of $X_2$ at the receiver $Y_1$.

***Proposition III.16)*** **[34]**, *The capacity region of the Gaussian CRC with* $|b| \leq 1$ *is given by:*

$$\bigcup_{0\leq\alpha\leq 1} \left\{ \begin{array}{l} (R_1,R_2) \in \mathbb{R}_+^2: \\ R_1 \leq \psi(\alpha P_1) \\ R_2 \leq \psi\left(\dfrac{(1-\alpha)b^2P_1 + 2|b|\sqrt{(1-\alpha)P_1P_2} + P_2}{1+\alpha b^2 P_1}\right) \end{array} \right\} \tag{III~47}$$

For the Gaussian CRC (III~17) with $|b| > 1$, the capacity region is unknown. Nonetheless, an outer bound was derived in [36] for this case which is given in the next proposition.

***Proposition III.17)*** **[36],** *The capacity region of the Gaussian CRC with* $|b| > 1$ *is contained into the following rate region:*

$$\bigcup_{0\le\alpha\le 1}\left\{\begin{array}{l}(R_1,R_2)\in\mathbb{R}_+^2:\\ R_2\le\psi(\alpha P_1)\\ R_1+R_2\le\psi\left(b^2P_1+P_2+2|b|\sqrt{(1-\alpha)P_1P_2}\right)\end{array}\right\}$$

(III~48)

Also, the capacity region for two classes of discrete CRCs has been recently derived in [75]: The semi-deterministic CRC and the less-noisy CRC (the latter channels are referred to as the CRCs with "cognitive better decoding" in [75]).

***Definition III.8:*** *The discrete CRC is said to be semi-deterministic when the received signal at one of the receivers is a deterministic function of the input signals; i.e., there exists either a deterministic function $f_1(.)$ such that $Y_1=f_1(X_1,X_2)$, or a deterministic function $f_2(.)$ such that $Y_2=f_2(X_1,X_2)$.*

Note that due to asymmetry of the channel, it is essential to determine which of $Y_1$ and $Y_2$ is a deterministic function of $(X_1,X_2)$. Each condition represents a specific situation that differs from the other.

***Proposition III.18)*** **[75, Th. VIII.1],** *The capacity region of the semi-deterministic CRC where the received signal at the cognitive receiver is a deterministic function of the input signals, i.e., $Y_1=f_1(X_1,X_2)$, is given by:*

$$\bigcup_{P_{VX_1X_2}(v,x_1,x_2)}\left\{\begin{array}{l}(R_1,R_2)\in\mathbb{R}_+^2:\\ R_1\le H(Y_1|X_2)\\ R_2\le I(V,X_2;Y_2)\\ R_1+R_2\le I(V,X_2;Y_2)+H(Y_1|V,X_2)\end{array}\right\}$$

(III~49)

***Definition III.9:*** *The CRC in Fig. 6 is said to be less-noisy (cognitive better decoding in* [75, Th. VII.1]*) if*

$$I(U,X_2;Y_2)\le I(U,X_2;Y_1)$$

(III~50)

*for all joint PDFs $P_{UX_1X_2}(u,x_1,x_2)$.*

The authors in [75, Th. VII.1] termed the condition (III~50) as the "better cognitive decoding" regime. We prefer to use the name of "*less-noisy CRC*" because for the special case of $\|\mathcal{X}_2\|=1$, it is reduced to the less-noisy BC in (III~6). The less-noisy CRCs include the channels for which capacity results have been previously derived in [34, Th. 3.4] and [31, Th. 5] as special cases. However, the Gaussian strong interference channels defined by (III~40) are not necessarily less-noisy in the sense of (III~50). To verify this, consider a Gaussian CRC satisfying the strong interference conditions (III~40) with $|b|$ strictly greater than one, i.e., $|b|>1$. Such a channel is not less-noisy in the sense of (III~50). To see this fact, first note that by setting $U\equiv X_1$ in (III~50), we have $I(X_1,X_2;Y_2)\le I(X_1,X_2;Y_1)$ for all joint PDFs $P_{X_1X_2}(x_1,x_2)$. Therefore if (III~50) holds, according to Lemma III.3, the following inequality must hold as well:

$$I(X_1;Y_2|X_2)\le I(X_1;Y_1|X_2)$$

(III~51)

for all joint PDFs $P_{X_1X_2}(x_1,x_2)$. But the inequality (III~51) implies that $|b|\le 1$ which is in contrast to the assumption $|b|>1$.

***Proposition III.19)*** **[75, Th. VII.1],** *The capacity region of the less-noisy CRC in (III~50) is given below:*

$$\bigcup_{P_{VX_1X_2}(v,x_1,x_2)}\left\{\begin{array}{l}(R_1,R_2)\in\mathbb{R}_+^2:\\ R_1\le I(X_1;Y_1|X_2)\\ R_2\le I(V,X_2;Y_2)\\ R_1+R_2\le I(X_1;Y_1|V,X_2)+I(V,X_2;Y_2)\end{array}\right\}$$

(III~52)

Finally, similar to the CIC, one may consider the one-sided CRCs as special cases of the general channel.

***Definition III.10:*** *The CRC is said to be one-sided if one of the following conditions hold:*

$$\begin{cases}\mathbb{P}(y_1,y_2|x_1,x_2)=\mathbb{P}(y_1|x_1,x_2)\mathbb{P}(y_2|x_2) & (a)\\ \mathbb{P}(y_1,y_2|x_1,x_2)=\mathbb{P}(y_1|x_1)\mathbb{P}(y_2|x_1,x_2) & (b)\end{cases}$$

(III~53)

Note that due to asymmetry of the channel, each of the conditions in (III~53) represents a specific situation that differs from the other. It should be mentioned that the one-sided Gaussian CRC satisfying (III~53-a) is derived by setting $b=0$ in (III~17); in this case, the capacity region is known and is given in Proposition III.16. The one-sided Gaussian CRC satisfying (III~53-b) is derived by setting $a=0$ in (III~17). This channel has been recently investigated in details in concurrent works [40], [41] and [76], where some new capacity results have been established.

*On the achievability schemes for the CRC:* Here, we intend to show that in the achievability scheme recently presented in [29] for the CRC, which yields the rate region $\mathfrak{R}_i^{RTD}$ given by (III~35), splitting the primary message into three messages not only is superfluous but also it may cause some rate loss. We present a new coding scheme which is simpler than that of [29] but yields a potentially larger rate region. This result is given in the following proposition.

***Proposition III.20)*** *Define the rate region $\mathfrak{R}_i^{CRC}$ as follows:*

$$\mathfrak{R}_i^{CRC}\triangleq\bigcup_{\mathcal{P}_i^{CRC}}\left\{\begin{array}{r}(R_1,R_2)\in\mathbb{R}_+^2:\ \exists\,(B_{10},B_{11},B_{12},R_{10},R_{11},R_{20},R_{22})\in\mathbb{R}_+^7:\\ R_1=R_{10}+R_{11}\\ R_2=R_{20}+R_{22}\\ B_{10}\geq I(V_2;W_1|W_2)\\ B_{10}+B_{11}\geq I(V_2;W_1|W_2)+I(V_2;U_1|W_2,W_1)\\ B_{10}+B_{11}+B_{12}\geq I(V_2;W_1|W_2)+I(V_2;U_1|W_2,W_1)\\ +I(U_1;V_1|W_2,V_2,W_1)\\ ,\\ R_{11}+B_{11}\leq I(U_1;Y_1|W_2,W_1)\\ R_{10}+B_{10}+R_{11}+B_{11}\leq I(W_1,U_1;Y_1|W_2)\\ R_{20}+R_{10}+B_{10}+R_{11}+B_{11}\leq I(W_2,W_1,U_1;Y_1)\\ R_{22}+B_{12}\leq I(V_2,V_1;Y_2|W_2,W_1)+I(V_2;W_1|W_2)\\ R_{22}+R_{10}+B_{10}+B_{12}\leq I(V_2,W_1,V_1;Y_2|W_2)+I(V_2;W_1|W_2)\\ R_{20}+R_{22}+R_{10}+B_{10}+B_{12}\leq I(W_2,V_2,W_1,V_1;Y_2)+I(V_2;W_1|W_2)\end{array}\right\}$$

(III~54)

*where $\mathcal{P}_i^{CRC}$ denotes the set of all joint PDFs $P_{W_2V_2W_1U_1V_1X_1X_2}(w_2,v_2,w_1,u_1,v_1,x_1,x_2)$ satisfying:*

$$P_{W_2V_2W_1U_1V_1X_1X_2}=P_{W_2V_2}P_{W_1U_1V_1|V_2W_2}P_{X_2|V_2W_2}P_{X_1|X_2V_1U_1W_1V_2W_2}$$

(III~55)

*The set $\mathfrak{R}_i^{CRC}$ is an achievable rate region for the CRC.*

***Remarks III.2***:

1. The achievable rate region $\mathfrak{R}_i^{CRC}$ in (III~54) contains $\mathfrak{R}_i^{RTD}$ in (III~35) as a subset, i.e., $\mathfrak{R}_i^{RTD}\subseteq\mathfrak{R}_i^{CRC}$. To prove this, in (III~54) we redefine:

$$W_1\cong U_{1c},\quad W_2\cong U_{2c},\quad U_1\cong U_{1pb},\quad V_1\cong U_{2pb}$$
$$B_{10}\cong \acute{R}_{1c},\quad B_{11}\cong \acute{R}_{1pb},\quad B_{12}\cong \acute{R}_{2pb},\quad R_{10}\cong R_{1c},\quad R_{11}\cong R_{1pb},\quad R_{20}\cong R_{2c}$$

Also, set $V_2\equiv X_2$ and $R_{22}=R_{2pa}+R_{2pb}$. Then, by a simple comparison of the resultant rate region with $\mathfrak{R}_i^{RTD}$, we obtain that $\mathfrak{R}_i^{RTD}$ has two additional constraints as follows:

$$\begin{array}{r}R_{1c}+\acute{R}_{1c}+R_{2pb}+\acute{R}_{2pb}\leq I\big(U_{2pb},U_{1c};Y_2\big|X_2,U_{2c}\big)+I(X_2;U_{1c}|U_{2c})\\ R_{2pb}+\acute{R}_{2pb}\leq I\big(U_{2pb};Y_2\big|U_{1c},X_2,U_{2c}\big)\end{array}$$

(III~56)

As stated before, the authors of [29] ignore some parts of the message at the primary transmitter (named $W_{2pb}$ in [29]), and indeed the constraints (III~56) are required to correctly decode this part of the message of the primary transmitter at its respective receiver. It should be noted that for the coding scheme of [29], without the constraints given in (III~56), the receiver $Y_2$ cannot correctly decode its respective message. Therefore, these constraints are unavoidable in the characterization of [29].

2. The rate region $\mathfrak{R}_i^{CRC}$ in (III~54) actually may be *strictly larger* than the rate region $\mathfrak{R}_i^{RTD}$ in (III~35). One may think that our achievability scheme may be extracted from that of [29] by setting $R_{2pb} = 0$ and removing the two constraints (III~56) from $\mathfrak{R}_i^{RTD}$; because for the choice of $R_{2pb} = 0$, these constraints are redundant (they are not required to correctly decode any desired message at the primary receiver). But this approach never proves that the rate region $\mathfrak{R}_i^{CRC}$ is a subset of $\mathfrak{R}_i^{RTD}$. In order to prove a rate region, e.g. $A$, includes another one, e.g. $B$, we need to fix both of them with their constraints and then show every point that belongs to $B$, it also belongs to $A$. In this case, during the process of proving the inclusion, one cannot remove certain constraints from either rate regions; note that the purpose is to show the inclusion of one *rate region* into another and not to show the extraction of one *achievability scheme* from another. Now, if we intend to prove that $\mathfrak{R}_i^{CRC} \subseteq \mathfrak{R}_i^{RTD}$, we can set $R_{2pb} = 0$ in $\mathfrak{R}_i^{RTD}$ but we cannot remove the constraints (III~56) from it. Thus, it is not clear whether $\mathfrak{R}_i^{CRC} \subseteq \mathfrak{R}_i^{RTD}$ is true? As a result, while our achievability scheme is simpler than that of [29], our achievable rate region may be larger. *This reveals the fact that unnecessarily complexities in the achievability schemes may cause rate loss*. Therefore, it is essential to design coding schemes intelligently based on reasonable criteria. In Part V of our multi-part papers [32], we propose a unique framework to design achievability schemes with satisfactory performances for any arbitrary interference network.

3. After the inclusion of the first version of this work on ArXiv [126], we were informed that in the paper [128, App. F] (which has been published concurrent to our work), it is claimed that "*without loss of generality one can take* $R_{2pb} = 0$ *in the rate region* $\mathfrak{R}_i^{RTD}$ *in (III~35) and then the resultant region is equal to* $\mathfrak{R}_i^{CRC}$ *in (III~54) because for the choice of* $R_{2pb} = 0$, *the constraints (III~56) are not required to correctly decode any desired message at the primary receiver*". However, this argument is not true, as discussed in details in previous remark. The fact is that the rate region $\mathfrak{R}_i^{CRC}$ in (III~54) may be *strictly larger* than the rate region $\mathfrak{R}_i^{RTD}$ in (III~35).

*Proof of Proposition III.20)* The achievable rate region $\mathfrak{R}_i^{CRC}$ can be derived as a consequence of the general achievability scheme which we will present in Part V of our multi-part papers [32] for the interference networks with general message sets (see also our previous paper [7] presented in [8]). Therefore, we do not provide a detailed proof and only give a brief summary.

The achievability scheme includes a rate splitting technique. Each of the messages $M_1$ and $M_2$ and thus their respective communication rates $R_1$ and $R_2$ are split into *two parts*:

$$\begin{cases} M_1 = (M_{10}, M_{11}), & R_1 = R_{10} + R_{11} \\ M_2 = (M_{20}, M_{22}), & R_2 = R_{20} + R_{22} \end{cases} \qquad \text{(III~57)}$$

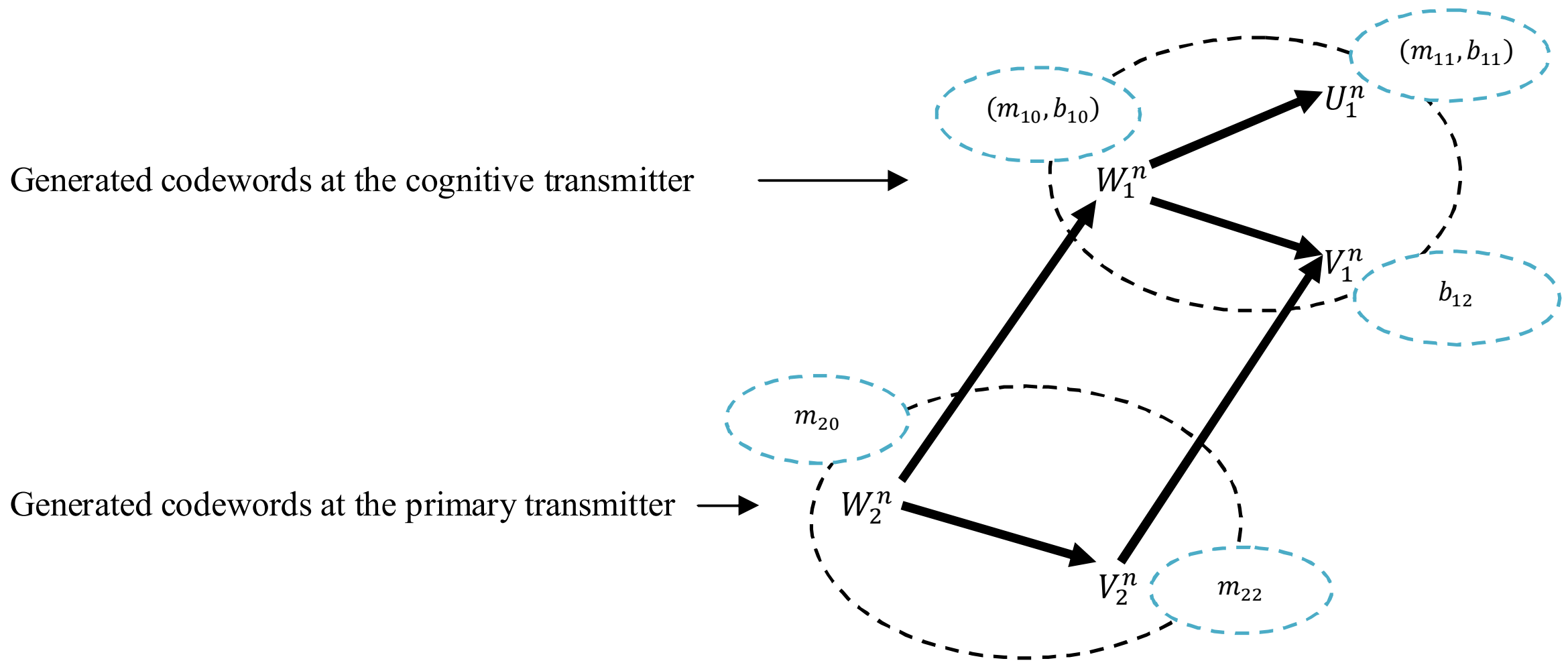

Figure 7. The achievability scheme for the CRC. This figure illustrates the superposition structures among the generated codewords. The ellipse beside each codeword shows what is conveyed by that codeword in addition to the ones in its cloud centers. Note that the codeword $V_1^n$ does not contain any part of the messages other than the ones conveyed by its cloud centers.

The split messages are then encoded by a random code of length-$n$. Roughly speaking, in our coding scheme for the CRC, the primary transmitter encodes its messages $M_{20}$ and $M_{22}$, respectively by the codewords $W_2^n$ and $V_2^n$ in a superposition fashion: $W_2^n$ serves as the cloud center and $V_2^n$ as the satellite codeword. At the cognitive transmitter, three bins of codewords are generated: $2^{nB_{10}}$ codewords $W_1^n$ conveying the message $M_{10}$ which are superimposed on $W_2^n$, $2^{nB_{11}}$ codewords $U_1^n$ conveying the message $M_{11}$ which are superimposed on $W_2^n, W_1^n$, and $2^{nB_{12}}$ codewords $V_1^n$ superimposed on the codewrods $(W_2^n, V_2^n, W_1^n)$. In fact, the codewords $V_1^n$ do not contain any part of the messages other than those conveyed by $W_2^n, V_2^n$ and $W_1^n$; these are satellite codewords (with the cloud centers $W_2^n, V_2^n, W_1^n$) which are only served to build a Marton's type encoding [47, 48] at the cognitive transmitter. Fig. 7 represents the generated codewords at the primary and the cognitive transmitters. In this graphical representation, every pair of codewords connected by a directed edge builds a superposition structure. The codeword at the beginning of the edge is the cloud center and that one at the end of the edge serves as the satellite codeword. The sizes of the bins, i.e., $B_{10}, B_{11}, B_{12}$ are selected to be large enough to guarantee that there exists a jointly typical 5-tuple of codewords $(W_2^n, V_2^n, W_1^n, U_1^n, V_1^n)$. Such a jointly typical 5-tuple is designated for transmission. The primary transmitter then generates a codeword $X_2^n$ superimposed on its related codewords, i.e., $W_2^n, V_2^n$, and sends it over the channel. The cognitive transmitter also generates a codeword $X_1^n$ superimposed on all other codewords, i.e., $W_2^n, V_2^n, W_1^n, U_1^n, V_1^n, X_2^n$, and sends it over the channel. Each receiver decodes its respective messages using a jointly typical decoder. Receiver $Y_1$ explores within the codewords $W_2^n, W_1^n, U_1^n$ to find its desired messages, and receiver $Y_2$ within the codewords $W_2^n, V_2^n, W_1^n, V_1^n$. The main difference between this achievability scheme and the one by [29] is that, unlike ours, in [29] the message of the primary transmitter, i.e., $M_2$ is split into *three parts*, one of them is ignored at this transmitter and its transmission is relegated to the cognitive transmitter. Our coding scheme does not contain this item and it results in a larger achievable rate region, as demonstrated in Remark III.2. ■

It is remarkable to note that our graphical illustration in Fig. 7 is helpful to extract achievable rate regions with satisfactory performance for channels with specific properties from the achievability scheme given for the general channel. For example, consider the Gaussian CRC formulated in (III~17). As mentioned before, the capacity region of this channel for $|b| \leq 1$ was determined in [34] which is given by (III~47). Therefore, it is sufficient to consider only the case of $|b| > 1$. In this case, the primary receiver perceives strong interference; accordingly, this receiver can decode both messages. We intend to derive an achievable rate region for the Gaussian CRC in this regime using $\mathfrak{R}_i^{CRC}$ given by (III~54). Note that due to presence of several auxiliary RVs in this rate region, its evaluation is difficult. To overcome this problem and reduce the corresponding computational complexity, we follow an intuitive approach. Let us review the achievability scheme in Fig. 7 which yields the rate region $\mathfrak{R}_i^{CRC}$ in (III~54) for the CRC. As we see from this figure, the codeword $U_1^n$ is used to transmit a part of the message $M_1$, i.e., $M_{11}$, which is only decoded at the cognitive receiver. On the one hand, for the channel with strong interference at the primary receiver, the message $M_1$ can be fully decoded at this receiver (in addition to $M_2$). Therefore, an appropriate choice for the channel with $|b| > 1$ is to set $U_1 \equiv \emptyset$ in (III~54); accordingly, we have $M_{11} \equiv \emptyset$ and $R_{11} = B_{11} = B_{12} = 0$. Then, in the resulting rate region, the random variable $V_1$ appears in the following expressions:

$$\begin{cases} I(V_2, V_1; Y_2|W_2, W_1) \\ I(V_2, W_1, V_1; Y_2|W_2) \\ I(W_2, V_2, W_1, V_1; Y_2) \end{cases}$$

As we see, in all these mutual information functions, $V_1$ appears before the conditioning operator. Therefore, intuitively we deduce that it is appropriate to set $V_1 \equiv X_1$. Moreover, we set $V_2 \equiv X_2$; using this choice, the rate region $\mathfrak{R}_i^{CRC}$ in (III~54) can be computed over all joint PDFs $P_{W_2 W_1 X_1 X_2}(w_2, w_1, x_1, x_2)$. By these assumptions, we derive the following achievable rate region:

$$\bigcup_{P_{W_2 W_1 X_1 X_2}} \begin{Bmatrix} (R_1, R_2) \in \mathbb{R}_+^2 : \\ R_1 \leq I(X_1, X_2; Y_2|W_2) \\ R_1 \leq I(W_1; Y_1|W_2) - I(W_1; X_2|W_2) \\ R_1 + R_2 \leq I(X_1, X_2; Y_2|W_2, W_1) + I(W_2, W_1; Y_1) \\ R_1 + R_2 \leq I(X_1, X_2; Y_2) \\ 2R_1 + R_2 \leq I(X_1, X_2; Y_2|W_2) + I(W_2, W_1; Y_1) - I(W_1; X_2|W_2) \end{Bmatrix}$$

(III~58)

Now, we see that the evaluation of the rate region (III~58) is substantially simpler than (III~54). If we set $W_2 \equiv X_2$ and $W_1 \equiv X_1$ in (III~58) and then compute it with Gaussian distributions, the resulting rate region achieves the capacity in the strong interference regime (III~40) as given in Proposition III.15. Moreover, one can check that this achievable rate region is optimal for all classes of Gaussian CRCs with $|b| > 1$ for which the capacity region has been recently derived in [40], [41] and [76]. However, it is not clear whether it is optimal for all channels with $|b| > 1$, which remains an open problem.

In the following subsections, we study the channels given here as the basic structures of the interference networks with new results.

## III.B) Novel Unified Outer Bounds for the Basic Structures with Two Receivers

In this subsection, we commence to establish a theory based on which capacity outer bounds are derived for the interference networks within a unified framework. From the view point of tightness, these are efficient outer bounds which are applicable to both discrete and Gaussian channels. They are also optimal in many cases which lead to explicit characterization of the capacity. Here, we only deal with the theory for the two-user CIC, the CRC, and the two-user BC. This theory will be extended to an arbitrary interference network in Part III of our multi-part papers [27].

First, we recall a useful lemma due to Csiszar and Korner [33] which is repeatedly used from now on.

***Lemma III.4)*** **[33, Lemma 7]**, *Let $X^n = \langle X_1, \dots, X_n \rangle$ and $Y^n = \langle Y_1, \dots, Y_n \rangle$ be two sequences of RVs where $X_t \in \mathcal{X}$ and $Y_t \in \mathcal{Y}, t = 1, \dots, n$. Then, for every arbitrary random variable $Z \in \mathcal{Z}$ and also for every arbitrary joint PDF $P_{X^n X^n Z}(x^n, y^n, z)$ defined on the set $\mathcal{X}^n \times \mathcal{Y}^n \times \mathcal{Z}$, we have the following identity:*

$$\sum_{t=1}^{n} I(X_{t+1}^n; Y_t | Z, Y^{t-1}) = \sum_{t=1}^{n} I(Y^{t-1}; X_t | Z, X_{t+1}^n) \tag{III~59}$$

Now, let us consider the interference network with two receivers in which some transmitters send two independent private messages to the receivers. Each receiver decodes its corresponding private message. At this time, the order of transmitting scheme is not considered. Therefore, the channel may be the two-user CIC, the CRC, or the two-user BC. Assume that the messages $M_1$ and $M_2$ are sent to the receivers $Y_1$ and $Y_2$, respectively. Fig. 8 illustrates this scenario.

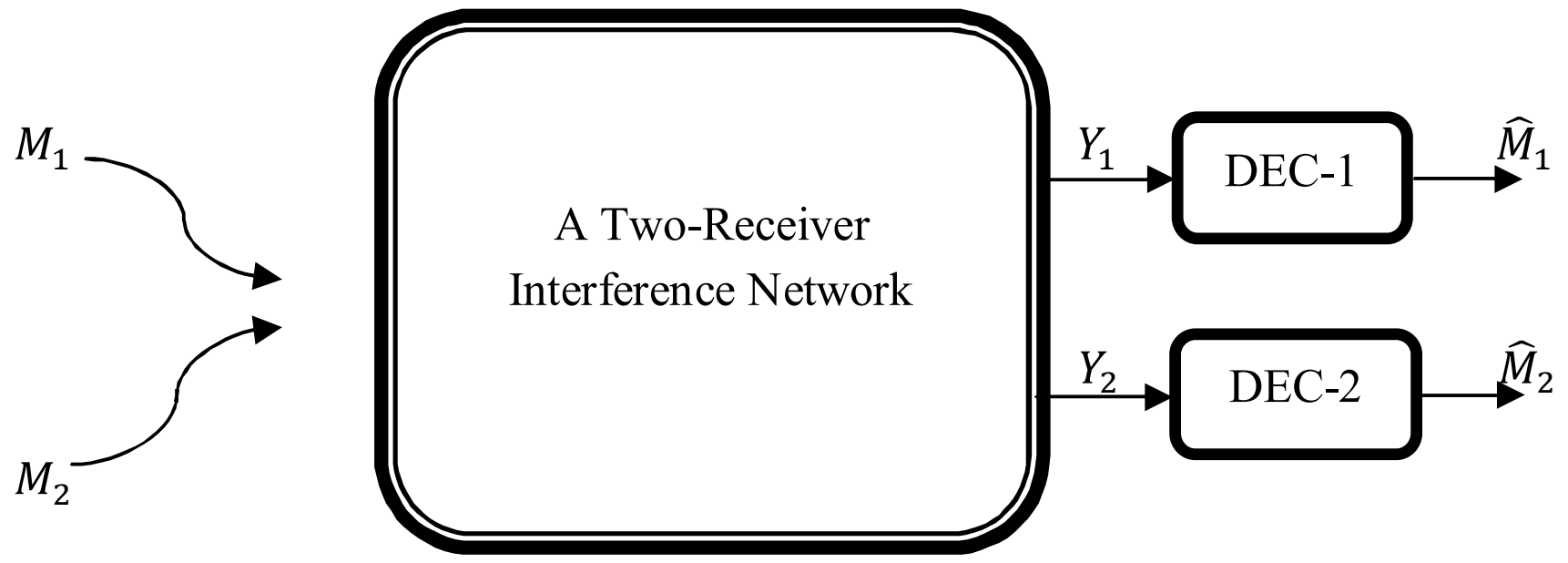

Figure 8. A two-receiver interference network in which the message $M_1$ is sent for the receiver $Y_1$ and the message $M_2$ to the receiver $Y_2$.

Next, we intend to derive an upper bound on the capacity region of this model using the Fano's inequality. Consider a length-$n$ code with the rate $(R_1, R_2)$ and with vanishing average error probability for the network. Define new RVs $W_t, t = 1, \dots, n$, as follows:

$$Z_t \triangleq \left(Y_1^{t-1}, Y_{2,t+1}^n\right) \tag{III~60}$$

Using the Fano's inequality, we can write:

$$\begin{cases} H(M_1|Y_1^n) \le n\epsilon_{1,n} \\ H(M_2|Y_2^n) \le n\epsilon_{2,n} \end{cases} \tag{III~61}$$

where $\epsilon_{1,n}, \epsilon_{2,n} \to 0$ as $n \to \infty$. Then, we proceed as follows. At first, suppose that $M_1^s \subseteq \{M_2\}$, i.e., $M_1^s$ may be $M_2$ or null. We have:

$$\begin{aligned} nR_1 &= H(M_1) = I(M_1; Y_1^n) + H(M_1|Y_1^n) \\ &\le I(M_1; Y_1^n, M_1^s) + n\epsilon_{1,n} \overset{(a)}{=} I(M_1; Y_1^n | M_1^s) + n\epsilon_{1,n} \\ &\overset{(b)}{=} \sum_{t=1}^{n} I\left(M_1; Y_{1,t} \middle| Y_1^{t-1}, M_1^s\right) + n\epsilon_{1,n} \end{aligned}$$

$$\leq \sum_{t=1}^{n} I\left(M_1, Y_1^{t-1}; Y_{1,t} \middle| M_1^s\right) + n\epsilon_{1,n}$$
$$\leq \sum_{t=1}^{n} I\left(M_1, Y_1^{t-1}, Y_{2,t+1}^n; Y_{1,t} \middle| M_1^s\right) + n\epsilon_{1,n}$$
$$= \sum_{t=1}^{n} I\left(M_1, Z_t; Y_{1,t} \middle| M_1^s\right) + n\epsilon_{1,n}$$

(III~62)

where equality (a) holds because $M_1^s$ is independent of $M_1$, and (b) is due to the chain rule [3]. Note that, $M_1^s$ can be considered as the side information virtually applied at the receiver $Y_1$. By another approach, we obtain:

$$nR_1 \overset{(a)}{\leq} \sum_{t=1}^{n} I\left(M_1; Y_{1,t} \middle| Y_1^{t-1}, M_1^s\right) + n\epsilon_{1,n}$$
$$\leq \sum_{t=1}^{n} I\left(M_1, Y_{2,t+1}^n; Y_{1,t} \middle| Y_1^{t-1}, M_1^s\right) + n\epsilon_{1,n}$$
$$= \sum_{t=1}^{n} I\left(M_1; Y_{1,t} \middle| Y_1^{t-1}, Y_{2,t+1}^n, M_1^s\right) + \sum_{t=1}^{n} I\left(Y_{2,t+1}^n; Y_{1,t} \middle| Y_1^{t-1}, M_1^s\right) + n\epsilon_{1,n}$$
$$\overset{(b)}{=} \sum_{t=1}^{n} I\left(M_1; Y_{1,t} \middle| Y_1^{t-1}, Y_{2,t+1}^n, M_1^s\right) + \sum_{t=1}^{n} I\left(Y_1^{t-1}; Y_{2,t} \middle| Y_{2,t+1}^n, M_1^s\right) + n\epsilon_{1,n}$$
$$\leq \sum_{t=1}^{n} I\left(M_1; Y_{1,t} \middle| Y_1^{t-1}, Y_{2,t+1}^n, M_1^s\right) + \sum_{t=1}^{n} I\left(Y_1^{t-1}, Y_{2,t+1}^n; Y_{2,t} \middle| M_1^s\right) + n\epsilon_{1,n}$$
$$= \sum_{t=1}^{n} I\left(M_1; Y_{1,t} \middle| Z_t, M_1^s\right) + \sum_{t=1}^{n} I\left(Z_t; Y_{2,t} \middle| M_1^s\right) + n\epsilon_{1,n}$$

(III~63)

where (a) is deduced by the lines of (III~62-b) and (b) is due to the Csiszar-Korner identity (III~59). Then, suppose that $M_2^s \subseteq \{M_1\}$, i.e., $M_2^s$ may be $M_1$ or null. Similar to the rate $R_1$ and by following a rather symmetrical approach, one can derive the following bounds on the rate $R_2$,:

$$nR_2 \leq \sum_{t=1}^{n} I\left(M_2, Z_t; Y_{2,t} \middle| M_2^s\right) + n\epsilon_{2,n}$$
$$nR_2 \leq \sum_{t=1}^{n} I\left(M_2; Y_{2,t} \middle| Z_t, M_2^s\right) + \sum_{t=1}^{n} I\left(Z_t; Y_{1,t} \middle| M_2^s\right) + n\epsilon_{2,n}$$

(III~64)

It can be easily verified that the relations included in (III~62) and (III~63) are still valid (except the last equalities in these relations due to the definition of $Z_t$ as in (III~60)) if we replace $Y_1^{t-1}$ by $Y_{1,t+1}^n$ and $Y_{2,t+1}^n$ by $Y_2^{t-1}$ in the related expressions. Therefore, to derive the bounds in (III~64), it is sufficient to perform this replacement first, and then exchange the indices 1 by 2 everywhere. Also, note that $M_2^s$ in the bounds (III~64) can be considered as virtual side information applied to the second receiver.

Next, we proceed to bound the sum rate $R_1 + R_2$, as follows. Assume that the virtual side information $M_2$ is provided to the receiver $Y_1$; we have:

$$n(R_1 + R_2) \leq I(M_1; Y_1^n, M_2) + I(M_2; Y_2^n) + n\left(\epsilon_{1,n} + \epsilon_{2,n}\right)$$
$$= I(M_1; Y_1^n | M_2) + I(M_2; Y_2^n) + n\left(\epsilon_{1,n} + \epsilon_{2,n}\right)$$
$$= \sum_{t=1}^{n} I\left(M_1; Y_{1,t} \middle| Y_1^{t-1}, M_2\right) + \sum_{t=1}^{n} I\left(M_2; Y_{2,t} \middle| Y_{2,t+1}^n\right) + n\left(\epsilon_{1,n} + \epsilon_{2,n}\right)$$
$$\leq \sum_{t=1}^{n} I\left(M_1, Y_{2,t+1}^n; Y_{1,t} \middle| Y_1^{t-1}, M_2\right) + \sum_{t=1}^{n} I\left(M_2, Y_1^{t-1}; Y_{2,t} \middle| Y_{2,t+1}^n\right) - \sum_{t=1}^{n} I\left(Y_1^{t-1}; Y_{2,t} \middle| Y_{2,t+1}^n, M_2\right) + n\left(\epsilon_{1,n} + \epsilon_{2,n}\right)$$
$$= \sum_{t=1}^{n} I\left(M_1; Y_{1,t} \middle| Y_1^{t-1}, Y_{2,t+1}^n, M_2\right) + \sum_{t=1}^{n} I\left(Y_{2,t+1}^n; Y_{1,t} \middle| Y_1^{t-1}, M_2\right) + \sum_{t=1}^{n} I\left(M_2, Y_1^{t-1}; Y_{2,t} \middle| Y_{2,t+1}^n\right)$$
$$- \sum_{t=1}^{n} I\left(Y_1^{t-1}; Y_{2,t} \middle| Y_{2,t+1}^n, M_2\right) + n\left(\epsilon_{1,n} + \epsilon_{2,n}\right)$$
$$\overset{(a)}{=} \sum_{t=1}^{n} I\left(M_1; Y_{1,t} \middle| Y_1^{t-1}, Y_{2,t+1}^n, M_2\right) + \sum_{t=1}^{n} I\left(M_2, Y_1^{t-1}; Y_{2,t} \middle| Y_{2,t+1}^n\right) + n\left(\epsilon_{1,n} + \epsilon_{2,n}\right)$$
$$\leq \sum_{t=1}^{n} I\left(M_1; Y_{1,t} \middle| Y_1^{t-1}, Y_{2,t+1}^n, M_2\right) + \sum_{t=1}^{n} I\left(M_2, Y_1^{t-1}, Y_{2,t+1}^n; Y_{2,t}\right) + n\left(\epsilon_{1,n} + \epsilon_{2,n}\right)$$
$$= \sum_{t=1}^{n} I\left(M_1; Y_{1,t} \middle| Z_t, M_2\right) + \sum_{t=1}^{n} I\left(M_2, Z_t; Y_{2,t}\right) + n\left(\epsilon_{1,n} + \epsilon_{2,n}\right)$$

(III~65)

where equality (a) is due to Csiszar-Korner identity (III~59). By another approach, we can derive:

$$\begin{aligned} n(R_1+R_2) &\le \sum_{t=1}^{n} I\left(M_1;Y_{1,t}\middle|Y_1^{t-1},Y_{2,t+1}^n,M_2\right)+\sum_{t=1}^{n} I\left(M_2,Y_1^{t-1};Y_{2,t}\middle|Y_{2,t+1}^n\right)+n\left(\epsilon_{1,n}+\epsilon_{2,n}\right) \\ &= \sum_{t=1}^{n} I\left(M_1;Y_{1,t}\middle|Y_1^{t-1},Y_{2,t+1}^n,M_2\right)+\sum_{t=1}^{n} I\left(M_2;Y_{2,t}\middle|Y_1^{t-1},Y_{2,t+1}^n\right)+\sum_{t=1}^{n} I\left(Y_1^{t-1};Y_{2,t}\middle|Y_{2,t+1}^n\right)+n\left(\epsilon_{1,n}+\epsilon_{2,n}\right) \\ &\overset{(a)}{=} \sum_{t=1}^{n} I\left(M_1;Y_{1,t}\middle|Y_1^{t-1},Y_{2,t+1}^n,M_2\right)+\sum_{t=1}^{n} I\left(M_2;Y_{2,t}\middle|Y_1^{t-1},Y_{2,t+1}^n\right)+\sum_{t=1}^{n} I\left(Y_{2,t+1}^n;Y_{1,t}\middle|Y_1^{t-1}\right)+n\left(\epsilon_{1,n}+\epsilon_{2,n}\right) \\ &\le \sum_{t=1}^{n} I\left(M_1;Y_{1,t}\middle|Y_1^{t-1},Y_{2,t+1}^n,M_2\right)+\sum_{t=1}^{n} I\left(M_2;Y_{2,t}\middle|Y_1^{t-1},Y_{2,t+1}^n\right)+\sum_{t=1}^{n} I\left(Y_1^{t-1},Y_{2,t+1}^n;Y_{1,t}\right)+n\left(\epsilon_{1,n}+\epsilon_{2,n}\right) \\ &= \sum_{t=1}^{n} I\left(M_1;Y_{1,t}\middle|Z_t,M_2\right)+\sum_{t=1}^{n} I\left(M_2;Y_{2,t}\middle|Z_t\right)+\sum_{t=1}^{n} I\left(Z_t;Y_{1,t}\right)+n\left(\epsilon_{1,n}+\epsilon_{2,n}\right) \end{aligned}$$

(III~66)

where the equality (a) is due to Csiszar-Korner identity (III~59). Again, one can see that the relations included in (III~65) and (III~66) are still valid (except the last equalities in these relations due to the definition of $Z_t$ as in (III~60)) if we replace $Y_1^{t-1}$ by $Y_{1,t+1}^n$ and $Y_{2,t+1}^n$ by $Y_2^{t-1}$ in the relevant expressions. By carrying out this replacement in (III~65) and (III~66) and exchanging the indices 1 by 2 everywhere, we obtain the following constraints on the sum-rate $R_1+R_2$ as well:

$$\begin{aligned} n(R_1+R_2) &\le \sum_{t=1}^{n} I\left(M_2;Y_{2,t}\middle|Z_t,M_1\right)+\sum_{t=1}^{n} I\left(M_1,Z_t;Y_{1,t}\right)+n\left(\epsilon_{1,n}+\epsilon_{2,n}\right) \\ n(R_1+R_2) &\le \sum_{t=1}^{n} I\left(M_2;Y_{2,t}\middle|Z_t,M_1\right)+\sum_{t=1}^{n} I\left(M_1;Y_{1,t}\middle|Z_t\right)+\sum_{t=1}^{n} I\left(Z_t;Y_{2,t}\right)+n\left(\epsilon_{1,n}+\epsilon_{2,n}\right) \end{aligned}$$

(III~67)

Note that the bounds in (III~67) are derived by giving the side information $M_1$ to the second receiver. Next, we introduce a time-sharing random variable $Q$ uniformly distributed over the set $\{1,\dots,n\}$ and independent of the other RVs. Define:

$$Z \triangleq Z_Q, \qquad Y_j \triangleq Y_{j,Q}, \qquad j=1,2$$

(III~68)

Therefore, we can write:

$$\sum_{t=1}^{n} I\left(M_1,Z_t;Y_{1,t}\middle|M_1^s\right)=n\left(\sum_{t=1}^{n}\frac{1}{n}I\left(M_1,Z_t;Y_{1,t}\middle|M_1^s,Q=t\right)\right)=nI(M_1,Z;Y_1|M_1^s,Q)$$

(III~69)

Other summations included in the bounds (III~62)-(III~67) can be similarly expressed in such a compact form similarly. Then, by letting $n\to\infty$, we derive the following set of constraints on the rates $R_1,R_2$, which is denoted by $\mathfrak{R}_o^{unified}$,

$$\mathfrak{R}_o^{unified} \triangleq \left\{\begin{array}{l} (R_1,R_2)\in\mathbb{R}_+^2: \\ R_1\le\min\left\{\begin{array}{c} I(M_1,Z;Y_1|Q),I(M_1,Z;Y_1|M_2,Q) \\ I(M_1;Y_1|Z,Q)+I(Z;Y_2|Q) \\ I(M_1;Y_1|Z,M_2,Q)+I(Z;Y_2|M_2,Q) \end{array}\right\}, \\ R_2\le\min\left\{\begin{array}{c} I(M_2,Z;Y_2|Q),I(M_2,Z;Y_2|M_1,Q) \\ I(M_2;Y_2|Z,Q)+I(Z;Y_1|Q) \\ I(M_2;Y_2|Z,M_1,Q)+I(Z;Y_1|M_1,Q) \end{array}\right\}, \\ R_1+R_2\le I(M_1;Y_1|Z,M_2,Q)+I(M_2,Z;Y_2|Q) \\ R_1+R_2\le I(M_2;Y_2|Z,M_1,Q)+I(M_1,Z;Y_1|Q) \\ R_1+R_2\le I(M_1;Y_1|Z,M_2,Q)+I(M_2;Y_2|Z,Q)+I(Z;Y_1|Q) \\ R_1+R_2\le I(M_2;Y_2|Z,M_1,Q)+I(M_1;Y_1|Z,Q)+I(Z;Y_2|Q) \end{array}\right\}$$

(III~70)

As indicated, up to now we have imposed no condition on the transmission order and the outer bound derived in (III~70) is valid for all channels in which two independent messages are sent to the receivers. Therefore, the set of constraints $\mathfrak{R}_o^{unified}$ constitutes a common framework based on which capacity outer bounds can be derived for the two-user BC, the two-user CIC, and the CRC as will be performed in the sequel.

First, consider the two-user BC in which the two messages $M_1$ and $M_2$ are encoded at the transmitter using the codeword $X^n$. By definition of the code for this channel, the random variable $X_t$, $t=1,\dots,n$, is a deterministic function of $(M_1,M_2)$. Define $X\triangleq X_Q$; thus, $X$ is a deterministic function of $(M_1,M_2,Q)$. Hence, we obtain the following outer bound for the two-user BC.

***Theorem III.1)*** *Define the rate region* $\mathfrak{R}_o^{BC}$ *as follows:*

$$\mathfrak{R}_o^{BC} = \bigcup_{\mathcal{P}_o^{BC}} \mathfrak{R}_o^{unified}$$

(III~71)

*where* $\mathcal{P}_o^{BC}$ *denotes the set of all joint PDFs* $P_{QM_1M_2ZX}(q, m_1, m_2, z, x)$ *satisfying:*

$$P_{QM_1M_2ZX} = P_Q P_{M_1} P_{M_2} P_{Z|M_1M_2Q} P_{X|M_1M_2Q}$$

(III~72)

*as well as, the PDFs* $P_{M_1}$ *and* $P_{M_2}$ *are uniformly distributed, and* $P_{X|M_1M_2Q} \in \{0,1\}$*; in other words,* $X$ *is a deterministic function of* $(M_1, M_2, Q)$*. The rate region* $\mathfrak{R}_o^{BC}$ *constitutes an outer bound on the capacity region of the two-user BC without common information.*

***Remarks III.3:***

1. The outer bound $\mathfrak{R}_o^{BC}$ given in (III~71) is tighter than the UV-outer bound given by (III~13). To see this, redefine $U \cong (M_1, Z, Q)$ and $V \cong (M_2, Z, Q)$. Note that $X$ is a deterministic function of $(U, V)$; we have:

$$\begin{aligned} R_1 &\le I(M_1, Z; Y_1|Q) \le I(M_1, Z, Q; Y_1) = I(U; Y_1) \\ R_1 + R_2 &\le I(M_2; Y_2|Z, M_1, Q) + I(M_1, Z; Y_1|Q) \le I(X; Y_2|U) + I(U; Y_1) \end{aligned}$$

(III~73)

   The two other bounds of the UV-outer bound $\mathfrak{R}_o^{UV\to BC}$ in (III~13) are derived similarly. It is not clear whether $\mathfrak{R}_o^{BC}$ in (III~71) is strictly tighter than $\mathfrak{R}_o^{UV\to BC}$.

2. Some of the constraints in the outer bound $\mathfrak{R}_o^{BC}$ in (III~71) are new. For example, the following constraints:

$$\begin{aligned} R_1 &\le I(M_1, Z; Y_1|M_2, Q) = I(X; Y_1|M_2, Q) \\ R_1 &\le I(M_1; Y_1|Z, Q) + I(Z; Y_2|Q) \end{aligned}$$

   Such constraints are not found in the previously derived outer bounds in [54] for the two-user BC, which are all equivalent to the UV-outer bound in (III~13), see [55]. As a result, our outer bound may be strictly tighter than the UV-outer bound.

Next, consider the two-user CIC where the messages $M_1$ and $M_2$ are separately sent by two transmitters over the channel: Transmitter $X_1$ encodes the message $M_1$ using the codeword $X_1^n$ and transmitter $X_2$ encodes the message $M_2$ using the codeword $X_2^n$. By definition of the code for the two-user CIC, the random variable $X_{1,t}$ is a deterministic function of $M_1$ and the random variable $X_{2,t}$ is a deterministic function of $M_2$ for $t = 1, \dots, n$. Now, define $X_1 \triangleq X_{1,Q}$ and $X_2 \triangleq X_{2,Q}$, (note that $Q$ is the time-sharing RV); thus, $X_1$ is a deterministic function of $(M_1, Q)$ and $X_2$ a deterministic function of $(M_2, Q)$. Therefore, the following result is derived.

***Theorem III.2)*** *Define the rate region* $\mathfrak{R}_o^{CIC}$ *as follows:*

$$\mathfrak{R}_o^{CIC} \triangleq \bigcup_{\mathcal{P}_o^{CIC}} \mathfrak{R}_o^{unified}$$

(III~74)

*where* $\mathcal{P}_o^{CIC}$ *denotes the set of all joint PDFs* $P_{QM_1M_2ZX_1X_2}(q, m_1, m_2, z, x_1, x_2)$ *satisfying:*

$$P_{QM_1M_2ZX_1X_2} = P_Q P_{M_1} P_{M_2} P_{Z|M_1M_2Q} P_{X_1|M_1Q} P_{X_2|M_2Q}$$

(III~75)

*as well as, the PDFs* $P_{M_1}$ *and* $P_{M_2}$ *are uniformly distributed, and* $P_{X_1|M_1Q}, P_{X_2|M_2Q} \in \{0,1\}$*; in other words,* $X_i$ *is a deterministic function of* $(M_i, Q)$*,* $i = 1,2$*. The rate region* $\mathfrak{R}_o^{CIC}$ *constitutes an outer bound on the capacity region of the two-user CIC.*

It should be noted that since $X_{1,t}$ and $X_{2,t}$ are given by deterministic functions of $M_1$ and $M_2$, respectively, $X_{1,t}, X_{2,t} \to (M_1, M_2) \to Z_t$ forms a Markov chain; hence, $X_1, X_2 \to (M_1, M_2, Q) \to Z$ also forms a Markov chain, justifying the factorization (III~75).

As a corollary of Theorem III.2, one can derive a new outer bound on the capacity region of the CIC, which is weaker than $\mathfrak{R}_o^{CIC}$ in (III~74) but with a simpler description.

***Corollary III.1) UV-outer bound for the two-user CIC***

*Define the rate region $\mathfrak{R}_o^{UV\to CIC}$ as follows:*

$$\mathfrak{R}_o^{UV\to CIC} \triangleq \bigcup_{\mathcal{P}_o^{UV\to CIC}} \left\{ \begin{array}{l} (R_1,R_2) \in \mathbb{R}_+^2 : \\ R_1 \leq \min\left\{ \begin{array}{c} I(U,X_1;Y_1|Q), I(X_1;Y_1|X_2,Q), \\ I(X_1;Y_1|V,X_2,Q) + I(V;Y_2|X_2,Q) \end{array} \right\}, \\ R_2 \leq \min\left\{ \begin{array}{c} I(V,X_2;Y_2|Q), I(X_2;Y_2|X_1,Q), \\ I(X_2;Y_2|U,X_1,Q) + I(U;Y_1|X_1,Q) \end{array} \right\}, \\ R_1 + R_2 \leq I(X_1;Y_1|V,X_2,Q) + I(V,X_2;Y_2|Q) \\ R_1 + R_2 \leq I(X_2;Y_2|U,X_1,Q) + I(U,X_1;Y_1|Q) \end{array} \right\}$$

(III~76)

*where $\mathcal{P}_o^{UV\to CIC}$ denotes the set of all joint PDFs $P_{QUVX_1X_2}(q,u,v,x_1,x_2)$ satisfying:*

$$P_{QUVX_1X_2} = P_Q P_{X_1|Q} P_{X_2|Q} P_{UV|X_1X_2Q}$$

(III~77)

*moreover, $P_{X_1|UQ}, P_{X_2|VQ} \in \{0,1\}$; in other words, $X_1$ is given by a deterministic function of $(U,Q)$ and $X_2$ is given by a deterministic function of $(V,Q)$. The set $\mathfrak{R}_o^{UV\to CIC}$ constitutes an outer bound on the capacity region of the two-user CIC.*

*Proof of Corollary III.1)* Refer to Appendix. ■

***Remark III.4:*** As we see, the outer bound $\mathfrak{R}_o^{UV\to CIC}$ in (III~76) is very similar to the UV-outer bound for the two-user BC given in (III~13). Similar to (III~13), the name "UV-outer bound" for $\mathfrak{R}_o^{UV\to CIC}$ is due to the auxiliary random variables $U$ and $V$ in (III~76). Note that the parameter $Q$ is the time-sharing RV.

Using the UV-outer bound (III~76) with the associated input power constraints (i.e., $\mathbb{E}[X_i^2] \leq P_i, i = 1,2$), one can establish capacity outer bounds also for the Gaussian channel (III~17). Specifically, in the weak interference regime where $|a|, |b| < 1$, and also in the mixed interference regime where $|b| < 1 \leq |a|$, this outer bound can be evaluated using the EPI. However, one can show that the resulting outer bounds are weaker than the ones recently established in [20-22]. In the strong interference regime, the UV-outer bound is optimal as will be pointed in the next subsection.

Finally, let the model depicted in Fig. 8 be the CRC where two transmitters send the two messages $M_1$ and $M_2$ over the channel: Transmitter 1 encodes $M_1$ and $M_2$ using the codeword $X_1^n$ and transmitter 2 encodes $M_2$ using the codeword $X_2^n$, see Fig. 6. By definition of the code for the CRC, the random variable $X_{1,t}$ is a deterministic function of $(M_1, M_2)$ and also the random variable $X_{2,t}$ is a deterministic function of $M_2$, for $t = 1, \ldots, n$. Continuing with the time-sharing scheme discussed in description of (III~70), we define $X_1 \triangleq X_{1,Q}$ and $X_2 \triangleq X_{2,Q}$; thus, $X_1$ is a deterministic function of $(M_1, M_2, Q)$ and $X_2$ a deterministic function of $(M_2, Q)$. Therefore, the following outer bound is derived for the CRC.

***Theorem III.3)*** *Define the rate region $\mathfrak{R}_o^{CRC}$ as follows:*

$$\mathfrak{R}_o^{CRC} = \bigcup_{\mathcal{P}_o^{CRC}} \mathfrak{R}_o^{unified}$$

(III~78)

*where $\mathcal{P}_o^{CRC}$ denotes the set of all joint PDFs $P_{QM_1M_2ZX_1X_2}(q,m_1,m_2,z,x_1,x_2)$ satisfying:*

$$P_{QM_1M_2ZX_1X_2} = P_Q P_{M_1} P_{M_2} P_{Z|M_1M_2Q} P_{X_1|M_1M_2Q} P_{X_2|M_2Q}$$

(III~79)

*also, the PDFs $P_{M_1}$ and $P_{M_2}$ are uniformly distributed, and $P_{X_1|M_1M_2Q}, P_{X_2|M_2Q} \in \{0,1\}$; in other words, $X_1$ is a deterministic function of $(M_1, M_2, Q)$ and $X_2$ is a deterministic function of $(M_2, Q)$. The rate region $\mathfrak{R}_o^{CRC}$ constitutes an outer bound on the capacity region of the CRC.*

***Remark III.5:*** The outer bound $\mathfrak{R}_o^{CRC}$ given in (III~78) for the CRC is included in $\mathfrak{R}_o^{MGKS}$ given by (III~37) as a subset. To see this, it is sufficient to redefine $U_1 \cong M_1$, $U_2 \cong M_2$, and $V \cong (Z, Q)$ in (III~78). Note that we have:

$$I(M_1, Z; Y_1|Q) \leq I(M_1, Z, Q; Y_1) = I(U_1, V; Y_1)$$
$$I(M_2, Z; Y_2|Q) \leq I(M_2, Z, Q; Y_2) \overset{(a)}{=} I(U_2, V, X_2; Y_2)$$

where equality (a) holds because $X_2$ is a deterministic function of $(U_2, V)$. It is not clear if this inclusion is strict.

Similar to the two-user CIC, one can derive a UV-outer bound on the capacity region of the CRC using the outer bound $\mathfrak{R}_o^{CRC}$ in (III~78).

***Corollary III.2)*** *Define the rate region $\mathfrak{R}_o^{UV\to CRC}$ as follows:*

$$\mathfrak{R}_o^{UV\to CRC} \triangleq \bigcup_{\mathcal{P}_o^{UV\to CRC}} \left\{\begin{array}{l} (R_1, R_2) \in \mathbb{R}_+^2 : \\ R_1 \leq \min\left\{\begin{array}{c} I(U; Y_1), I(X_1; Y_1|X_2), \\ I(X_1; Y_1|V, X_2) + I(V; Y_2|X_2) \end{array}\right\}, \\ R_2 \leq I(V, X_2; Y_2) \\ R_1 + R_2 \leq I(X_1; Y_1|V, X_2) + I(V, X_2; Y_2) \\ R_1 + R_2 \leq I(X_1, X_2; Y_2|U) + I(U; Y_1) \end{array}\right\}$$

(III~80)

*where $\mathcal{P}_o^{UV\to CRC}$ denotes the set of all joint PDFs $P_{UVX_1X_2}(u, v, x_1, x_2)$ satisfying:*

$$P_{UVX_1X_2} = P_{X_1|UV} P_{X_2|V}$$

(III~81)

*moreover, $P_{X_1|UV}, P_{X_2|V} \in \{0,1\}$; in other words, $X_1$ is a deterministic function of $(U, V)$ and $X_2$ a deterministic function of $V$. The set $\mathfrak{R}_o^{UV\to CRC}$ constitutes an outer bound on the capacity region of the two-user CRC.*

*Proof of Corollary III.2)* Refer to Appendix. ■

Now, consider the following rate region.

$$\bigcup_{P_{VX_1X_2}} \left\{\begin{array}{l} (R_1, R_2) \in \mathbb{R}_+^2 : \\ R_1 \leq I(X_1; Y_1|X_2) \\ R_2 \leq I(V, X_2; Y_2) \\ R_1 + R_2 \leq I(X_1; Y_1|V, X_2) + I(V, X_2; Y_2) \end{array}\right\}$$

(III~82)

It is clear that the rate region (III~82) is also a capacity outer bound for the CRC, which is weaker than $\mathfrak{R}_o^{UV\to CRC}$ in (III~80). This outer bound, first established in [34], can be evaluated using the EPI for the Gaussian channel (the associated input power constraints, i.e., $\mathbb{E}[X_i^2] \leq P_i, i = 1,2$, should be additionally considered) with weak interference at the primary receiver, i.e., $|b| \leq 1$. In this case, it is optimal and leads to the capacity result of Proposition III.16. Also, for the Gaussian channel with strong interference at the primary receiver where $|b| > 1$, the result of Proposition III.17 is a direct consequence of the outer bound (III~82) as given in [36, p. 413].

Note that the outer bounds derived in Theorems III.1, III.2 and III.3 for the two-user BC, the two-user CIC and the CRC have a common framework given by (III~70). This framework is constituted by the following parameters:

1. The RVs representing the receiver signals, i.e., $Y_1, Y_2$.
2. The RVs representing the messages $M_1, M_2$.
3. The parameter "$Q$" which is the time-sharing random variable.
4. The auxiliary random variable " $Z$ ". In the process of deriving the common framework, this RV denotes $Z \triangleq Z_Q \triangleq \left(Y_1^{Q-1}, Y_{2,Q+1}^n\right)$.

Such representation of the outer bounds indeed enables us to perceive how the structure of their constraints varies by network topology. This fact can be easily understood by comparing the UV-outer bounds in (III~13), (III~76) and (III~80) for the two-user BC, the two-user CIC and the CRC, respectively. We extend this theory to arbitrary interference networks in Part III of our multi-part papers [27] where we develop new outer bounds on the capacity region of all interference networks based on such unified frameworks. As we will see, the outer bounds derived by this approach are very helpful to study the behavior of information flow in large multi-user networks.

Lastly, it is remarkable to mention that in some special cases, certain properties of the given channel can be exploited to establish capacity outer bounds tighter than those discussed above. For example, consider the one-sided CIC in (III~31-a) or the one-sided CRC in (III~53-a). For both of these channels, the received signal at the second receiver, i.e., $Y_2$, is independent of the message $M_1$. This is the key to derive the following results.

***Theorem III.4)*** *Define the rate region $\mathfrak{R}_o^{ZIC}$ as follows:*

$$\mathfrak{R}_o^{ZIC} \triangleq \bigcup_{\mathcal{P}_o^{CIC}} \mathfrak{R}_o^{unified\rightarrow ch_{OS}}$$

(III~83)

*where $\mathcal{P}_o^{CIC}$ is defined similar to Theorem III.2 (see (III~75)), and $\mathfrak{R}_o^{unified\rightarrow ch_{OS}}$ is given similar to $\mathfrak{R}_o^{unified}$ in (III~70) but with an additional constraint as:*

$$R_1 \leq I(M_1; Y_1|Z,Q) - I(M_1; Y_2|Z,Q)$$

(III~84)

*The set $\mathfrak{R}_o^{ZIC}$ constitutes an outer bound on the capacity region of the two-user ZIC defined by (III~31-a).*

*Proof of Theorem III.4)* It is only required to derive the constraint (III~84). Let us assume that the channel model depicted in Fig. 8 be a ZIC. Again, consider a length-$n$ code with the rate $(R_1, R_2)$ and vanishing average error probability for the channel. Let also the RVs $Z_t, t = 1, \dots, n$, be given as (III~60). First note that for the ZIC, one can show that $Y_2^n$ is independent of $M_1$. Now, using Fano's inequality we have:

$$\begin{aligned}
nR_1 &\leq I(M_1; Y_1^n) + n\epsilon_{1,n} \\
&= \textstyle\sum_{t=1}^n I\left(M_1; Y_{1,t}\middle|Y_1^{t-1}\right) + n\epsilon_{1,n} \\
&= \textstyle\sum_{t=1}^n I\left(M_1, Y_{2,t+1}^n; Y_{1,t}\middle|Y_1^{t-1}\right) - \sum_{t=1}^n I\left(Y_{2,t+1}^n; Y_{1,t}\middle|M_1, Y_1^{t-1}\right) + n\epsilon_{1,n} \\
&= \textstyle\sum_{t=1}^n I\left(M_1; Y_{1,t}\middle|Y_1^{t-1}, Y_{2,t+1}^n\right) + \sum_{t=1}^n I\left(Y_{2,t+1}^n; Y_{1,t}\middle|Y_1^{t-1}\right) - \sum_{t=1}^n I\left(Y_{2,t+1}^n; Y_{1,t}\middle|M_1, Y_1^{t-1}\right) + n\epsilon_{1,n} \\
&\overset{(a)}{=} \textstyle\sum_{t=1}^n I\left(M_1; Y_{1,t}\middle|Z_t\right) + \sum_{t=1}^n I\left(Y_1^{t-1}; Y_{2,t}\middle|Y_{2,t+1}^n\right) - \sum_{t=1}^n I\left(Y_1^{t-1}; Y_{2,t}\middle|M_1, Y_{2,t+1}^n\right) + n\epsilon_{1,n} \\
&\overset{(b)}{=} \textstyle\sum_{t=1}^n I\left(M_1; Y_{1,t}\middle|Z_t\right) + \sum_{t=1}^n I\left(Y_1^{t-1}; Y_{2,t}\middle|Y_{2,t+1}^n\right) - \sum_{t=1}^n I\left(Y_1^{t-1}, M_1; Y_{2,t}\middle|Y_{2,t+1}^n\right) + n\epsilon_{1,n} \\
&= \textstyle\sum_{t=1}^n I\left(M_1; Y_{1,t}\middle|Z_t\right) - \sum_{t=1}^n I\left(M_1; Y_{2,t}\middle|Y_1^{t-1}, Y_{2,t+1}^n\right) + n\epsilon_{1,n} \\
&= \textstyle\sum_{t=1}^n I\left(M_1; Y_{1,t}\middle|Z_t\right) - \sum_{t=1}^n I\left(M_1; Y_{2,t}\middle|Z_t\right) + n\epsilon_{1,n} \\
&= nI(M_1; Y_1|Z,Q) - nI(M_1; Y_2|Z,Q) + n\epsilon_{1,n}
\end{aligned}$$

(III~85)

where equality (a) is due to Csiszar-Korner identity (III~59), and equality (b) holds because $Y_2^n$ is independent of $M_1$. Also, to derive the last equality, we have exploited the definitions (III~68). Then, by letting $n \rightarrow 0$, we obtain the constraint (III~84). ■

***Theorem III.5)*** *Define the rate region* $\mathfrak{R}_o^{CRC_{OS}}$ *as follows:*

$$\mathfrak{R}_o^{CRC_{OS}} \triangleq \bigcup_{\mathcal{P}_o^{CRC}} \mathfrak{R}_o^{unified \to ch_{OS}}$$

(III~86)

*where* $\mathcal{P}_o^{CRC}$ *is given similar to Theorem III.3 (see (III~79));* $\mathfrak{R}_o^{unified \to ch_{OS}}$ *is also given similar to Theorem III.4, i.e., it is* $\mathfrak{R}_o^{unified}$ *in (III~70) with the additional constraint (III~84). The set* $\mathfrak{R}_o^{CRC_{OS}}$ *constitutes an outer bound on the capacity region of the one-sided CRC defined by (III~53-a).*

*Proof of Theorem III.5)* The derivation of the additional constraint (III~84) is exactly similar to Theorem III.4. It should be noted that for code of length-$n$ for the one-sided CRC in (III~53-a), similar to ZIC in (III~31-a), $Y_2^n$ is independent of $M_1$ and therefore all steps in establishing (III~85) are also valid here. ■

***Remark III.6:*** As mentioned in the introduction, the one-sided CRC in (III~53-a) was considered in [39] where the capacity region of the channel with one noiseless component was established. In the next subsection, we will re-derive the result of [39] using $\mathfrak{R}_o^{CRC_{OS}}$ in (III~86); therefore, justifying that such type of outer bounds are useful to establish capacity results for special channels.

In the next subsection, we make use of the outer bounds derived here to establish new capacity results for the two-user CIC and the CRC.

## III.C) New Capacity Results for the Two-User CIC and CRC

Now, we obtain new capacity results for the two-user CIC and CRC using the derived outer bounds in previous subsection. We identify a mixed interference (or less noisy) regime for any two-user CIC where the successive decoding scheme achieves the sum-rate capacity. Also, we provide a full characterization of the capacity region for the *more-capable* CRC. Moreover, we re-derive some of the previously known results for these channels.

### III.C.1) The Two-User CIC

First consider the two-user CIC in Fig. 5. As mentioned before, the capacity region of the CIC in the strong interference regime was established in [14] and [15]. Subsequently, we show that the outer bound $\mathfrak{R}_o^{UV \to CIC}$ given in (III~76) is optimal in this regime which yields an alternative but simpler proof for the problem.

***Corollary III.3)*** *The UV-outer bound for the two-user CIC,* $\mathfrak{R}_o^{UV \to CIC}$ *given in (III~76), is optimal in the strong interference regime (III~21)-(III~22).*

*Proof of Corollary III.3)* Consider the rate region (III~23). The constraints on the individual rates, i.e., $R_1$ and $R_2$, are directly found in the UV-outer bound $\mathfrak{R}_o^{UV \to CIC}$. To derive the sum-rate constraints, first note that according to the strong interference conditions (III~21) and also by the result of Lemma III.1, we have:

$$I(X_1;Y_1|X_2,V) \leq I(X_1;Y_2|X_2,V)$$

(III~87)

for every $P_{VX_1X_2}(v,x_1,x_2)$. Now, using the UV-outer bound we obtain:

$$R_1 + R_2 \leq I(X_1;Y_1|V,X_2,Q) + I(V,X_2;Y_2|Q)$$
$$\overset{(a)}{\leq} I(X_1;Y_2|V,X_2,Q) + I(V,X_2;Y_2|Q) = I(X_1,X_2;Y_2|Q)$$

where (a) is derived from (III~87) by replacing $V$ by $(V,Q)$ and considering the inequality for all PDFs $P_Q P_{X_1|Q} P_{X_2|Q} P_{V|X_1X_2Q}$. Similarly, one can derive: $R_1 + R_2 \leq I(X_1,X_2;Y_1|Q)$. Note that the proof is also valid for the Gaussian CIC with strong interference

where $|a| \geq 1$ and $|b| \geq 1$. This is due to Lemma III.2 based on which for the Gaussian channel the conditions (III~21) are equivalent to (III~22). ■

***Remarks III.7:***

1. To derive the capacity region of the strong interference channel using the UV-outer bound $\mathfrak{R}_o^{UV\to CIC}$ in (III~76), the result of Lemma III.1, based on which the strong interference condition (III~21) is extended as in (III~26) for all joint PDFs $P_{VX_1X_2}$, is crucial. Note that according to the joint PDFs of the UV-outer bound (III~76), we need the condition (III~87) to hold for all PDFs $P_Q P_{X_1|Q} P_{X_2|Q} P_{V|X_1X_2Q}$; this is essentially guaranteed by Lemma III.1.

2. To derive the capacity region of the two-user CIC with strong interference (III~21), we only required the sum-rate constraints of the UV-outer bound $\mathfrak{R}_o^{UV\to CIC}$ in (III~76). Now, let us consider the following constraints of this outer bound:

$$\begin{array}{l} R_1 \leq I(X_1;Y_1|V,X_2,Q) + I(V;Y_2|X_2,Q) \overset{(a)}{\leq} I(X_1;Y_2|V,X_2,Q) + I(V;Y_2|X_2,Q) = I(X_1;Y_2|X_2,Q) \\ R_2 \leq I(X_2;Y_2|U,X_1,Q) + I(U;Y_1|X_1,Q) \overset{(b)}{\leq} I(X_2;Y_1|U,X_1,Q) + I(U;Y_1|X_1,Q) = I(X_2;Y_1|X_1,Q) \end{array} \tag{III~88}$$

where inequalities (a) and (b) are derived from the strong interference conditions. These constraints are interesting. They state that in the strong interference regime, we necessarily have $R_1 \leq I(X_1;Y_2|X_2,Q)$ and $R_2 \leq I(X_2;Y_1|X_1,Q)$. However, such constraints are not required for establishing the capacity region in the strong interference regime because the explicit constraints $R_1 \leq I(X_1;Y_1|X_2,Q)$ and $R_2 \leq I(X_2;Y_2|X_1,Q)$ are stricter (due to the conditions (III~21)). Therefore, the constraints (III~88) are indeed redundant. Nevertheless, as shown in [115], constraints such as those in the left side of (III~88) are useful to establish capacity results for the networks with security issues.

In [20] the authors established the sum-rate capacity of the two-user Gaussian CIC (III~17) with $|b| < 1 \leq |a|$. In this case, the interference at the first receiver is strong while it is weak at the second receiver. The authors in [20] called this case the *mixed interference regime*. Now, we intend to extend this result to any arbitrary two-user CIC (both discrete and Gaussian channels).

***Definition III.11)*** *The two-user CIC in Fig. 5 is said to have mixed interference provided that:*

$$\begin{cases} I(X_2;Y_2|X_1) \leq I(X_2;Y_1|X_1) \\ I(V;Y_2|X_2) \leq I(V;Y_1|X_2) \end{cases} \tag{III~89}$$

*for all joint PDFs $P_{VX_1}(v,x_1)P_{X_2}(x_2)$.*

Note that an alternative mixed in interference regime is derived by exchanging the indices 1 and 2 in (III~89) everywhere. One can show that for the two-user Gaussian CIC (III~17), the conditions (III~89) are equivalent to the regime $|b| < 1 \leq |a|$. The proof is similar to Lemma III.2, see also Part III [27, Sec. II] for a detailed proof. Also, as an instance for discrete channels satisfying (III~89), we can mention the class of degraded CICs where

$$X_1, X_2 \to Y_1 \to Y_2 \tag{III~90}$$

forms a Markov chain. It should be noted that mixed interference channels satisfying (III~89) strictly contains the degraded channels in (III~90). For example, the Gaussian channel (III~17) has mixed interference in the sense of (III~89) if $|b| < 1 \leq |a|$; while in order to be degraded in the sense of (III~90), it is required that the conditions $|b| < 1 \leq |a|$ and $ab = 1$ hold simultaneously.

***Theorem III.6)*** *The sum-rate capacity of the two-user CIC (see Fig. 5) in the mixed interference regime (III~89), denoted by $\mathcal{C}_{sum}^{CIC_{mi}}$, is given by:*

$$\mathcal{C}_{sum}^{CIC_{mi}} = \max_{P_Q P_{X_1|Q} P_{X_2|Q}} \min \begin{Bmatrix} I(X_1,X_2;Y_1|Q) \\ I(X_1;Y_1|X_2,Q) + I(X_2;Y_2|Q) \end{Bmatrix} \tag{III~91}$$

*Proof of Theorem III.6)* The achievability is derived by a successive interference cancelation scheme. Precisely, the messages $M_1$ and $M_2$ are encoded by codewords constructed by $X_1$ and $X_2$ based on $P_{X_1}(x_1)$ and $P_{X_2}(x_2)$, respectively. The receiver $Y_1$ first decodes the interference $X_2(M_2)$ and cancels it from its received signal and then decodes its own signal $X_1(M_1)$. The receiver $Y_2$ treats interference as noise and only decodes its respective signal $X_2(M_2)$. A simple analysis of this scheme leads to the sum-rate (III~91) for $Q \equiv \emptyset$. Therefore, a time-shared version of the scheme achieves the entire (III~91). For the converse part, first note that the conditions (III~89) can be extended to:

$$\begin{cases} I(X_2;Y_2|U,X_1,Q) \leq I(X_2;Y_1|U,X_1,Q) & \text{for all} \quad P_{QUX_1X_2} \\ I(V;Y_2|X_2,Q) \leq I(V;Y_1|X_2,Q) & \text{for all} \quad P_{QVX_1X_2} \end{cases}$$

(III~92)

This point that the second condition of (III~91) is extended to the form given in (III~92) can be proved by following the same approach as Lemma III.1; see also Part III [27, Sec. II, Lemma 4]. Therefore, from the UV-outer bound $\mathfrak{R}_o^{UV\to CIC}$ in (III~76), we obtain:

$$\begin{aligned} R_1 + R_2 &\leq I(X_1;Y_1|V,X_2,Q) + I(V,X_2;Y_2|Q) \\ &= I(X_1;Y_1|V,X_2,Q) + I(V;Y_2|X_2,Q) + I(X_2;Y_2|Q) \\ &\leq I(X_1;Y_1|V,X_2,Q) + I(V;Y_1|X_2,Q) + I(X_2;Y_2|Q) \\ &= I(X_1;Y_1|X_2,Q) + I(X_2;Y_2|Q) \end{aligned}$$

and also,

$$\begin{aligned} R_1 + R_2 &\leq I(X_2;Y_2|U,X_1,Q) + I(U,X_1;Y_1|Q) \\ &= I(X_2;Y_1|U,X_1,Q) + I(U,X_1;Y_1|Q) \\ &= I(X_1,X_2;Y_1|Q) \end{aligned}$$

The proof is thus complete. ■

***Remarks III.8:***

1. Theorem III.6 can be directly applied to the Gaussian channel in the mixed interference regime with $|b| < 1 \leq |a|$. Hence, establishing a new proof for the result of [20].
2. Consider the mixed interference conditions (III~89). Our result in Theorem III.6 reveals that this regime indeed represents a situation where a receiver ($Y_1$ in (III~89)) is *less noisy* than the other one ($Y_2$ in (III~89)) such that the optimal coding scheme to achieve the sum-rate capacity is that the stronger (less noisy) receiver decodes the message of the weaker (more noisy) receiver as well. Therefore, from the viewpoint of optimal decoding order, the channel has a characteristic similar to the less-noisy BCs in (III~6). Hence, we can refer to the channels in (III~89) as the *less-noisy CIC*. In Part IV of our multi-part papers [28], we will identify classes of interference networks with less-noisy receivers for which successive interference cancelation is sum-rate optimal.
3. Consider a class of CICs which satisfy the following Markov chain as well as the first inequality of (III~89).

   $$X_1 \to Y_1, X_2 \to Y_2$$

   In a concurrent paper [127], using a different approach, the authors prove that the sum-rate capacity for such channels is given by (III~91). Note that the above Markov chain necessarily implies the second inequality of (III~89). Therefore, the result of [127] is a special case of Theorem III.6. It should be remarked that the authors of [127] are motivated that the sum-rate capacity of the channels (III~89) is also given by (III~91); however, it remains a conjecture in [127, Conclusion].

***Corollary III.4)*** *The sum-rate capacity of the two-user degraded CIC, in the sense that $X_1, X_2 \to Y_1 \to Y_2$ forms a Markov chain, is given by:*

$$\mathcal{C}_{sum}^{CIC_{deg}} = \max_{P_Q P_{X_1|Q} P_{X_2|Q}} \big(I(X_1;Y_1|X_2,Q) + I(X_2;Y_2|Q)\big)$$

(III~93)

*Proof of Corollary III.4)* For the degraded channel, (III~91) is reduced to (III~93). ■

Now, let us deal with the two-user ZIC given by (III~31-a). One can also define the strong interference regime for this scenario. However, here this regime differs from the one given in (III~21). In fact, the first condition of (III~21) can never be realized for the ZIC. In the following, we determine the strong interference regime for this channel.

***Definition III.12)*** *The ZIC (one-sided CIC) in (III~31-a) is said to have strong interference provided that:*

$$I(X_2;Y_2) \leq I(X_2;Y_1|X_1) \tag{III~94}$$

*for all joint PDFs* $P_{X_1}(x_1)P_{X_2}(x_2)$.

For the Gaussian ZIC (III~32), the strong interference regime (III~94) is equivalent to $|a| \geq 1$. The capacity region of this channel was established in [109]. In the next proposition, we derive the capacity region for any ZIC satisfying (III~94).

***Proposition III.21)*** *The capacity region of the ZIC (III~31-a) with strong interference (III~94) is given by:*

$$\bigcup_{P_Q P_{X_1|Q} P_{X_2|Q}} \left\{ \begin{array}{l} (R_1,R_2) \in \mathbb{R}_+^2: \\ R_1 \leq I(X_1;Y_1|X_2,Q) \\ R_2 \leq I(X_2;Y_2|Q) \\ R_1 + R_2 \leq I(X_1,X_2;Y_1|Q) \end{array} \right\} \tag{III~95}$$

*Proof of Proposition III.21)* The achievability is simply derived by decoding both messages at the receiver $Y_1$. For the converse part, it is sufficient to derive the bound on the sum-rate. First note that, as the channel is one-sided, $Y_2$ is independent of $X_2$. Thereby, the condition (III~94) can be rewritten as follows:

$$I(X_2;Y_2|X_1) \leq I(X_2;Y_1|X_1) \quad \text{for all joint PDFs} \quad P_{X_1}(x_1)P_{X_2}(x_2)$$

Now, using the above inequality, its extension in Lemma III.1, and also the UV-outer bound $\mathfrak{R}_o^{UV\to CIC}$ in (III~76), we get:

$$\begin{aligned} R_1 + R_2 &\leq I(X_2;Y_2|U,X_1,Q) + I(U,X_1;Y_1|Q) \\ &\leq I(X_2;Y_1|U,X_1,Q) + I(U,X_1;Y_1|Q) \\ &= I(X_1,X_2;Y_1|Q) \end{aligned}$$

The proof is thus complete. ■

As we see, for the ZIC (III~31-a) in the strong interference regime (III~94), the capacity is achieved by requiring the first receiver (the receiver that depends on both channel inputs) to decode both transmitted messages. This differs from the general two-user CIC with strong interference (III~21) where both receivers decode both transmitted messages. This leads us to the following insight;

***Note:*** *For an interference network, in the strong interference regime, each receiver can decode all transmitted messages by its connected transmitters. In Part III of our multi-part papers [27], we will generalize this fact to any arbitrary interference networks.*

The Gaussian ZIC has noisy interference if $|a| < 1$. In this case, it was shown in [109] that treating interference as noise achieves the sum-rate capacity. In the following, we shall present such a regime for any two-user ZIC.

***Definition III.13)*** *The two-user ZIC in (III~31-a) is said to have noisy interference provided that:*

$$I(U;Y_1|X_1) \leq I(U;Y_2|X_1) \tag{III~96}$$

*for all joint PDFs* $P_{X_1}(x_1)P_{UX_2}(u,x_2)$.

One can easily show that for the Gaussian channel (III~32), the condition (III~96) is equivalent to $|a| < 1$. Also, note that similar to Lemma III.1, one can show that if (III~96) holds for all product PDFs $P_{X_1}(x_1)P_{UX_2}(u,x_2)$, it holds for all joint PDFs $P_{UX_1X_2}(u,x_1,x_2)$ as well.

***Theorem III.7)*** *The sum-rate capacity of the two-user ZIC (III~31-a) with the condition (III~96) is given by:*

$$\mathcal{C}_{sum}^{ZIC_{ni}} = \max_{P_Q P_{X_1|Q} P_{X_2|Q}} \left( I(X_2;Y_2|Q) + I(X_1;Y_1|Q) \right)$$

(III~97)

*The above sum-rate capacity is achieved by treating interference as noise.*

*Proof of Theorem III.7)* The achievability is immediately derived by treating interference as noise. Therefore, we need to prove the converse part. Using the UV-outer bound, we can write:

$$\begin{aligned} R_1 + R_2 &\leq I(X_2;Y_2|U,X_1,Q) + I(U,X_1;Y_1|Q) \\ &= I(X_2;Y_2|U,X_1,Q) + I(U;Y_1|X_1,Q) + I(X_1;Y_1|Q) \\ &\overset{(a)}{\leq} I(X_2;Y_2|U,X_1,Q) + I(U;Y_2|X_1,Q) + I(X_1;Y_1|Q) \\ &= I(X_2;Y_2|X_1,Q) + I(X_1;Y_1|Q) \\ &\overset{(b)}{=} I(X_2;Y_2|Q) + I(X_1;Y_1|Q) \end{aligned}$$

where inequality (a) follows from (III~96) and equality (b) follows from the fact that for the ZIC (III~31-a), $Y_2$ is independent of $X_1$ when conditioned on $Q$. This completes the proof. ■

Note that the consequence of Theorem III.7 can be directly applied to the Gaussian ZIC (III~32) with $|a| < 1$. We thus establish a new proof for the result of [109]. Also, a special case of Theorem III.7, where the one-sided channel satisfies the Markov chain $X_2 \to Y_2, X_1 \to Y_1$, is concurrently derived in [127] though using a different approach.

## III.C.2) The CRC

Now consider the CRC, as depicted in Fig. 6. In what follows, we derive new capacity results for some specials cases. The first result is for a class of *more-capable CRCs*. Clearly, inspired by the more-capable BC (III~7), one can define the more-capable CRCs as follows:

***Definition III.14:*** **[41]** *The CRC is said to be more-capable if any of the following conditions holds:*

$$\begin{cases} I(X_1,X_2;Y_2) \leq I(X_1,X_2;Y_1) & (a) \\ I(X_1,X_2;Y_1) \leq I(X_1,X_2;Y_2) & (b) \end{cases}$$

(III~98)

*for all joint PDFs* $P_{X_1X_2}(x_1,x_2)$.

It is easily seen that the more capable CRC (III~98-a) includes the CRC with strong interference (III~21) and the less-noisy CRC (III~50) as special cases. But the inverse is not true. In fact, the class of more-capable CRCs in (III~98-a) is *strictly* larger than the class of strong interference channels (III~21). For example, consider the Gaussian CRC with $ab = 1$ and $a > 1$; this channel is more-capable in the sense of (III~98-a) because it is degraded. But it does not satisfy the second condition of (III~21). The more-capable CRCs in (III~98-a) also *strictly* contain the class of less-noisy CRCs in (III~50). To verify this, consider a CRC with $\|\mathcal{X}_2\| = 1$ which indeed is reduced to a BC. In this case, the condition (III~50) represents the class of less-noisy BCs in (III~6) while the condition (III~98-a) represents the class of more-capable BCs in (III~7). On the other hand, it was shown in [49] that the more-capable BCs strictly include the less-noisy ones as a subset. In the next theorem, we establish the capacity region of the more-capable CRC in (III~98-a). For the more-capable channel in (III~98-b), the capacity still remains an open problem.

***Theorem III.8)*** *The capacity region of the more-capable CRC (III~98-a) is given by:*

$$\mathcal{C}^{CRC_{mc}} \triangleq \bigcup_{P_{VX_1X_2}(v,x_1,x_2)} \left\{ \begin{array}{l} (R_1,R_2) \in \mathbb{R}_+^2 : \\ R_1 \leq I(X_1;Y_1|X_2) \\ R_2 \leq I(V,X_2;Y_2) \\ R_1 + R_2 \leq I(X_1;Y_1|V,X_2) + I(V,X_2;Y_2) \\ R_1 + R_2 \leq I(X_1,X_2;Y_1) \end{array} \right\}$$

(III~99)

*Proof of Theorem III.8)* The achievability is derived using superposition coding; in fact, the rate region (III~99) is the capacity for the case where both messages are required at the cognitive receiver [88, Th. 4]. The converse part is derived using the UV-outer bound $\mathfrak{R}_o^{UV\to CRC}$ given by (III~80). Specifically, for the sum-rate we have:

$$\begin{aligned} R_1 + R_2 &\leq I(X_1, X_2; Y_2|U) + I(U; Y_1) \\ &\overset{(a)}{\leq} I(X_1, X_2; Y_1|U) + I(U; Y_1) \\ &= I(X_1, X_2; Y_1) \end{aligned}$$

where (a) is due to the more-capable condition in (III~98-a) and its extension in Lemma III.3. ■

***Remarks III.9:***

1. As discussed before, the class of more-capable CRC (III~98-a) *strictly* contains the class of CRCs satisfying the "better cognitive decoding" condition in (III~50) for which the capacity region was recently derived in [75]. Therefore, Theorem III.8 establishes the capacity region for a new class of CRCs that strictly includes all the ones in [34, Th. 3.4], [25, Th.5], [75, Th. VII.1].
2. Comparing (III~99) and (III~52), we see that the capacity region of the more-capable CRC is described with an additional constraint than that of the channel in "better cognitive decoding regime", i.e.,:

   $$R_1 + R_2 \leq I(X_1, X_2; Y_1)$$

   Note that for the less-noisy CRCs or the CRCs with "better cognitive decoding" in (III~50), this constraint is redundant because the other sum-rate constraint in (III~99) is stricter.
3. Theorem III.8 shows that in the more-capable regime (III~98-a), the cognitive receiver can decode both transmitted messages. The same conclusion holds for the more-capable BC (III~7).
4. The capacity region of the more-capable CRC (III~98-a) is also derived in [129] by a rather different approach, concurrently.

We next consider the CRC with strong interference at the primary receiver. In such channels, we have:

$$I(X_1; Y_1|X_2) \leq I(X_1; Y_2|X_2) \tag{III~100}$$

for all PDFs $P_{X_1}(x_1)P_{X_2}(x_2)$. In the next proposition, we establish the sum-rate capacity of these channels.

***Proposition III.22)*** *The sum-rate capacity of the CRC with strong interference at the primary receiver (III~100) is given as follows:*

$$\mathcal{C}_{sum}^{CRC_{sip}} = \max_{P_{X_1 X_2}} I(X_1, X_2; Y_2) \tag{III~101}$$

*Proof of Proposition III.22)* To derive the achievability, both transmitters cooperatively transmit the primary message $M_2$ with its maximum rate (the message $M_1$ is withdrawn from transmission) that is given by (III~101). In fact, under the strong interference condition at the primary receiver (III~100), the sum-rate capacity is achieved at a corner point of the capacity region, i.e., the cut-off point on the $R_2$-axis. For the converse part, we can use the UV-outer bound (III~80) for the CRC. We have:

$$\begin{aligned} R_1 + R_2 &\leq I(X_1; Y_1|V, X_2) + I(V, X_2; Y_2) \\ &\overset{(a)}{\leq} I(X_1; Y_2|V, X_2) + I(V, X_2; Y_2) = I(X_1, X_2; Y_2) \end{aligned}$$

where (a) is due to (III~100) and its extension in Lemma III.1. ■

Also, as a direct consequence of Proposition III.22, we have the following result.

***Corollary III.5)*** *The sum-rate capacity of the Gaussian CRC formulated in (III~17) with strong interference at the primary receiver where* $|b| > 1$ *is given by:*

$$\mathcal{C}_{sum}^{CRC_{sip}\sim G} = \psi\left(b^2 P_1 + P_2 + 2|b|\sqrt{P_1 P_2}\right) \tag{III~102}$$

Finally, consider the one-sided CRC in (III~53-a). As discussed in Subsection III.B, this channel was considered in [39] where capacity inner and outer bounds were derived. The capacity region of the channel with one noiseless component was also obtained. In the following, we re-derive the result of [39] where we show that for the considered scenario, the inner bound $\mathfrak{R}_i^{CRC}$ in (III~54) and the outer bound derived in Theorem III.5 coincide which yield the capacity region.

***Proposition III.23)*** **[39],** *The capacity region of the semi-deterministic one-sided CRC in (III~53-a) with* $Y_2 = X_2$ *is given by*:

$$\bigcup_{P_{WUX_1X_2}} \left\{ \begin{array}{l} (R_1, R_2) \in \mathbb{R}_+^2 : \\ R_1 \leq I(U;Y_1|W) - I(U;X_2|W) \\ R_2 \leq H(X_2) \\ R_2 \leq H(X_2|W) + I(W;Y_1) \end{array} \right\}$$

(III~103)

*Proof of Proposition III.23)* For the direct part, consider the achievable rate region $\mathfrak{R}_i^{CRC}$ in (III~54). Substitute $W_1 \equiv V_1 \equiv \emptyset$. By this choice, we can set $R_{10} = B_{10} = B_{12} = 0$ and $B_{11} = I(V_2;U_1|W_2)$. Therefore, by redefining $U_1 \cong U$, $W_2 \cong W$ and $V_2 \cong X_2$, we obtain the following rate region:

$$\bigcup_{P_{WUX_1X_2}} \left\{ \begin{array}{l} (R_1, R_2) \in \mathbb{R}_+^2 : \\ R_1 \leq I(U;Y_1|W) - I(U;X_2|W) \\ R_2 \leq I(X_2,W;Y_2) \\ R_1 + R_2 \leq I(U,W;Y_1) + I(X_2;Y_2|W) - I(U;X_2|W) \end{array} \right\}$$

(III~104)

Now consider the one-sided semi-deterministic channel with $Y_2 \equiv X_2$. One can easily see that the rate region (III~103) is a subset of (III~104). For the converse part, using the outer bound $\mathfrak{R}_o^{CRCos}$ in (III~86), we have:

$$\begin{aligned} R_1 &\leq I(M_1;Y_1|Z,Q) - I(M_1;Y_2|Z,Q) \\ &= I(M_1;Y_1|Z,Q) - I(M_1;X_2|Z,Q) \end{aligned}$$

(III~105)

Also,

$$R_2 \leq I(M_2,Z;Y_2|Q) \leq H(Y_2) = H(X_2)$$

(III~106)

$$R_2 \leq I(M_2;Y_2|Z,Q) + I(Z;Y_1|Q) \leq H(Y_2|Z,Q) + I(Z,Q;Y_1) = H(X_2|Z,Q) + I(Z,Q;Y_1)$$

(III~107)

Then, by substituting $W \equiv (Z,Q)$ and $U \equiv (M_1,Z,Q)$ in (III~105)-(III~107), we derive (III~103). The proof is complete. ■

Note that our method to derive the converse part for Proposition III.23 is different from [39]. Our approach indeed demonstrates the usefulness of the capacity outer bound we established in Theorem III.5 for the one-sided channels.

## III.D) The Basic Structures with One-Sided Receiver Side Information

In this subsection, we intend to show that the outer bounds derived in Subsection III.B are also useful for the channels with One-Sided Receiver Side Information (OSRSI); scenarios wherein one receiver has access to the message with respect to the other receiver as side information. The results of this subsection give us some useful insights to understand the nature of information flow in the basic structures (and accordingly in all other network topologies); for example, this investigation enables us to highlight some interesting similarities/differences between the less-noisy and the more-capable concepts for the BC and the CRC. The results also shed light on the essence of the UV-outer bounds given in Subsection III.B.

It is noteworthy to mention that the study of interference networks for receivers with some knowledge of non-corresponding messages has a long history. Such communication scenarios were introduced by Shannon [60] where he considered a two-way channel and derived capacity inner and outer bounds; this was the first multi-terminal model studied in network information theory. This channel in the case of no cooperation between transmitters, which is also called the restricted two-way channel [111], is nothing but a two-user CIC where each receiver has access to the message with respect to the other. Also, the two-user BC with receiver side information was investigated in [112]. The capacity region was established for two cases: 1) The case where each receiver knows the message designated to the other, 2) The BC with degraded message sets where the receivers have only partial message information.

Here, we study the cases where only one receiver has knowledge of the non-corresponding message. As indicated in [111], such problems arise in several settings, e.g., when one receiver decodes its message first and then transmits it to the other one before its decoding.

## III.D.1) The Two-User BC with OSRSI

First, we consider the two-user BC with OSRSI. The channel model is depicted in Fig. 9.

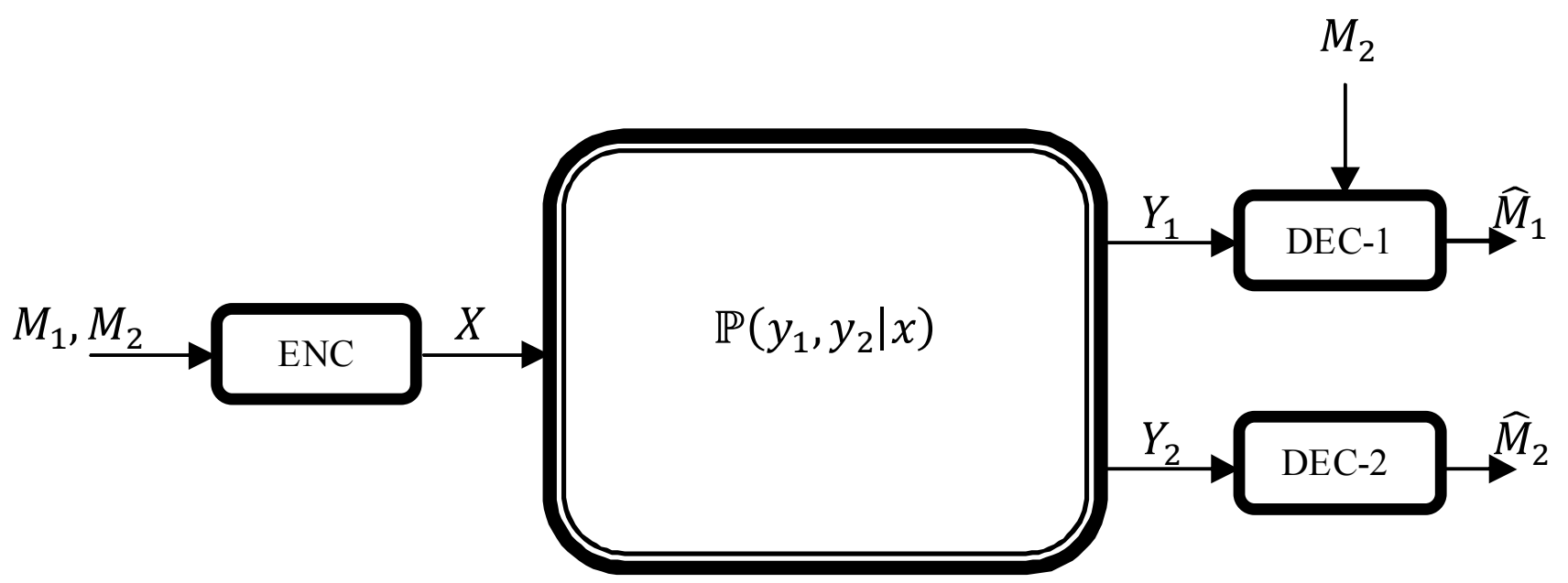

Figure 9. The two-user BC with One-Sided Reciever Side Information (OSRSI).

In this model the receiver $Y_1$ has access to the message with respect to the receiver $Y_2$, i.e., $M_2$. A length-$n$ code for this channel is defined similar to the case without OSRSI except in the decoding rule at receiver $Y_1$; the decoder function at the receiver $Y_1$ is given by:

$$\mathfrak{D}_1: \mathcal{Y}_1^n \times \mathcal{M}_2 \to \mathcal{M}_1 \tag{III~108}$$

where the message $M_1$ is estimated as $\hat{M}_1 = \mathfrak{D}_1(Y_1^n, M_2)$.

The essential nature of such scenario is that since receiver $Y_1$ knows $M_2$, it can cancel the interference effect of this message on its received signal perfectly, while receiver $Y_2$ still receives an interfered signal. In the next proposition, we derive a full characterization of the capacity region for this channel.

***Proposition III.24)*** *The capacity region of the two-user BC with OSRSI in Fig. 9 is given by:*

$$\mathcal{C}^{BC/OSRSI} \triangleq \bigcup_{P_{VX}} \left\{ \begin{array}{l} (R_1, R_2) \in \mathbb{R}_+^2 : \\ R_1 \leq I(X; Y_1) \\ R_2 \leq I(V; Y_2) \\ R_1 + R_2 \leq I(X; Y_1|V) + I(V; Y_2) \end{array} \right\} \tag{III~109}$$

*Proof of Proposition III.24)* The result of this proposition is actually a special case of Theorem III.9 which will be proved later; see Remark III.11. Therefore, we omit the proof here to avoid repetition. ■

Note that one can symmetrically derive the capacity region of the two-user BC with OSRSI for the case where the message $M_1$ is available at the receiver $Y_2$ as follows:

$$\bigcup_{P_{UX}} \left\{ \begin{array}{l} (R_1, R_2) \in \mathbb{R}_+^2 : \\ R_1 \leq I(U; Y_1) \\ R_2 \leq I(X; Y_2) \\ R_1 + R_2 \leq I(X; Y_2|U) + I(U; Y_1) \end{array} \right\}$$

(III~110)

Now consider the rate regions (III~109) and (III~110). It is clear that each of these rate regions and thereby the intersection of them constitutes an outer bound on the capacity region of the BC without OSRSI. The UV-outer bound in (III~13) for the two-user BC is given by collecting the constraints of the two bounds (III~109) and (III~110), although, as shown in [53], it is strictly tighter than the outer bound derived by the intersection of these bounds. It is well-known [113] that the UV-outer bound does not extend for the multi-user BC in a straight-forward manner. In fact, it seems that to derive useful capacity outer bounds for multi-receiver BCs, it is required to first obtain capacities of the corresponding channels in which the receivers are aware of non-corresponding messages (say, for example, in a degraded order).

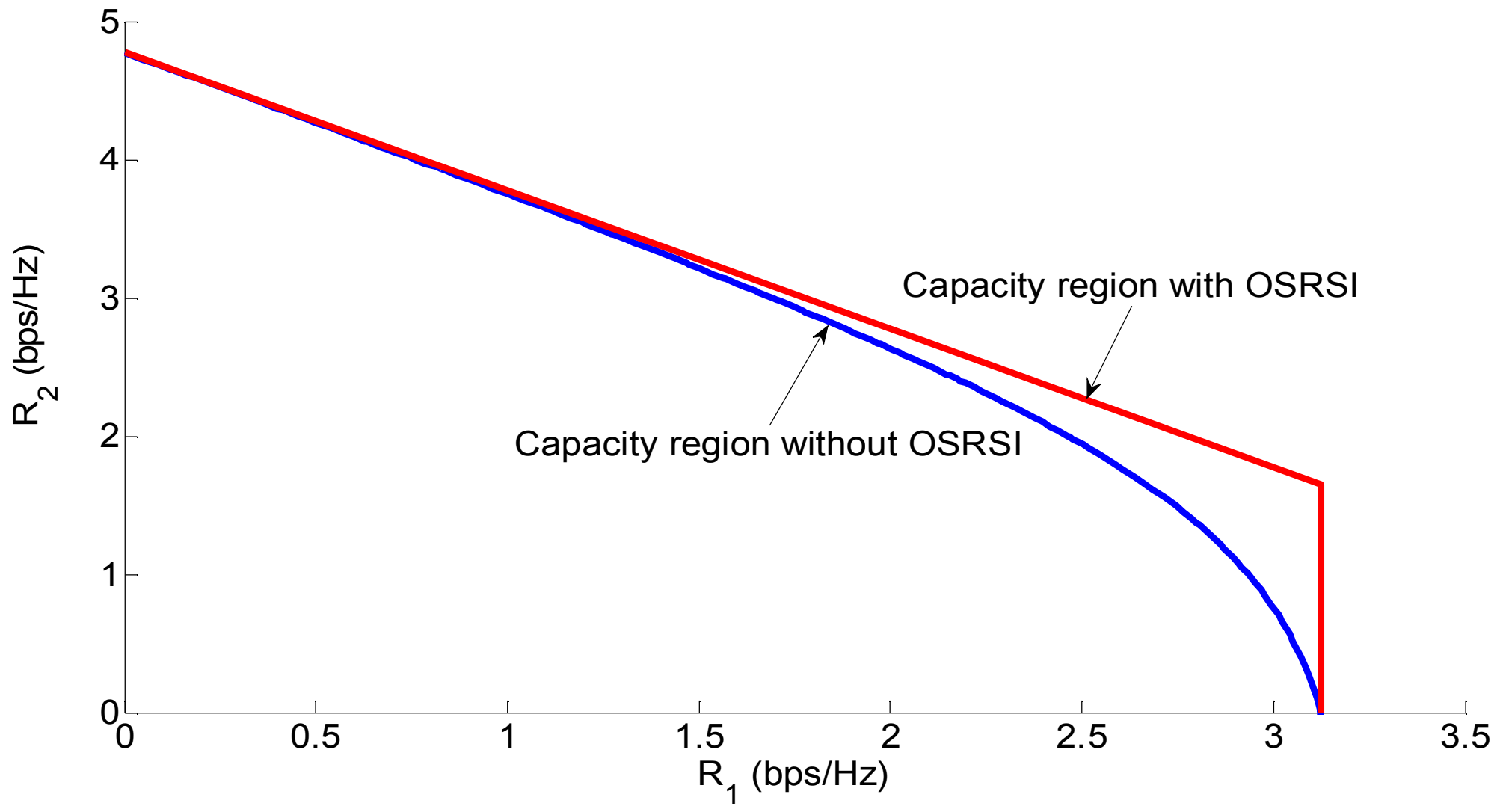

Figure 10. The capacity region of the two-user Gaussian BC (III~1) with and without OSRSI for $a = \sqrt{5},\ b = \sqrt{50},\ P = 15$.

Now, let us consider the Gaussian BC in (III~1). As discussed in Subsection III.A.2, this channel is essentially degraded. For the case of $|a| \geq |b|$, the channel is degraded in the sense of (III~3), i.e., the receiver $Y_1$ is the stronger receiver and $Y_2$ is the weaker. In this setup, the capacity region is not increased by providing the message $M_2$ at the receiver $Y_1$ as in Fig. 9; it is still given by (III~8). For the case of $|a| \leq |b|$, the channel is degraded in the sense that the receiver $Y_2$ is the stronger receiver and the receiver $Y_1$ is the weaker. For this case, one can easily show that the rate region (III~109) is optimized for $V \equiv X$. Therefore, the capacity region is given by:

$$\mathcal{C}_{|a|\leq|b|}^{BC/OSRSI\sim G} \triangleq \bigcup \left\{ \begin{array}{l} (R_1, R_2) \in \mathbb{R}_+^2 : \\ R_1 \leq \psi(a^2 P) \\ R_1 + R_2 \leq \psi(b^2 P) \end{array} \right\}$$

(III~111)

Figure 10 depicts the capacity region of the Gaussian BC (III~1) with and without OSRSI ($M_2$ at the receiver $Y_1$) for the case where $|a| \leq |b|$. As we observe, the OSRSI strictly increases the capacity region for this case.

## III.D.2) The Two-User CIC with OSRSI

Next, we consider the two-user CIC with OSRSI as shown in Fig. 11.

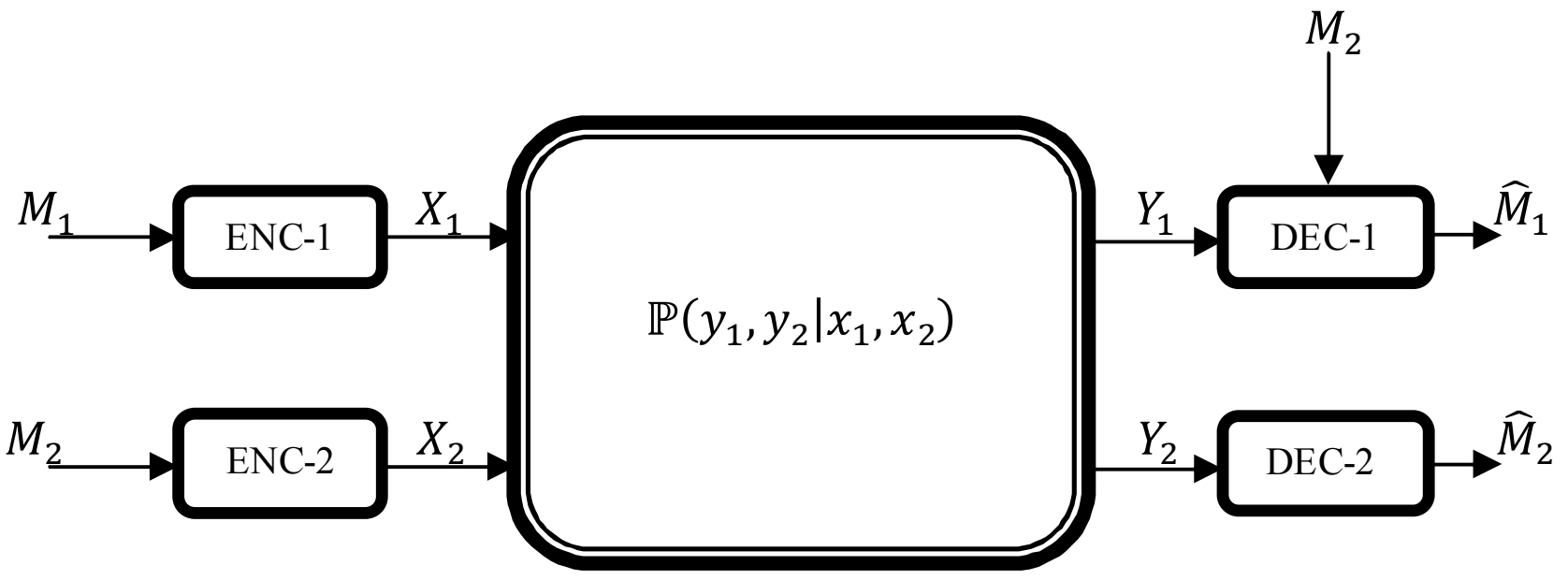

Figure 11. The two-user CIC with OSRSI.

Similar to the BC, we consider the case where receiver $Y_1$ knows perfectly the message with respect to the receiver $Y_2$, i.e., $M_2$. Accordingly, the decoding rule at receiver $Y_1$ is given by (III~108). In this scenario, since receiver $Y_1$ knows $M_2$, it can cancel the interference effect of $X_2$ on its received signal perfectly, while receiver $Y_2$ still receives a signal interfered by $X_1$. Subsequently, we provide inner and outer bounds on the capacity region of this model.

***Proposition III.25)*** *Define the rate region* $\mathfrak{R}_i^{CIC/OSRSI}$ *as follows:*

$$\mathfrak{R}_i^{CIC/OSRSI} \triangleq \bigcup_{P_Q P_{X_1 W_1|Q} P_{X_2|Q}} \left\{\begin{array}{l}(R_1, R_2) \in \mathbb{R}_+^2 : \\ R_1 \leq I(X_1; Y_1|X_2, Q) \\ R_2 \leq I(X_2; Y_2|W_1, Q) \\ R_1 + R_2 \leq I(X_1; Y_1|X_2, W_1, Q) + I(X_2, W_1; Y_2|Q)\end{array}\right\}$$

(III~112)

*The set* $\mathfrak{R}_i^{CIC/OSRSI}$ *constitutes an inner bound on the capacity region of the two-user CIC with OSRSI in Fig.11.*

*Proof of Proposition III.25)* The proof is based on a random coding scheme. As a brief outline, we mention that at the transmitter $X_1$, the message $M_1$ is split into two parts as $M_1 \triangleq (M_{10}, M_{11})$; these sub-messages are encoded in a superposition structure such that $M_{10}$ is conveyed by the cloud center and $M_{11}$ by the satellite. At the transmitter $X_2$, the message $M_2$ is simply encoded without message splitting. Indeed, since $M_2$ is perfectly known at the receiver $Y_1$, it is not required to split $M_2$, unlike the HK scheme [67] for the channel without OSRSI. At the receiver $Y_1$, the messages $M_{10}$ and $M_{11}$ are jointly decoded using the received sequence as well as the known codeword conveying $M_2$. The decoding procedure at the receiver $Y_2$ is similar to its counterpart in [68, Lemma 3]. ■

***Proposition III.26)*** *Define the rate region* $\mathfrak{R}_o^{CIC/OSRSI}$ *as follows:*

$$\mathfrak{R}_o^{CIC/OSRSI} \triangleq \bigcup_{P_Q P_{X_1|Q} P_{X_2|Q} P_{V|X_1 X_2 Q}} \left\{\begin{array}{l}(R_1, R_2) \in \mathbb{R}_+^2 : \\ R_1 \leq \min\left\{\begin{array}{c} I(X_1; Y_1|X_2, Q), \\ I(X_1; Y_1|V, X_2, Q) + I(V; Y_2|X_2, Q)\end{array}\right\} \\ , \\ R_2 \leq \min\{I(V, X_2; Y_2|Q), I(X_2; Y_2|X_1, Q)\} \\ , \\ R_1 + R_2 \leq I(X_1; Y_1|V, X_2, Q) + I(V, X_2; Y_2|Q)\end{array}\right\}$$

(III~113)

*The set* $\mathfrak{R}_o^{CIC/OSRSI}$ *constitutes an outer bound on the capacity region of the two-user CIC with OSRSI in Fig.11.*

*Proof of Proposition III.26)* From the approach we followed in Subsection III.B to derive the constraints (III~70), it is directly deduced that the following subset of these constraints also holds for the case where $M_2$ is available at the receiver $Y_1$ as side information.

$$\left\{\begin{array}{l} R_1 \leq \min\left\{\begin{array}{c} I(M_1, Z; Y_1|M_2, Q) \\ I(M_1; Y_1|Z, M_2, Q) + I(Z; Y_2|M_2, Q)\end{array}\right\} \\ , \\ R_2 \leq \min\{I(M_2, Z; Y_2|Q), I(M_2, Z; Y_2|M_1, Q)\} \\ , \\ R_1 + R_2 \leq I(M_1; Y_1|Z, M_2, Q) + I(M_2, Z; Y_2|Q)\end{array}\right.$$

(III~114)

For the two-user CIC, the joint PDF of the RVs $M_1, M_2, Z, Q, X_1, X_2$ is given by (III~75). Now, the bounds (III~113) are derived from (III~114) similar to derivation of their counterparts in the UV-outer bound given by (III~76). ■

We next derive some capacity results for the system. First, consider the channel with strong interference at receiver $Y_2$ which satisfies:

$$I(X_1;Y_1|X_2) \leq I(X_1;Y_2|X_2)$$

(III~115)

for all joint PDFs $P_{X_1}(x_1)P_{X_2}(x_2)$. We have the following result.

***Proposition III.27)*** *The capacity region of the two-user CIC with OSRSI in Fig. 11, which satisfies the strong interference condition at the receiver $Y_2$ in (III~115), is given as follows:*

$$\bigcup_{P_Q P_{X_1|Q} P_{X_2|Q}} \left\{\begin{array}{l} (R_1,R_2) \in \mathbb{R}_+^2 : \\ R_1 \leq I(X_1;Y_1|X_2,Q) \\ R_2 \leq I(X_2;Y_2|X_1,Q) \\ R_1 + R_2 \leq I(X_1,X_2;Y_2|Q) \end{array}\right\}$$

(III~116)

*Proof of Proposition III.27)* The achievability is derived from (III~112) by setting $W_1 \equiv X_1$. The converse part is obtained from the outer bound $\mathfrak{R}_o^{CIC/OSRSI}$ given in (III~113). Note that we have:

$$\begin{aligned} R_1 + R_2 &\leq I(X_1;Y_1|V,X_2,Q) + I(V,X_2;Y_2|Q) \\ &\overset{(a)}{\leq} I(X_1;Y_2|V,X_2,Q) + I(V,X_2;Y_2|Q) \\ &= I(X_1,X_2;Y_2|Q) \end{aligned}$$

where the inequality (a) is due to (III~115) and its extension in Lemma III.1. ■

***Remark III.10:*** Comparing (III~116) and (III~23), we see that the OSRSI increases the capacity region of the two-user CIC even for the channel with strong interference at both receivers (III~21). In other words, the optimality of decoding both messages at each receiver in the strong interference regime does not imply that providing a non-respective message at a receiver is useless. Later, we demonstrate that the improvement on the capacity region due to the OSRSI can be strict.

We next consider the channel with weak interference at the receiver $Y_2$ such that:

$$I(V;Y_2|X_2) \leq I(V;Y_1|X_2) \quad \text{for all joint PDFs} \quad P_{VX_1}(v,x_1)P_{X_2}(x_2)$$

(III~117)

In the following proposition, we derive the sum-rate capacity of this channel.

***Proposition III.28)*** *The sum-rate capacity of the two-user CIC with OSRSI in Fig. 11, which satisfies the weak interference condition at the receiver $Y_2$ in (III~117), is given as follows:*

$$\max_{P_Q P_{X_1|Q} P_{X_2|Q}} \big(I(X_1;Y_1|X_2,Q) + I(X_2;Y_2|Q)\big)$$

(III~118)

*Proof of Proposition III.28)* The achievability is derived from (III~112) by setting $W_1 \equiv \emptyset$. Indeed, to achieve this sum-rate, the receiver $Y_2$ treats interference as noise. The converse part is obtained from the outer bound $\mathfrak{R}_o^{CIC/OSRSI}$ given by (III~113); we have:

$$\begin{aligned} R_1 + R_2 &\leq I(X_1;Y_1|V,X_2,Q) + I(V,X_2;Y_2|Q) \\ &= I(X_1;Y_1|V,X_2,Q) + I(V;Y_2|X_2,Q) + I(X_2;Y_2|Q) \\ &\overset{(a)}{\leq} I(X_1;Y_1|V,X_2,Q) + I(V;Y_1|X_2,Q) + I(X_2;Y_2|Q) \\ &= I(X_1;Y_1|X_2,Q) + I(X_2;Y_2|Q) \end{aligned}$$

where inequality (a) is due to (III~117). ■

We see that for the channel with OSRSI depicted in Fig. 11, a single-letter characterization of the capacity region and the sum-rate capacity is derived for both the cases of strong and weak interference at the receiver $Y_2$, respectively. In fact, the capacity region does not depend on the interference level (strong or weak) at the receiver $Y_1$, because this receiver knows the interference perfectly. Now, let us investigate the two-user CIC with OSRSI in the mixed interference regime. We have two different cases:

- In the first case, the interference at the receiver $Y_1$ is weak while it is strong at receiver $Y_2$. In other words, the channel satisfies:

$$\begin{cases} I(X_1;Y_1|X_2) \leq I(X_1;Y_2|X_2) \\ I(U;Y_1|X_1) \leq I(U;Y_2|X_1) \end{cases}$$

(III~119)

  for all joint PDFs $P_{X_1}(x_1)P_{UX_2}(u,x_2)$. Based on the result of Proposition III.27, the sum-rate capacity is given by:

$$\min_{P_Q P_{X_1|Q} P_{X_2|Q}} \begin{Bmatrix} I(X_1,X_2;Y_2|Q), \\ I(X_1;Y_1|X_2,Q) + I(X_2;Y_2|X_1,Q) \end{Bmatrix}$$

(III~120)

  Also, note that the sum-rate capacity of the channel without OSRSI is given below (see Theorem III.6 wherein the indices 1 and 2 should be exchanged everywhere):

$$\min_{P_Q P_{X_1|Q} P_{X_2|Q}} \begin{Bmatrix} I(X_1,X_2;Y_2|Q), \\ I(X_1;Y_1|Q) + I(X_2;Y_2|X_1,Q) \end{Bmatrix}$$

(III~121)

  Comparing (III~120) and (III~121), we deduce that the OSRSI increases the sum-rate capacity.

- In the second case, the interference at the receiver $Y_1$ is strong while it is weak at the receiver $Y_2$, i.e., the channel satisfies:

$$\begin{cases} I(X_2;Y_2|X_1) \leq I(X_2;Y_1|X_1) \\ I(V;Y_2|X_2) \leq I(V;Y_1|X_2) \end{cases}$$

(III~122)

  for all joint PDFs $P_{VX_1}(v,x_1)P_{X_2}(x_2)$. Based on Proposition III.28, the sum-rate capacity is given by (III~118). Also, note that the sum-rate capacity of the channel without OSRSI is given by (III~91). Comparing (III~118) and (III~91), we see that the OSRSI also increases the sum-rate capacity for this case.

Thus, the OSRSI increases the sum-rate capacity of the channel with mixed interference.

Then, consider the Gaussian channel (III~17) with OSRSI. In this scenario, the interference $X_2$ is subtracted from $Y_1$ as it is known at this receiver a priori. Therefore, the channel is equivalent to the one-sided CIC in (III~31-b). In other words, we have the following equivalence:

$$\begin{cases} Y_1 = X_1 + aX_2 + Z_1 \\ Y_2 = bX_1 + X_2 + Z_2 \end{cases} \overset{Y_1 \text{ knows } M_2}{\approx} \begin{cases} Y_1 = X_1 + Z_1 \\ Y_2 = bX_1 + X_2 + Z_2 \end{cases}$$

(III~123)

Therefore, regardless of the value of $a$, the capacity region of the channel with $|b| \geq 1$ and the sum-rate capacity for the channel with $|b| < 1$ are derived from Proposition III.12.

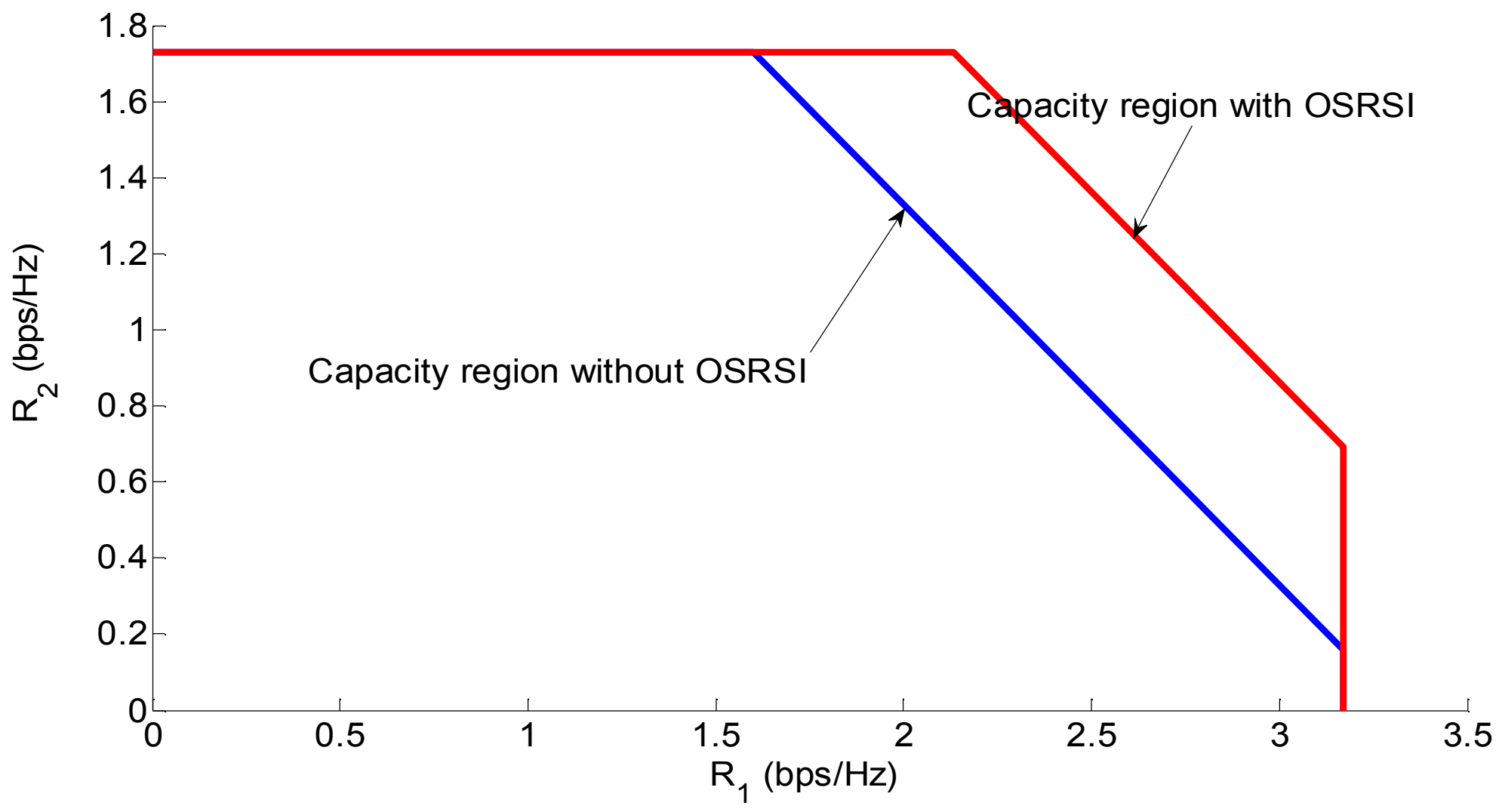

Figure 12. The capacity region of the two-user Gaussian CIC (III~17) with and without OSRSI in the strong interference regime; $a=\sqrt{2},\ b=\sqrt{2.5},\ P_1=80,\ P_2=10$.

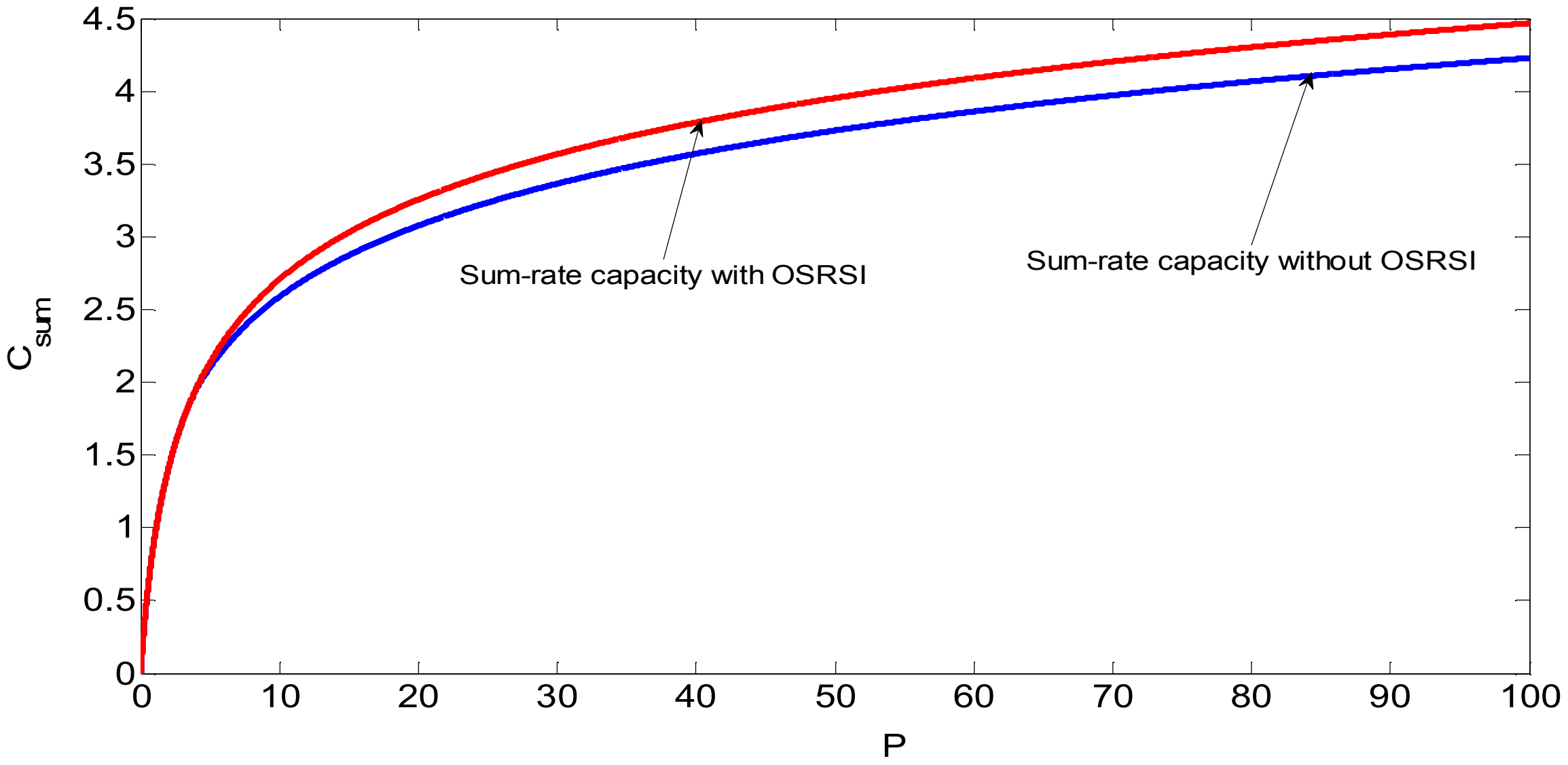

Figure 13. The sum-rate capacity of the two-user Gaussian CIC (III~17) with and without OSRSI in the mixed interference regime for $P_1=P_2=P$, $a=\sqrt{2.5},\ b=\sqrt{.25}$.

Figure 12 depicts the capacity region of the two-user Gaussian CIC (III~17) with and without OSRSI in the strong interference regime. As we see, in spite of the optimality of decoding both messages at both receivers in this regime, the availability of the OSRSI strictly improves the capacity region. In Fig. 13, we observe the sum-rate capacity of the Gaussian channel with and without OSRSI in the mixed interference regime in terms of the power $P$, where $P=P_1=P_2$. As shown, a strict improvement is achieved by the OSRSI for this case as well.

### III.D.3) The CRC with OSRSI

Lastly, we consider the CRC with OSRSI. In this case, due to asymmetric property of the channel, two different scenarios arise. These two situations are represented by Cases 1 and 2: In Case 1, the cognitive receiver ($Y_1$ in Fig. 14) has access to the message designated to the primary receiver ($M_2$ in Fig. 14), while in Case 2, the primary receiver ($Y_2$ in Fig. 14) has access to the message designated to the cognitive receiver ($M_1$ in Fig. 14).

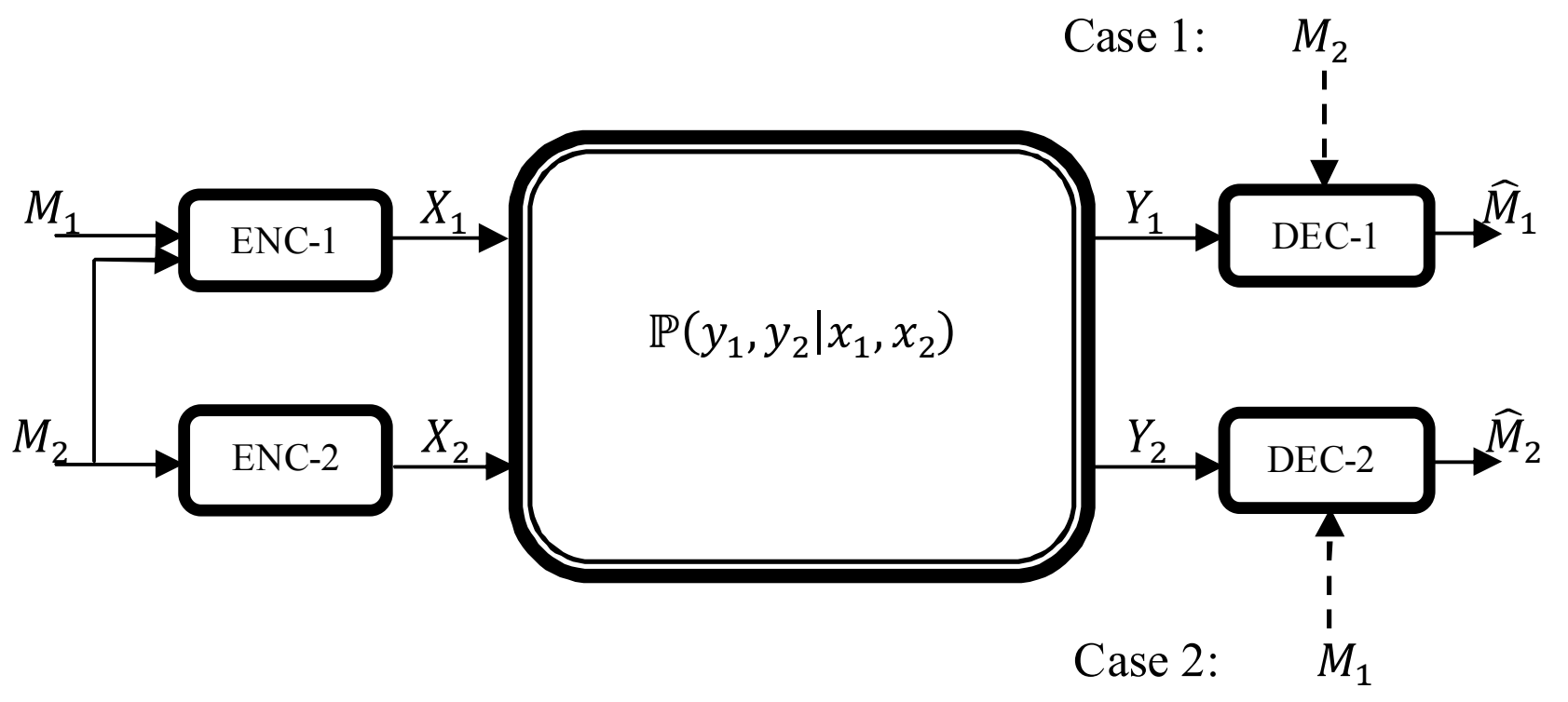

Figure 14. The CRC with OSRSI: In Case 1, $M_2$ is known at the cognitive reciever; in Case 2, $M_1$ is known at the primary receiver.

For Case 1, the decoding rule at the receiver $Y_1$ is given by (III~108). In Case 2, the decoder function at the receiver $Y_2$ for a code of length $n$ is given as follows:

$$\mathfrak{D}_2: \mathcal{Y}_2^n \times \mathcal{M}_1 \to \mathcal{M}_2$$

(III~124)

where the message $M_2$ is estimated as $\widehat{M}_2 = \mathfrak{D}_2(Y_2^n, M_1)$. In both cases, other elements of the coding scheme are identical to the CRC without OSRSI.

First, we investigate Case 1. In the following, we prove that for this case a full characterization of the capacity region can be derived using the superposition coding.

***Theorem III.9)*** *Define the rate region $\mathcal{C}_{case\ 1}^{CRC/OSRSI}$ as follows:*

$$\mathcal{C}_{case\ 1}^{CRC/OSRSI} \triangleq \bigcup_{P_{VX_1X_2}} \left\{ \begin{array}{l} (R_1, R_2) \in \mathbb{R}_+^2 : \\ R_1 \leq I(X_1; Y_1|X_2) \\ R_2 \leq I(V, X_2; Y_2) \\ R_1 + R_2 \leq I(X_1; Y_1|V, X_2) + I(V, X_2; Y_2) \end{array} \right\}$$

(III~125)

*The set $\mathcal{C}_{case\ 1}^{CRC/OSRSI}$ is the capacity region for the CRC with OSRSI where the cognitive receiver has access to the message designated for the primary receiver, i.e., Case 1 in Fig. 14.*

*Proof of Theorem III.9)* To prove the achievability, we use rate splitting as well as random superposition coding. Consider transmission of messages $M_1, M_2$ with the rates $R_1, R_2$, respectively. Fix the distribution $P_{VX_1X_2}$. The message $M_1$ and its corresponding rate $R_1$ are split in two parts as:

$$M_1 = (M_{10}, M_{11}), \qquad R_1 = R_{10} + R_{11}$$

(III~126)

The messages $M_{10}, M_{11}, M_2$ are then encoded by random codewords of length $n$ in a superposition fashion as follows:

1. Generate at random $2^{nR_2}$ independent codewords $X_2^n$ according to $Pr(x_2^n) = \prod_{t=1}^{n} P_{X_2}(x_{2,t})$. Label these codewords as $X_2^n(m_2)$, where $m_2 \in [1:2^{nR_2}]$.
2. For each $X_2^n(m_2)$, randomly generate $2^{nR_{10}}$ independent codewords $V^n$ according to $Pr(v^n) = \prod_{t=1}^{n} P_{V|X_2}(v_t|x_{2,t})$. Label these codewords as $V^n(m_2, m_{10})$, where $m_{10} \in [1:2^{nR_{10}}]$.
3. For each pair $(X_2^n(m_2), V^n(m_2, m_{10}))$, generate at random $2^{nR_{11}}$ codewords $X_1^n$ according to $Pr(x_1^n) = \prod_{t=1}^{n} P_{X_1|VX_2}(x_{1,t}|v_t, x_{2,t})$. Label these codewords as $X_1^n(m_2, m_{10}, m_{11})$.

Given a message triple $(m_2, m_{10}, m_{11})$, the primary transmitter and the cognitive transmitter transmit $X_2^n(m_2)$ and $X_1^n(m_2, m_{10}, m_{11})$, respectively. The decoding steps are also as follows:

1. At the receiver $Y_1$, the message $m_2$ is known a priori; the decoder tries to find a unique pair $(\hat{m}_{10},\hat{m}_{11})$ such that:

$$(X_1^n(m_2,\hat{m}_{10},\hat{m}_{11}),V^n(m_2,\hat{m}_{10}),X_2^n(m_2),Y_1^n)\in\mathcal{T}_\epsilon^n(P_{X_2VX_1Y_1})$$

Decoding at this step can be done reliably provided that:

$$\begin{aligned} R_{11}+R_{10}&<I(X_1,V;Y_1|X_2)=I(X_1;Y_1|X_2)\\ R_{11}&<I(X_1;Y_1|V,X_2) \end{aligned}$$

(III~127)

2. At the receiver $Y_2$, the decoder tries to find a unique $\hat{m}_2$ for which there exists some $\bar{m}_{10}$ such that:

$$(V^n(\hat{m}_2,\bar{m}_{10}),X_2^n(\hat{m}_2),Y_2^n)\in\mathcal{T}_\epsilon^n(P_{VX_2Y_2})$$

Decoding at this step can be done reliably provided that:

$$R_2+R_{10}<I(V,X_2;Y_2)$$

(III~128)

Then, by collecting (III~126)-(III~128) and applying a simple Fourier-Motzkin elimination (see, e.g. [114]) to remove $R_{10},R_{11}$, we derive (III~125). For the converse part, as mentioned in the proof of Proposition III.26, the constraints (III.114) hold for case where $M_2$ is available at the receiver $Y_1$ as side information. For the CRC, the joint PDF of the random variables $M_1,M_2,Z,Q,X_1,X_2$ is given by (III~79). Now, the bounds in (III~125) are derived similarly to (A~9), (A~11), and (A~12). This completes the proof. ■

***Remark III.11:*** For the case where $\|\mathcal{X}_2\|=1$, the CRC is reduced to a two-user BC. Accordingly, Theorem III.9 yields Proposition III.24 as a special case.

Here, we recall the rate region (III~99). As discussed before, this rate region, which is derived by superposition coding, achieves the capacity for the case where both messages are required to be decoded the cognitive receiver (see [88, Th. 4]). Hence, it is also achievable for the CRC (in Fig. 6) without OSRSI. Specifically, this achievable rate region is optimal for the more-capable CRC ((III~98-a) without OSRSI) as shown in Theorem III.8. In fact, for the more-capable CRC without OSRSI, it is optimal to decode both messages at the cognitive receiver. Let us compare (III~125) with the rate region (III~99). We see that the latter has one additional constraint given by:

$$R_1+R_2\leq I(X_1,X_2;Y_1)$$

Thus, we deduce that the knowledge of the primary message at the cognitive receiver is beneficial and increases the capacity region even for the more-capable CRC for which decoding both messages at the cognitive receiver is the optimal strategy. Then, consider the less-noisy CRC (the CRC in the "better cognitive decoding regime") defined in (III~50). As proved in Subsection III.C, these channels are strictly included in the class of more-capable CRCs in (III~98-a). By comparing (III~125) and (III~52), we see that unlike the more-capable channel, providing the message of the primary receiver at the cognitive receiver does not increase the capacity region of the less-noisy CRC. Based on this discussion, we infer the following fact regarding the "more-capable" and the "less-noisy" concepts;

***Note:*** *For both more-capable (III~98-a) and less-noisy (III~50) CRCs, the superposition coding with decoding both messages at the cognitive receiver is the optimal coding scheme; however, the knowledge of the message of the primary receiver at the cognitive receiver has no benefit for the less-noisy CRC, but this strategy increases the capacity region of the more-capable channel.*

Let us now consider the Gaussian channel formulated in (III~17) with OSRSI (Case 1). Based on Theorem III.9, we can characterize the capacity region of this channel as follows.

***Corollary III.6)*** Consider *the Gaussian CRC formulated in (III~17) with OSRSI where the cognitive receiver has access to the message intended to the primary receiver, i.e., Case 1 in Fig. 14. The capacity region for the case of* $|b|\leq 1$ *is similar to the channel without OSRSI and is given by (III~47). Also, for the case where* $|b|>1$, *the capacity region is given by (III~48).*

*Proof of Corollary III.6)* For the channel with $|b|\leq 1$, using the EPI, one can show that the rate region (III~125) is optimal in consideration of Gaussian distributions and coincides with (III~47); see also [34]. For the channel with $|b|>1$, the interference at the primary receiver is strong, i.e., (III~100) holds. Therefore, we have:

$$I(X_1;Y_1|V,X_2)+I(V,X_2;Y_2)\leq I(X_1;Y_2|V,X_2)+I(V,X_2;Y_2)=I(X_1,X_2;Y_2)$$

Thus, the rate region (III~125) is optimized in consideration of $V\equiv X_1$. By this substitution, the resulting rate region is equivalent to (III~48). ■

***Remarks III.12:***

1. The reason that for the case of $|b|\leq 1$, the knowledge of the primary message at the cognitive receiver does not increase the capacity region is justified as follows: As discussed in Proposition III.16, for this channel the capacity region is achieved by the dirty paper coding [34]. This coding technique is used to pre-cancel the known interference at the cognitive receiver. Note that for the dirty paper channel [10], the knowledge of the interference at the transmitter is equivalent to knowing it at the receiver; in other words, when the transmitter is informed of the additive interference, its availability at the receiver does not affect the capacity region. Similarly, for the Gaussian CRC with $|b|\leq 1$, the availability of the interference (known to the cognitive transmitter) at the cognitive receiver does not increase the capacity region.
2. Corollary III.6 reveals that for the Gaussian CRC with $|b|>1$, the outer bound (III~48) for the channel without OSRSI is indeed the capacity of the corresponding channel with OSRSI where the cognitive receiver has access to the message intended to the primary receiver, i.e., Case 1 in Fig. 14.
3. For the case of $|b|>1$, the knowledge of the primary message at the cognitive receiver can increase the capacity region. This is because the outer bound (III~48) is not optimal in general for the Gaussian CRC with $|b|>1$ without OSRSI.
4. Unlike the two-user CIC for which the OSRSI is beneficial even in the strong interference regime, for the CRC satisfying the strong interference conditions in (III~39)-(III~40), the knowledge of the primary message at the cognitive receiver does not increase the capacity region, as it is still given by (III~41)-(III~42).

Next, let us consider Case 2 where the primary receiver knows the message of the cognitive receiver. In the following, we derive inner and outer bounds on the capacity region of this scenario. We also establish the capacity region for some special cases; however, the capacity region of the general case remains unresolved.

***Theorem III.10)*** *Define the rate region* $\mathfrak{R}_{i,case\ 2}^{CRC/OSRSI}$ *as follows:*

$$\mathfrak{R}_{i,case\ 2}^{CRC/OSRSI}\triangleq\bigcup_{P_{W_2W_1X_1X_2}(w_2,w_1,x_1,x_2)}\left\{\begin{array}{l}(R_1,R_2)\in\mathbb{R}_+^2:\\ R_1\leq I(W_1;Y_1|W_2)-I(X_2;W_1|W_2)\\ R_2\leq I(X_1,X_2;Y_2)\\ R_1+R_2\leq I(X_1,X_2;Y_2|W_2,W_1)+I(W_2,W_1;Y_1)\\ R_1+R_2\leq I(X_1,X_2;Y_2|W_2)+I(W_2,W_1;Y_1)-I(X_2;W_1|W_2)\end{array}\right\}$$

(III~129)

*The set* $\mathfrak{R}_{i,case\ 2}^{CRC/OSRSI}$ *constitutes an inner bound on the capacity region of the CRC with OSRSI where the primary receiver has access to the message of the cognitive receiver, i.e., Case 2 in Fig. 14.*

*Proof of Theorem III.10)* Refer to Appendix. ■

***Proposition III.29)*** *Define the rate region* $\mathfrak{R}_{o,case\ 2}^{CRC/OSRSI}$ *as follows:*

$$\mathfrak{R}_{o,case\ 2}^{CRC/OSRSI}\triangleq\bigcup_{P_{UX_1X_2}(u,x_1,x_2)}\left\{\begin{array}{l}(R_1,R_2)\in\mathbb{R}_+^2:\\ R_1\leq\min\{I(U;Y_1),I(X_1;Y_1|X_2)\}\\ R_2\leq I(X_1,X_2;Y_2)\\ R_1+R_2\leq I(X_1,X_2;Y_2|U)+I(U;Y_1)\end{array}\right\}$$

(III~130)

*The set* $\mathfrak{R}_{o,case\ 2}^{CRC/OSRSI}$ *constitutes an outer bound on the capacity region of the CRC with OSRSI where the primary receiver has access to the message of the cognitive receiver, i.e., Case 2 in Fig. 14.*

*Proof of Proposition III.29)* From the approach we established the constraints (III~70), it is directly derived that the following subset of these constraints also holds for the case when $M_1$ is available at the receiver $Y_2$ as side information:

$$\begin{cases} R_1 \leq \min\{I(M_1,Z;Y_1|Q), I(M_1,Z;Y_1|M_2,Q)\} \\ R_2 \leq I(M_2,Z;Y_2|M_1,Q) \\ R_1+R_2 \leq I(M_2;Y_2|Z,M_1,Q)+I(M_1,Z;Y_1|Q) \end{cases}$$

(III~131)

For the CRC, the joint PDF of the RVs $M_1, M_2, Z, Q, X_1, X_2$ is given by (III~79). Now, with the exception of the second constraint of (III~130), the others are derived from (III~131) similar to the derivation of their counterparts in the UV-outer bound given in (III~80). To derive the second constraint of (III~130), from (III~131) we have:

$$R_2 \leq I(M_2,Z;Y_2|M_1,Q) \leq I(M_1,M_2,Z,Q;Y_2) \overset{(a)}{=} I(X_1,X_2;Y_2)$$

where equality (a) holds because $X_1$ is a deterministic function of $(M_1,M_2,Q)$ and $X_2$ is a deterministic function of $(M_2,Q)$. The proof is thus complete. ■

In the following, we provide special cases where the inner bound (III~129) and the outer bound (III~130) coincide which establish the capacity region. The first class is the more-capable CRCs.

***Proposition III.30)*** *The capacity region of the more-capable CRC (III~98-a) with OSRSI where the primary receiver has access to the message of the cognitive receiver, Case 2 in Fig. 14, is given by:*

$$\mathcal{C}_{case\ 2}^{CRC_{mc}/OSRSI} \triangleq \bigcup_{P_{X_1X_2}} \left\{ \begin{array}{l} (R_1,R_2) \in \mathbb{R}_+^2 : \\ R_1 \leq I(X_1;Y_1|X_2) \\ R_2 \leq I(X_1,X_2;Y_2) \\ R_1+R_2 \leq I(X_1,X_2;Y_1) \end{array} \right\}$$

(III~132)

*Also, the capacity region of the Gaussian channel (III~17) with* $|b| \leq 1 \leq |a|$ *and* $ab=1$ *is given by:*

$$\mathcal{C}_{case\ 2}^{CRC_{mc}/OSRSI\sim G} \triangleq \bigcup_{0\leq\alpha\leq 1} \left\{ \begin{array}{l} (R_1,R_2) \in \mathbb{R}_+^2 : \\ R_1 \leq \psi(\alpha P_1) \\ R_2 \leq \psi\left(b^2P_1 + 2|b|\sqrt{(1-\alpha)P_1P_2} + P_2\right) \\ R_1+R_2 \leq \psi\left(P_1 + 2|a|\sqrt{(1-\alpha)P_1P_2} + a^2P_2\right) \end{array} \right\}$$

(III~133)

*Proof of Proposition III.30)* The achievability is derived from $\mathfrak{R}_{i,case\ 2}^{CRC/OSRSI}$ in (III~129) by setting $W_1 \equiv X_1$ and $W_2 \equiv X_2$. For the converse part, using the outer bound $\mathfrak{R}_{o,case\ 2}^{CRC/OSRSI}$ given in (III~130), we have:

$$\begin{aligned} R_1+R_2 &\leq I(X_1,X_2;Y_2|U) + I(U;Y_1) \\ &\overset{(a)}{\leq} I(X_1,X_2;Y_1|U) + I(U;Y_1) \\ &\leq I(X_1,X_2;Y_1) \end{aligned}$$

where inequality (a) is due to the more-capable condition in (III~98-a) and its extension in Lemma III.3. Also, consider the Gaussian channel (III~17) with $|b| \leq 1 \leq |a|$ and $ab=1$. Define:

$$\tilde{Y}_2 \triangleq bY_1 + \sqrt{(1-b^2)}\tilde{Z}_2$$

(III~134)

where $\tilde{Z}_2$ is a Gaussian RV independent of $Z_1$ with zero mean and unit variance. Note that (III~134) implies the following Markov relation:

$$X_1, X_2 \to Y_1 \to \tilde{Y}_2$$

Moreover, $\tilde{Y}_2 \cong Y_2$ for all distributions $P_{X_1X_2}$. Therefore, we have:

$$I(X_1,X_2;Y_2) = I(X_1,X_2;\tilde{Y}_2) \leq I(X_1,X_2;Y_1) \qquad \text{for all joint PDFs} \qquad P_{X_1X_2}(x_1,x_2)$$

In other words, the channel satisfies the more-capable condition (III~98-a). Thereby, its capacity is given by (III~132). ■

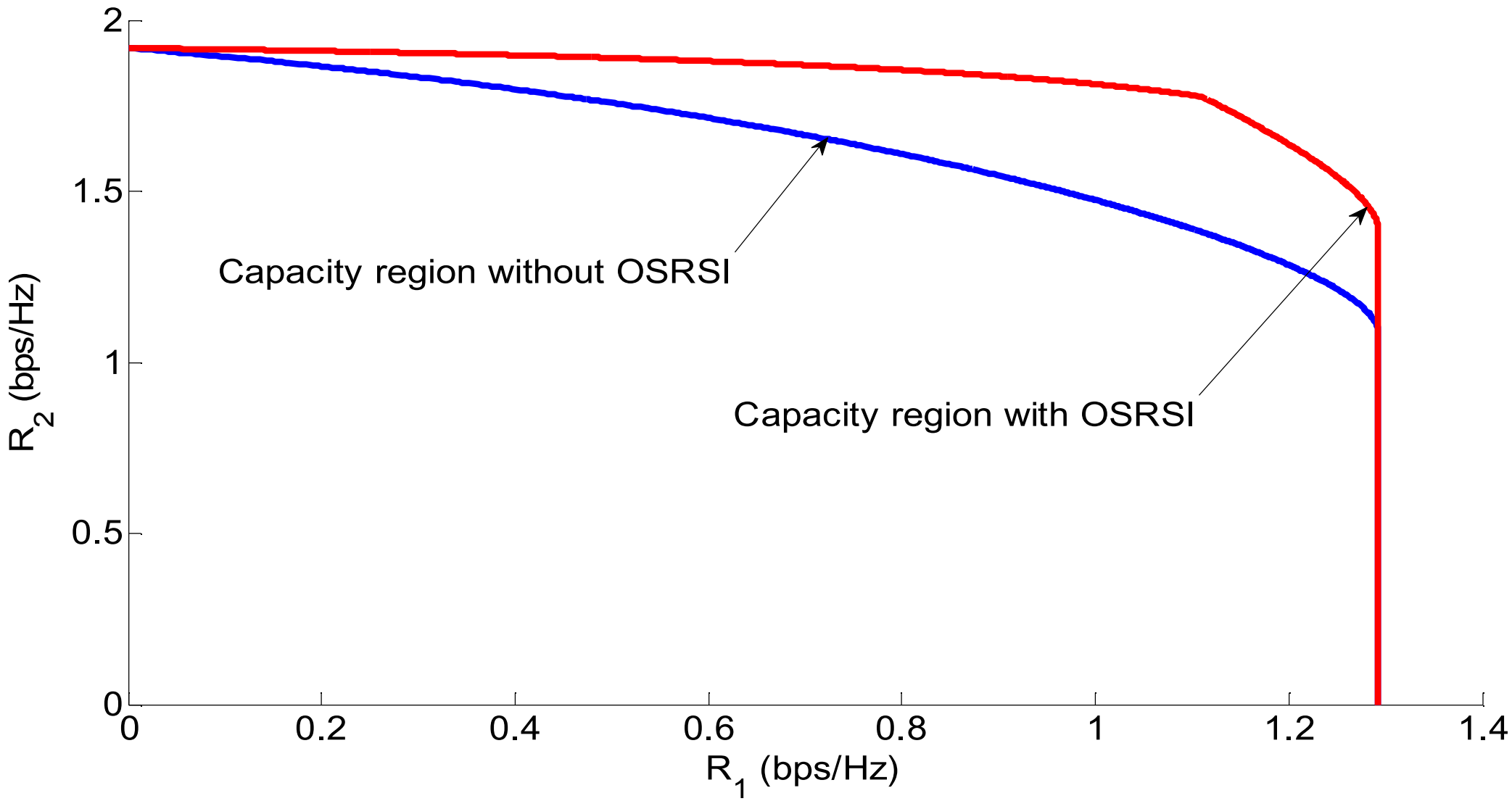

Figure 15. The capacity region of the Gaussian degraded CRC (III~17) with and without OSRSI (Case 2 of Fig. 14) for $a=\sqrt{5},\ b=\frac{1}{\sqrt{5}},\ P_1=5,\ P_2=7$.

***Remarks III.13:***

1. Comparing (III~132) and (III~99), we see that the availability of $M_1$ at the receiver $Y_2$ increases the capacity region for the more-capable CRC (III~98-a). In Fig. 15, we have plotted the capacity region of a Gaussian CRC (III~17) with and without OSRSI ($M_1$ at receiver $Y_2$) where $|b|\leq 1\leq |a|$ and $ab=1$. As we observe, the capacity region is strictly increased with OSRSI for this channel.
2. The capacity result in Proposition III.30 also holds for the CRC satisfying the strong interference conditions (III~39) because this channel is also more-capable in the sense of (III~98-a). By comparing (III~132) and (III~41), we see that the OSRSI ($M_1$ at the receiver $Y_2$) can increase the capacity region for the strong interference channel (III~39).
3. The Gaussian CRC in the strong interference regime (III~40) does not necessarily satisfy the more-capable condition (III~98-a). Therefore, it is not clear that how much the availability of $M_1$ at the receiver $Y_2$ increases the capacity region for this channel.

Another case for which the inner bound (III~129) and the outer bound (III~130) coincide is a class of semi-deterministic CRCs as given in the next proposition.

***Proposition III.31)*** *Consider the semi-deterministic CRC with $Y_1=f(X_1,X_2)$ where $f(.)$ is a deterministic function. The capacity region of the channel with OSRSI where the primary receiver has access to the message of the cognitive receiver, i.e., Case 2 in Fig. 14, is given by:*

$$\mathcal{C}_{case\ 2}^{CRC_{sd}/OSRSI} \triangleq \bigcup_{P_{X_1X_2}} \left\{ \begin{array}{l} (R_1,R_2)\in\mathbb{R}_+^2 : \\ R_1\leq H(Y_1|X_2) \\ R_2\leq I(X_1,X_2;Y_2) \\ R_1+R_2\leq I(X_1,X_2;Y_1,Y_2) \end{array} \right\}$$

(III~135)

*Proof of Proposition III.31)* The achievability is derived from $\mathfrak{R}_{i,case\ 2}^{CRC/OSRSI}$ in (III~129) by setting $W_1\equiv Y_1$ and $W_2\equiv\emptyset$. For the converse part using the outer bound $\mathfrak{R}_{o,case\ 2}^{CRC/OSRSI}$ given in (III~130), we have:

$$\begin{aligned} R_1+R_2 &\leq I(X_1,X_2;Y_2|U)+I(U;Y_1) \\ &\leq I(X_1,X_2;Y_1,Y_2|U)+I(U;Y_1,Y_2) \\ &= I(X_1,X_2;Y_1,Y_2) \end{aligned}$$

The proof is thus complete. ■

Finally, it is noteworthy to mention that establishing the capacity region of an interference network with OSRSI is sometimes more difficult than that of its counterpart without OSRSI. For example, for the case of no OSRSI, we know (see Proposition III.16) that the dirty paper coding achieves the capacity region of the Gaussian CRC (III~17) where $|b| \leq 1$; however, the capacity region is not known when $M_1$ is available at the receiver $Y_2$ except for the special case of $ab = 1$. Another example in this regard is the Gaussian CRC (III~17) in the strong interference regime (III~40) as discussed in Remarks III.13. The difficulty of establishing the capacity region for these cases is due to the lack of the following constraints:

$$R_2 \leq I(V, X_2; Y_2)$$
$$R_1 + R_2 \leq I(X_1; Y_1|V, X_2) + I(V, X_2; Y_2)$$

(III~136)

subject to the joint PDFs $P_{VX_1X_2}(v, x_1, x_2)$. For the channel with no OSRSI, these constraints hold for the rates $R_1$ and $R_2$, which are used to establish the capacity region in the latter Gaussian scenarios. Nonetheless, this is not the case when $M_1$ is available at the receiver $Y_2$, i.e., Case 2 in Fig. 14.

# CONCLUSION

We tried to build a systematic study of fundamental limits of communications for the interference networks with arbitrary configurations. We launched this study by investigating capacity limits for the basic building blocks in this part. We proposed the MAC, the BC, the CIC and the CRC as the main building blocks for all interference networks. First, we briefly reviewed the existing results regarding these channels. We then established new capacity outer bounds for the basic structures with two receivers. These outer bounds are all derived based on a unified framework. Using the derived outer bounds, we identified a "*mixed interference regime*" for the two-user CIC and obtained the sum-rate capacity for this regime. This result contains the previously known one regarding the Gaussian mixed interference channel as a special case. Also, we presented a noisy interference regime for the one-sided CIC in which treating interference as noise is sum-rate optimal. For the CRC, we derived a full characterization of the capacity region for a class of "*more-capable*" channels. Finally, we studied capacity bounds for the two-user BC, the two-user CIC and the CRC with one-sided receiver side information where one receiver has access to the non-intended message. Our results lead to new insights regarding the nature of information flow in the basic interference networks.

Our systematic study will be continued by considering large multi-user/multi-message networks in the next parts.

# APPENDIX

➢ ***Proof of Corollary III.1***

Consider the outer bound $\mathfrak{R}_o^{CIC}$ given by (III~74) and (III~75). We recall that in this bound, $X_i$ is a deterministic function of $(M_i,Q)$, $i=1,2$, and also $M_1,M_2,Z,Q \to X_1,X_2 \to Y_1,Y_2$ forms a Markov chain. Define new random variables $U$ and $V$ as follows:

$$U \triangleq (M_1,Z), \qquad V \triangleq (M_2,Z) \tag{A~1}$$

We have:

$$R_1 \leq I(M_1,Z;Y_1|Q) = I(M_1,Z,X_1;Y_1|Q) = I(U,X_1;Y_1|Q) \tag{A~2}$$

$$R_1 \leq I(M_1,Z;Y_1|M_2,Q) = I(X_1;Y_1|X_2,M_2,Q) \leq I(X_1;Y_1|X_2,Q) \tag{A~3}$$

$$\begin{aligned} R_1 &\leq I(M_1;Y_1|Z,M_2,Q) + I(Z;Y_2|M_2,Q) \\ &= I(X_1;Y_1|Z,M_2,X_2,Q) + I(Z;Y_2|M_2,X_2,Q) \\ &\leq I(X_1;Y_1|V,X_2,Q) + I(Z,M_2;Y_2|X_2,Q) \\ &= I(X_1;Y_1|V,X_2,Q) + I(V;Y_2|X_2,Q) \end{aligned} \tag{A~4}$$

Similarly, one can derive the desired bounds on the rate $R_2$. For the sum-rate we have:

$$\begin{aligned} R_1+R_2 &\leq I(M_1;Y_1|Z,M_2,Q) + I(M_2,Z;Y_2|Q) \\ &= I(X_1;Y_1|Z,M_2,X_2,Q) + I(M_2,Z,X_2;Y_2|Q) \\ &= I(X_1;Y_1|V,X_2,Q) + I(V,X_2;Y_2|Q) \end{aligned} \tag{A~5}$$

Similarly, one can derive:

$$R_1+R_2 \leq I(X_2;Y_2|U,X_1,Q) + I(U,X_1;Y_1|Q) \tag{A~6}$$

Finally, by definition (A~1), $X_1$ is a deterministic function of $(U,Q)$ and $X_2$ is a deterministic function of $(V,Q)$. This completes the proof. ■

➢ ***Proof of Corollary III.2***

Consider the outer bound $\mathfrak{R}_o^{CRC}$ given by (III~78) and (III~79). Define new random variables $U$ and $V$ as follows:

$$U \triangleq (M_1,Z,Q), \qquad V \triangleq (M_2,Z,Q) \tag{A~7}$$

Note that, by this definition, $X_1$ is a deterministic function of $(U,V)$ and $X_2$ is a deterministic function of $V$. We have:

$$R_1 \leq I(M_1,Z;Y_1|Q) \leq I(M_1,Z,Q;Y_1) = I(U;Y_1) \tag{A~8}$$

Also,

$$R_1 \leq I(M_1,Z;Y_1|M_2,Q) = I(X_1;Y_1|X_2,M_2,Q) \leq I(X_1;Y_1|X_2) \tag{A~9}$$

$$
\begin{aligned}
R_1 &\leq I(M_1;Y_1|Z,M_2,Q) + I(Z;Y_2|M_2,Q) \\
&= I(X_1;Y_1|Z,M_2,Q,X_2) + I(Z;Y_2|M_2,Q,X_2) \\
&\leq I(X_1;Y_1|X_2,Z,M_2,Q) + I(M_2,Z,Q;Y_2|X_2) \\
&= I(X_1;Y_1|X_2,V) + I(V;Y_2|X_2)
\end{aligned}
$$

(A~10)

For the rate $R_2$, we have:

$$R_2 \leq I(M_2,Z;Y_2|Q) \leq I(M_2,Z,Q,X_2;Y_2) = I(V,X_2;Y_2)$$

(A~11)

Lastly, for the sum-rate we have:

$$
\begin{aligned}
R_1 + R_2 &\leq I(M_1;Y_1|Z,M_2,Q) + I(M_2,Z;Y_2|Q) \\
&= I(X_1;Y_1|Z,M_2,Q,X_2) + I(M_2,Z,X_2;Y_2|Q) \\
&\leq I(X_1;Y_1|V,X_2) + I(V,X_2;Y_2)
\end{aligned}
$$

(A~12)

$$
\begin{aligned}
R_1 + R_2 &\leq I(M_2;Y_2|Z,M_1,Q) + I(M_1,Z;Y_1|Q) \\
&= I(X_1,X_2;Y_2|Z,M_1,Q) + I(M_1,Z;Y_1|Q) \\
&\leq I(X_1,X_2;Y_2|U) + I(U;Y_1)
\end{aligned}
$$

(A~13)

By collecting (A~8)-(A~13), we derive (III~80). ■

- ***Proof of Theorem III.10***

The coding scheme to achieve the rate region (III~129) is rather similar to that one we described for the channel without OSRSI with the exception that since the receiver $Y_2$ knows $M_1$ perfectly, it is not required to split the corresponding rate of this message, i.e., $R_1$. Only the message $M_2$ and its corresponding rate $R_2$ are split into two parts as follows:

$$M_2 = (M_{20}, M_{22}), \qquad R_2 = R_{20} + R_{22}$$

(A~14)

Consider the set of all joint PDFs $P_{W_2V_2W_1V_1X_2X_1}(w_2,v_2,w_1,v_1,x_2,x_1)$ satsifying:

$$P_{W_2V_2W_1V_1X_2X_1} = P_{W_2V_2W_1V_1}P_{X_2|V_2W_2}P_{X_1|X_2V_1W_1V_2W_2}$$

(A~15)

The messages $M_1, M_{20}, M_{22}$ are then encoded by random codewords of length $n$ as follows:

*Encoding at the primary transmitter:*

1. Generate at random $2^{nR_{20}}$ independent codewords $W_2^n$ according to $Pr(w_2^n) = \prod_{t=1}^n P_{W_2}(w_{2,t})$. Label these codewords as $W_2^n(m_{20})$, where $m_{20} \in [1:2^{nR_{20}}]$.
2. For each $W_2^n(m_{20})$, randomly generate $2^{nR_{22}}$ independent codewords $V_2^n$ according to $Pr(v_2^n) = \prod_{t=1}^n P_{V_2|W_2}(v_{2,t}|w_{2,t})$. Label these codewords as $V_2^n(m_{20},m_{22})$, where $m_{22} \in [1:2^{nR_{22}}]$.
3. For each pair $(W_2^n(m_{20}), V_2^n(m_{20},m_{22}))$, generate at random a codeword $X_2^n$ according to $Pr(x_2^n) = \prod_{t=1}^n P_{X_2|V_2W_2}(x_{2,t}|v_{2,t},w_{2,t})$. Label this codeword by $X_2^n(m_{20},m_{22})$.

Given the messages $(m_{20},m_{22})$, the primary transmitter sends $X_2^n(m_{20},m_{22})$ over the channel.

*Encoding at the cognitive transmitter:*

1. For each $W_2^n$, randomly generate $2^{n(R_1+B_1)}$ independent codewords $W_1^n$ according to $Pr(w_1^n) = \prod_{t=1}^n P_{W_1|W_2}(w_{1,t}|w_{2,t})$. Label these codwords by $W_1^n(m_{20},m_1,b_1)$, where $(m_1,b_1) \in [1:2^{nR_1}] \times [1:2^{nB_1}]$.

Given $(m_{20}, m_{22}, m_1)$, let $b_1^T$ denotes the smallest $b_1 \in [1:2^{nB_1}]$ with:

$$\left(W_2^n(m_{20}), V_2^n(m_{20}, m_{22}), W_1^n(m_{20}, m_1, b_1)\right) \in \mathcal{T}_\epsilon^n\left(P_{W_2V_2W_1}\right)$$

If there is no such $b_1$, then let $b_1^T \triangleq 1$.

2. For each triple $\left(W_2^n(m_{20}), V_2^n(m_{20}, m_{22}), W_1^n(m_{20}, m_1, b_1^T)\right)$ generate at random a codword $V_1^n$ according to $Pr(v_1^n) = \prod_{t=1}^n P_{V_1|W_1V_2W_2}\left(v_{1,t}\middle|w_{1,t}, v_{2,t}, w_{2,t}\right)$. Label this codeword by $V_1^n(m_{20}, m_{22}, m_1, b_1^T)$.
3. For each 5-tuple $\left(W_2^n(m_{20}), V_2^n(m_{20}, m_{22}), W_1^n(m_{20}, m_1, b_1^T), V_1^n(m_{20}, m_{22}, m_1, b_1^T), X_2^n(m_{20}, m_{22})\right)$ randomly generate a codeword $X_1^n$ according to $Pr(x_1^n) = \prod_{t=1}^n P_{X_1|X_2V_1W_1V_2W_2}\left(x_{1,t}\middle|x_{2,t}, v_{1,t}, w_{1,t}, v_{2,t}, w_{2,t}\right)$. Label this codeword as $X_1^n(m_{20}, m_{22}, m_1, b_1^T)$.

Given the messages $(m_{20}, m_{22}, m_1)$, the cognitive transmitter sends $X_1^n(m_{20}, m_{22}, m_1, b_1^T)$ over the channel.

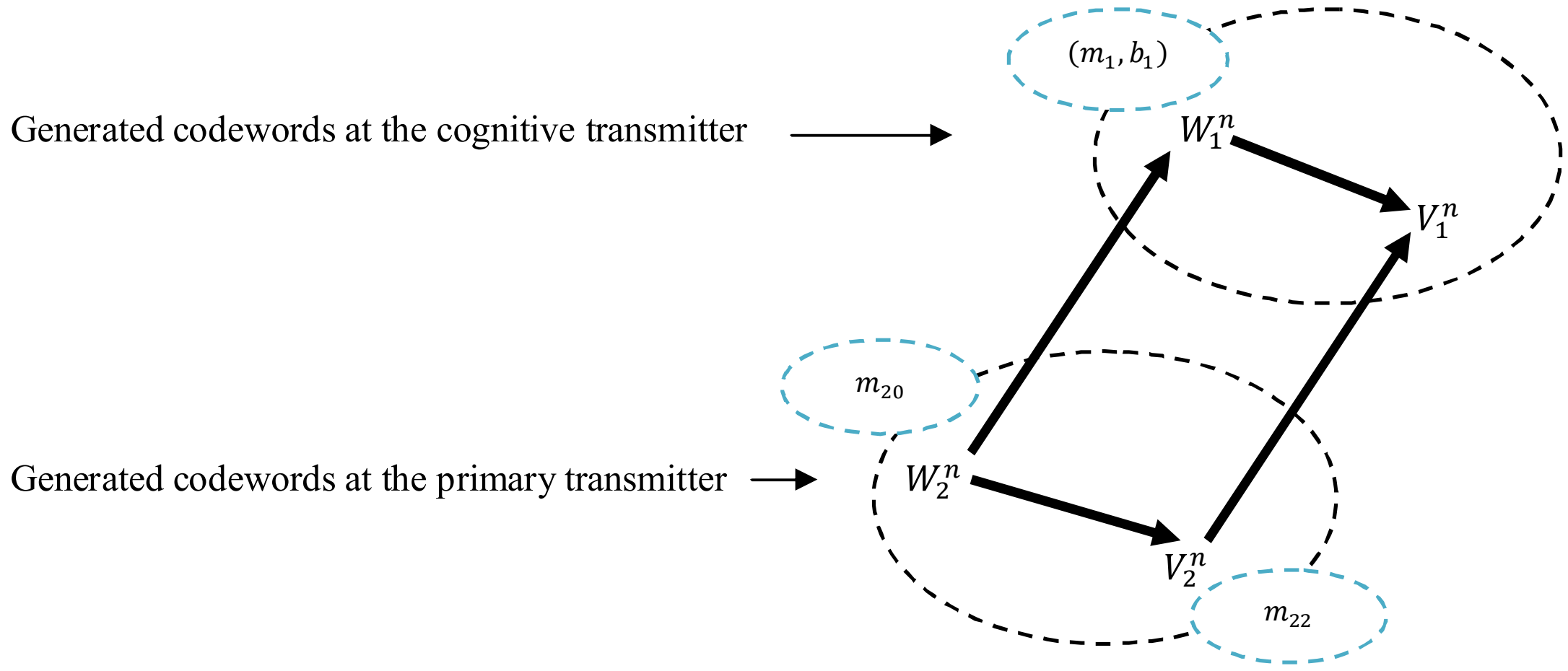

Figure 16. The graphical illustration of the generated codewords for the acheivability scheme in Theorem III.10.

The graphical illustration with respect to the coding scheme has been depicted in Fig. 16. This graphical representation indicates the superposition structures among the generated codewords. Comparing Fig. 16 and Fig. 7, we see that the superposition structures among the generated codewords in both schemes are the same; however, here because the message $M_1$ is not split, the codewords $U_1^n$ do not appear in the scheme.

Also, the decoding steps are as follows:

1. At the receiver $Y_1$, the decoder tries to find a unique triple $\left(\hat{m}_{20}, \hat{m}_1, \hat{b}_1\right)$ such that:

$$\left(W_2^n(\hat{m}_{20}), W_1^n\left(\hat{m}_{20}, \hat{m}_1, \hat{b}_1\right), Y_1^n\right) \in \mathcal{T}_\epsilon^n\left(P_{W_2W_1Y_1}\right)$$

If there exists such a triple, the decoder estimates its respective message by $\hat{m}_1$. If there is no such triple or there is more than one, a decoding error is declared.

2. At the receiver $Y_2$, the message $m_1$ is known in advance; the decoder tries to find a unique pair $(\hat{m}_{20}, \hat{m}_{22})$ for which there exists some $\bar{b}_1$ such that:

$$\left(W_2^n(\hat{m}_{20}), V_2^n(\hat{m}_{20}, \hat{m}_{22}), W_1^n\left(\hat{m}_{20}, m_1, \bar{b}_1\right), V_1^n\left(\hat{m}_{20}, \hat{m}_{22}, m_1, \bar{b}_1\right), Y_2^n\right) \in \mathcal{T}_\epsilon^n\left(P_{W_2V_2W_1V_1Y_2}\right)$$

If there exists such a pair, the decoder estimates its respective messages by $(\hat{m}_{20}, \hat{m}_{22})$. If there is no such pair or there is more than one, it declares a decoding error.

*Analysis of probability of error:* Let $0 < \epsilon < p_{min}\left(P_{W_2V_2W_1V_1X_2X_1}\right)$. Given the triple $(m_{20}, m_{22}, m_1) \in [1:2^{nR_{20}}] \times [1:2^{nR_{22}}] \times [1:2^{nR_1}]$, the error events $E_1, E_2, \dots, E_9$ are defined as follows:

- Encoding errors:

$$E_1 \triangleq \left\{\begin{matrix} \forall\, b_1 \in [1:2^{nB_1}]: \\ \left(W_2^n(m_{20}), V_2^n(m_{20}, m_{22}), W_1^n(m_{20}, m_1, b_1)\right) \notin \mathcal{T}_\epsilon^n\left(P_{W_2V_2W_1}\right) \end{matrix}\right\}$$

$$E_2 \triangleq \left\{\begin{pmatrix} W_2^n(m_{20}), V_2^n(m_{20}, m_{22}), \\ W_1^n(m_{20}, m_1, b_1^T), V_1^n(m_{20}, m_{22}, m_1, b_1^T), \\ X_2^n(m_{20}, m_{22}), X_1^n(m_{20}, m_{22}, m_1, b_1^T) \end{pmatrix} \notin \mathcal{T}_\epsilon^n\left(P_{W_2V_2W_1V_1X_2X_1}\right)\right\}$$

- Decoding errors at receiver 1:

$$E_3 \triangleq \left\{\left(W_2^n(m_{20}), W_1^n(m_{20}, m_1, b_1^T), Y_1^n\right) \notin \mathcal{T}_\epsilon^n\left(P_{W_2W_1Y_1}\right)\right\}$$

$$E_4 \triangleq \left\{\begin{matrix} \exists\left(\tilde{m}_1, \tilde{b}_1\right) \neq (m_1, b_1^T): \\ \left(W_2^n(m_{20}), W_1^n\left(m_{20}, \tilde{m}_1, \tilde{b}_1\right), Y_1^n\right) \in \mathcal{T}_\epsilon^n\left(P_{W_2W_1Y_1}\right) \end{matrix}\right\}$$

$$E_5 \triangleq \left\{\begin{matrix} \exists\left(\tilde{m}_{20}, \tilde{m}_1, \tilde{b}_1\right), \tilde{m}_{20} \neq m_{20}: \\ \left(W_2^n(\tilde{m}_{20}), W_1^n\left(\tilde{m}_{20}, \tilde{m}_1, \tilde{b}_1\right), Y_1^n\right) \in \mathcal{T}_\epsilon^n\left(P_{W_2W_1Y_1}\right) \end{matrix}\right\}$$

- Decoding errors at receiver 2:

$$E_6 \triangleq \left\{\begin{pmatrix} W_2^n(m_{20}), V_2^n(m_{20}, m_{22}), \\ W_1^n(m_{20}, m_1, b_1^T), \\ V_1^n(m_{20}, m_{22}, m_1, b_1^T), Y_2^n \end{pmatrix} \notin \mathcal{T}_\epsilon^n\left(P_{W_2V_2W_1V_1Y_2}\right)\right\}, \quad E_7 \triangleq \left\{\begin{matrix} \exists\, \tilde{m}_{22} \neq m_{22}: \\ \begin{pmatrix} W_2^n(m_{20}), V_2^n(m_{20}, \tilde{m}_{22}), \\ W_1^n(m_{20}, m_1, b_1^T), \\ V_1^n(m_{20}, \tilde{m}_{22}, m_1, b_1^T), Y_2^n \end{pmatrix} \in \mathcal{T}_\epsilon^n\left(P_{W_2V_2W_1V_1Y_2}\right) \end{matrix}\right\}$$

$$E_8 \triangleq \left\{\begin{matrix} \exists\left(\tilde{m}_{22}, \tilde{b}_1\right), \tilde{m}_{22} \neq m_{22}, \tilde{b}_1 \neq b_1^T: \\ \begin{pmatrix} W_2^n(m_{20}), V_2^n(m_{20}, \tilde{m}_{22}), \\ W_1^n\left(m_{20}, m_1, \tilde{b}_1\right), \\ V_1^n\left(m_{20}, \tilde{m}_{22}, m_1, \tilde{b}_1\right), Y_2^n \end{pmatrix} \in \mathcal{T}_\epsilon^n\left(P_{W_2V_2W_1V_1Y_2}\right) \end{matrix}\right\}, \quad E_9 \triangleq \left\{\begin{matrix} \exists\left(\tilde{m}_{20}, \tilde{m}_{22}, \tilde{b}_1\right), \tilde{m}_{20} \neq m_{20}: \\ \begin{pmatrix} W_2^n(\tilde{m}_{20}), V_2^n(\tilde{m}_{20}, \tilde{m}_{22}), \\ W_1^n\left(\tilde{m}_{20}, m_1, \tilde{b}_1\right), \\ V_1^n\left(\tilde{m}_{20}, \tilde{m}_{22}, m_1, \tilde{b}_1\right), Y_2^n \end{pmatrix} \in \mathcal{T}_\epsilon^n\left(P_{W_2V_2W_1V_1Y_2}\right) \end{matrix}\right\}$$

Then, the error probability of the code denoted by $P_e^n$ is given as follows:

$$\begin{aligned} P_e^n &= \frac{1}{2^{n(R_1+R_{20}+R_{22})}} \sum_{m_1, m_{20}, m_{22}} Pr^*(E_1 \cup \dots \cup E_9) \\ &\le \frac{1}{2^{n(R_1+R_{20}+R_{22})}} \sum_{m_1, m_{20}, m_{22}} \begin{pmatrix} Pr^*(E_1) + Pr^*(E_2|E_1^c) + Pr^*(E_3|E_2^c) + Pr^*(E_4) + Pr^*(E_5) \\ +Pr^*(E_6|E_2^c) + Pr^*(E_7) + Pr^*(E_8) + Pr^*(E_9) \end{pmatrix} \end{aligned} \tag{A~16}$$

where $Pr^*(.) \triangleq Pr(.|m_{20}, m_{22}, m_1)$ and $A^c$ denotes the complement of the set $A$. Now, we bound the summands in (A~16). In the following analysis, $O(\epsilon)$ denotes a deterministic function of $\epsilon$ with $O(\epsilon) \to 0$ as $\epsilon \to 0$. Regarding the error event $E_1$, using the covering lemma [110], we directly derive $Pr^*(E_1) \to 0$ provided that:

$$B_1 > I(V_2; W_1|W_2) + O(\epsilon) \tag{A~17}$$

According to the factorization (A~15), the Markov chain $W_1, V_1 \to W_2, V_2 \to X_2$ holds. Considering this point and using the Markov lemma [110], we have $Pr^*(E_2|E_1^c) \to 0$. Also, since $W_2, V_2, W_1, V_1 \to X_1, X_2 \to Y_1$ form a Markov chain and there is no encoding error given $E_2^c$, we have $Pr^*(E_3|E_2^c) \to 0$. Then consider the error events $E_4$ and $E_5$. Note that the codewords $W_2^n$ and $W_1^n$ form a superposition structure. Using the packing lemma [110], we can readily derive $Pr^*(E_4) \to 0$ provided that:

$$R_1 + B_1 < I(W_1; Y_1|W_2) - O(\epsilon) \tag{A~18}$$

and also, $Pr^*(E_5) \to 0$ if:

$$R_{20}+R_1+B_1<I(W_2,W_1;Y_1)-O(\epsilon)$$

(A~19)

Now, let us evaluate the decoding errors at the primary receiver. Similar to $E_3$, we have $Pr^*(E_6|E_2^c)\to 0$. To bound the probability of the error events[2] $E_7,E_8$ and $E_9$, we use the general formula derived in our previous paper [7, p. 12]. Based on this general formula we have:

$$\sum_{E_l}\begin{pmatrix}\text{Rates with respect to incorrect decoded}\\ \text{messages and bin indices}\end{pmatrix}<I_{E_l}-O(\epsilon),\qquad l=7,8,9,\qquad \Rightarrow\qquad Pr^*(E_l)\to 0$$

(A~20)

where,

$$I_{E_l}\triangleq\sum_{\substack{A^n\ is\ incorrctly\\ decoded\ in\ E_l}}H\left(A\middle|\left\{B^n\begin{array}{c}B:\\ \text{is a cloud center for}\end{array}A^n\right\}\right)-H\left(\left\{C^n\begin{array}{c}C:\\ \text{is incorrectly decoded}\\ \text{in }E_l\end{array}\right\}\middle|\left\{D^n\begin{array}{c}D:\\ \text{is correctly decoded}\\ \text{in }E_l\end{array}\right\},Y_2\right)$$

(A~21)

As mentioned in [7], this general formula can be used to evaluate the decoding error probabilities at the receivers for any given achievability scheme for a memoryless network. Now using the graphical illustration in Fig. 16, which depicts the superposition structures among the generated codewords, one can easily check that for vanishing the probability of the events $E_7,E_8,E_9$, it is required:

$$\begin{aligned}R_{22}&<I(V_2,V_1;Y_2|W_2,W_1)+I(V_2;W_1|W_2)-O(\epsilon)\\ R_{22}+B_1&<I(V_2,W_1,V_1;Y_2|W_2)+I(V_2;W_1|W_2)-O(\epsilon)\\ R_{20}+R_{22}+B_1&<I(W_2,V_2,W_1,V_1;Y_2)+I(V_2;W_1|W_2)-O(\epsilon)\end{aligned}$$

(A~22)

By collecting (A~17)-(A~19) and (A~22), we derive the following achievable rate region:

$$\bigcup_{\substack{P_{W_2V_2W_1V_1}\\ \times P_{X_2|V_2W_2}P_{X_1|X_2V_1W_1V_2W_2}}}\left\{\begin{array}{r}(R_1,R_2)\in\mathbb{R}_+^2:\ \exists\,(B_1,R_1,R_{20},R_{22})\in\mathbb{R}_+^4:\\ R_2=R_{20}+R_{22}\\ B_1\geq I(V_2;W_1|W_2)\\ R_1+B_1\leq I(W_1;Y_1|W_2)\\ R_{20}+R_1+B_1\leq I(W_2,W_1;Y_1)\\ R_{22}\leq I(V_2,V_1;Y_2|W_2,W_1)+I(V_2;W_1|W_2)\\ R_{22}+B_1\leq I(V_2,W_1,V_1;Y_2|W_2)+I(V_2;W_1|W_2)\\ R_{20}+R_{22}+B_1\leq I(W_2,V_2,W_1,V_1;Y_2)+I(V_2;W_1|W_2)\end{array}\right\}$$

(A~23)

Now, similar to the arguments presented in Subsection III.A.4 to derive the rate region (III~58) for the CRC with strong interference at the primary receiver, one can deduce that a suitable choice for the rate region (A~23) is to set $V_1\equiv X_1$ and $V_2\equiv X_2$. Then by applying a simple Fourier-Motzkin elimination to remove $B_1,R_{20},R_{22}$ from the description (A~23), we obtain (III~129). The proof is thus complete. ■

## ACKNOWLEDGEMENT

The author would like to appreciate Dr. H. Estekey, head of School of Cognitive Sciences, IPM, Tehran, Iran, and Dr. R. Ebrahimpour, faculty member at School of Cognitive Sciences, for their kindly support of the first author during this research. He also greatly thanks his mother whose love made this research possible for him. Lastly, F. Marvasti is acknowledged whose editing comments improved the language of this work.

[2] The decoding error events $E_4$ and $E_5$ at the receiver $Y_1$ could also be evaluated using this general formula.